\documentclass[reprint, pra,twocolumn,superscriptaddress,floatfix,nofootinbib,amsmath,amssymb]{revtex4-1}

\usepackage[T1]{fontenc}
\usepackage{float}
\usepackage{graphicx}
\usepackage{bm}
\usepackage{braket}
\usepackage{xcolor}
\usepackage{amsfonts}
\usepackage{comment}
\usepackage{dirtytalk}
\usepackage{bibunits}

\usepackage{url}
\usepackage{bbold}
\usepackage{realboxes}
\usepackage{minitoc}

\usepackage{enumitem}
\usepackage[toc,page,titletoc]{appendix}

\usepackage{tocloft}
\definecolor{dark-red}{rgb}{0.4,0.15,0.15}
\definecolor{dark-blue}{rgb}{0.15,0.15,0.4}
\definecolor{medium-blue}{rgb}{0,0,0.5}
\usepackage[hypertexnames=false]{hyperref}
\usepackage[all]{hypcap}%fix problems with figures and tables of the 'hyperref' package
\hypersetup{
    colorlinks, linkcolor={dark-red},
    citecolor={dark-blue}, urlcolor={medium-blue}
}
\usepackage{cleveref}

\usepackage{algpseudocode}

\usepackage{listings}
\usepackage{xcolor}

\definecolor{bkgd}{RGB}{240,242,246}
\definecolor{ceruleanblue}{rgb}{0.16, 0.32, 0.75}
\definecolor{orange-red}{rgb}{1.0, 0.27, 0.0}
\definecolor{anotherblue}{RGB}{37,92,243}
\definecolor{blackblue}{RGB}{46,60,85}
\definecolor{goldyellow}{RGB}{199,146,12}
\lstdefinestyle{altstyle2}{
    backgroundcolor=\color{bkgd},
    basicstyle=\ttfamily\footnotesize\color{blackblue},
    breakatwhitespace=false,
    breaklines=true,
    captionpos=b,
    commentstyle=\color{goldyellow},
    keepspaces=true,
    keywordstyle=\color{orange-red},
    language=Python,
    numbersep=5pt,
    numberstyle=\tiny\color{ceruleanblue},
    showspaces=false,
    showstringspaces=false,
    showtabs=false,
    stringstyle=\color{anotherblue},
    tabsize=2
}
\begin{document}

\doparttoc % Tell to minitoc to generate a toc for the parts
\faketableofcontents 
\part{} % Start the document part

\title{Learning Noise via Dynamical Decoupling of Entangled Qubits
}

\author{Trevor McCourt}
\affiliation{Department of Electrical Engineering and Computer Science, Massachusetts Institute of Technology, Cambridge, MA 02139, USA
}
\affiliation{Department of Physics, Co-Design Center for Quantum Advantage, Massachusetts Institute of Technology, Cambridge, Massachusetts 02139, USA}

\author{Charles Neill}
\affiliation{Google Quantum AI, Santa Barbara, CA
}

\author{Kenny Lee}
\affiliation{Google Quantum AI, Santa Barbara, CA
}

\author{Chris Quintana}
\affiliation{Google Quantum AI, Santa Barbara, CA
}

\author{Yu Chen}
\affiliation{Google Quantum AI, Santa Barbara, CA
}

\author{Julian Kelly}
\affiliation{Google Quantum AI, Santa Barbara, CA
}

\author{V. N. Smelyanskiy}
\affiliation{Google Quantum AI, Santa Barbara, CA
}

\author{M. I. Dykman}
\affiliation{%Google Quantum AI, Santa Barbara, CA
Department of Physics and Astronomy, Michigan State University, East Lansing, MI 48824, USA}

\author{Alexander Korotkov}
\affiliation{Google Quantum AI, Santa Barbara, CA
}

\author{Isaac L. Chuang}
\affiliation{Department of Electrical Engineering and Computer Science, Massachusetts Institute of Technology, Cambridge, MA 02139, USA
}
\affiliation{Department of Physics, Massachusetts Institute of Technology, Cambridge, MA 02139, USA
}

\author{A. G. Petukhov}
\affiliation{Google Quantum AI, Santa Barbara, CA
}

\date{\today}

%\begin{bibunit}[ieeetr]

%TC:ignore
\begin{abstract}
Noise in entangled quantum systems is difficult to characterize due to many-body effects involving multiple degrees of freedom. This noise poses a challenge to quantum computing, where two-qubit gate performance is critical. Here, we develop and apply multi-qubit dynamical decoupling sequences that characterize noise that occurs during two-qubit gates. In our superconducting system comprised of Transmon qubits with tunable couplers, we observe noise that is consistent with flux fluctuations in the coupler that simultaneously affects both qubits and induces noise in their entangling parameter. The effect of this noise on the qubits is very different from the well-studied single-qubit dephasing. Additionally, steps are observed in the decoupled signals, implying the presence of non-Gaussian noise. 

\end{abstract}
                              
%TC:endignore

\maketitle

Producing interesting, large-scale, quantum dynamics in engineered systems is being made increasingly possible by the advancement of superconducting qubits. 
Transmon qubits that use frequency tunable couplers to realize inter-qubit interactions have been successful at this task in the areas of quantum simulation \cite{neill_2020, zhang_2020, tan2021}, 
quantum chemistry \cite{rubin_2020}, and theoretical computer science \cite{kelly_2015, chen_2021, Arute2019, huang2021quantum}. 
Imperative to this is the ability to generate entanglement using high-fidelity two-qubit gates \cite{oliver_2021, foxen_2020}. 
As control of these gates is improved, their performance will start to become limited by system-environment interaction. The characterization and eventual mitigation of this noise producing interaction is therefore critical to continual forward progress. 

Traditionally, low-frequency noise characterization in qubits has been dedicated to the study of single-qubit dephasing noise. 
This is modeled as either a qubit coupling to external quantum degrees of freedom or as classical stochastic fluctuations in the qubit frequency \cite{Uhrig2008}. Most often, the noise is assumed to have Gaussian statistics. In this Gaussian scenario, sophisticated tools based on dynamical decoupling have been developed 
to characterize the power spectral density of the noise \cite{Cywinski2008, Bylander2011, Biercuk2011}. 
There have also been efforts to characterize noise outside of this regime. 
These have been focused on measuring the higher-order moments of single-qubit non-Gaussian dephasing \cite{Norris2016, Sung2019} as well as 
characterizing spatially correlated Gaussian dephasing noise \cite{viola_2017, euros_2016, euros_2019}.
\begin{figure*}
    \centering
    \includegraphics[width=\textwidth]{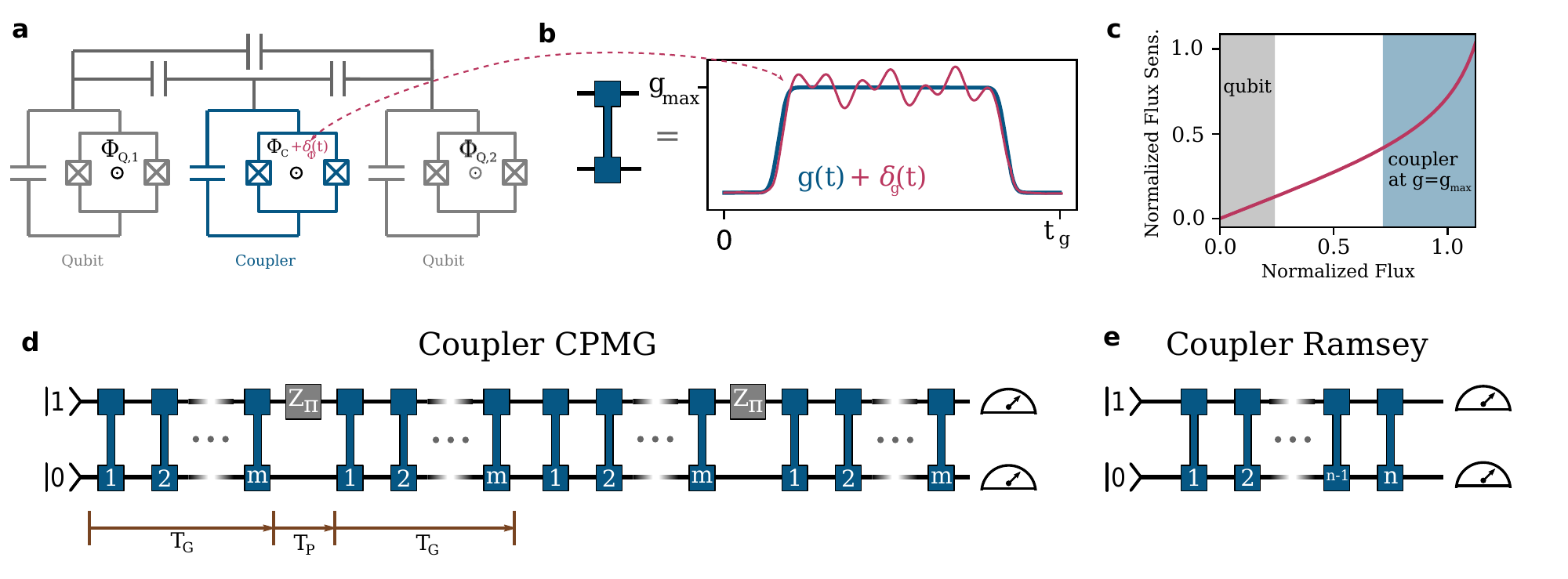}
    \caption{\textbf{Circuits for entangled noise metrology a} Simplified circuit diagram for two qubits and the tunable coupler. The qubit frequencies $\omega_j$ are modulated by changing $\Phi_{Q,j}$. The coupler frequency is changed significantly during two-qubit gates via $\Phi_C$. \textbf{b} Schematic of the time-dependent coupling $g(t)$ enacted during two-qubit gates. The coupler flux noise $\delta_{\Phi}(t)$ generates coupling fluctuations $\delta_g(t)$ according to  Eq.~\ref{eqn:flux_sens_xi}. \textbf{c} Flux sensitivity $\chi$ (Eq.~\ref{eqn:flux_sens_chi}) vs external flux. The qubits are generally operated at frequencies with much lower flux sensitivity than the coupler. \textbf{d} Circuit diagram showing the Coupler CPMG sequence. Shown here are $n=2$ repetitions of a pulse sequence involving $2m$ two-qubit gates that are separated by a qubit frequency $\pi$ pulse. The two-qubit gates serve to expose the qubits to g-noise, which is refocused by the frequency pulse. The decay of the pseudo-qubit $\langle \sigma_z \rangle$ observable is measured at the end of the circuit, which can be used to characterize the noise. See supplementary material section \ref{sec:add_cpmg} for further examples. \textbf{e} Circuit diagram showing the Coupler Ramsey sequence involving $n$ two-qubit gates, which can be used to measure the response of the qubits to $g$-noise in the absence of refocusing pulses.}
    \label{fig:circuit_pulse_diagrams}
\end{figure*}

Studies of single-qubit dephasing may be sufficient to understand the behavior of small systems involving only one or a few qubits. However, large systems have many degrees of freedom, and therefore many channels through which noise can enter. For example, noise that occurs during two-qubit gates may lead to collective noise that affects two qubits simultaneously. Understanding these collective noise mechanisms in the context of quantum computing will be important for implementing NISQ algorithms in the near term and building a fault-tolerant quantum computer in the long term. The difficulty in characterizing noise in larger systems stems from the fact that measurement of a particular kind of noise may be confounded by competing error mechanisms, as larger systems are generally more difficult to control precisely than the small ones.

In this Letter, we characterize noise that occurs during two-qubit gates. The gate we study is performed using a tunable coupler that modulates the qubit-qubit coupling.  Our key observation is that the primary source of noise is frequency fluctuations of this coupler. These fluctuations lead to noise in the entangling parameter $g$, the coupling strength between the two qubits. The noise is therefore turned on during a gate operation and affects two qubits simultaneously, in qualitative distinction from single-qubit dephasing. We show that this fundamentally two-qubit noise can be studied by driving pairs of qubits through two-qubit pulse sequences with interleaved coupler and qubit frequency control. We find that in many samples this noise is composed of Gaussian $1/f$ noise, similar to the noise dominating single qubit dephasing, and a signal from a few random telegraph fluctuators with correlation times on the order of 100 $\mu$s. These findings are significant because both the collective and non-Gaussian nature of the observed noise demand new error mitigation techniques. Additionally, the clean signatures of non-Gaussian noise that we see are a significant departure from what is typically assumed and observed in condensed matter systems, where Gaussian $1/f$ noise is ubiquitous \cite{dutta_1981, Schoelkopf1998, An2019, Yoshihara2014}.

We begin by introducing the theory of flux noise entering through the coupler and a technique for measuring it. We then present the measurement results and show that while they match well what would be expected for coupler flux noise, they do not agree well with Gaussian theory. Finally, we generalize to a non-Gaussian model of the noise and validate it with further experiments. 

The single excitation subspace of two qubits is spanned by the states $\ket{01}$ and $\ket{10}$ and forms a pseudo-qubit with the Hamiltonian
\begin{equation}
\label{Eq:SpinHam}
H=\frac{1}{2}\left(\omega(t) + \delta\omega(t) \right)\sigma_z+\left(g(t)+\delta g(t)\right)\sigma_{x} \: ,
\end{equation}
where $\sigma_z=\ket{01}\bra{01} - \ket{10}\bra{10}$ and $\sigma_x=\ket{01}\bra{10} + \ket{10}\bra{01}$. Here $\omega(t)$ and $\delta\omega(t)$ are the control and noise contributions to the difference of the qubit frequencies, respectively, while $g(t)$ and $\delta g(t)$ are the control and noise contributions to the inter-qubit coupling. 

During two-qubit gates the two qubits are on resonance, $\omega(t) = 0$. In this case, $\delta\omega(t)$ and $g(t)+\delta g(t)$ can be considered respectively as $z$ and $x$ components of an effective magnetic field. The Bloch vector of our effective two-level system undergoes Larmor precession around the instantaneous axis, which is almost parallel to the $x$-field, with the instantaneous Larmor frequency given by
\begin{equation}
\label{Eq:Larmor}
\omega_L(t)
\simeq 2g(t)+2\delta g(t)+\frac{\delta\omega^2(t)}{4g(t)} \: .
\end{equation}

From this, we can see that coupler noise will dominate during two-qubit gates: $\delta g(t)$ shows up to first order in the dynamics while $\delta\omega(t)$ only shows up to second-order and is suppressed by a factor of $g(t)$.

\begin{figure*}
    \centering
    \includegraphics[width=\textwidth]{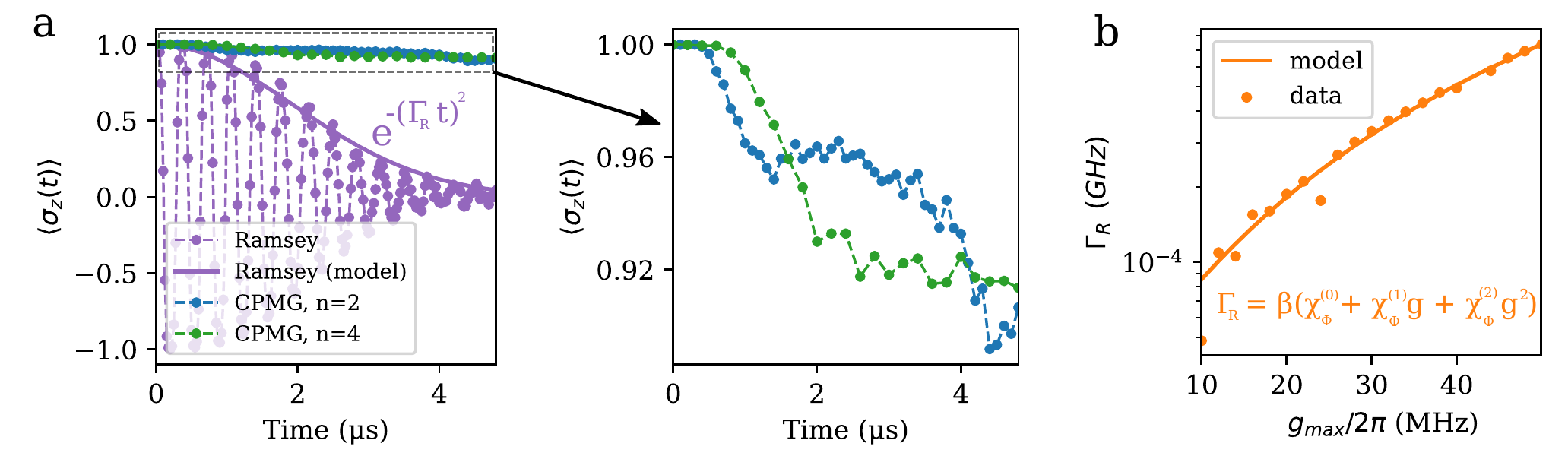}
    \caption{\textbf{Experimentally observed Ramsey and CPMG dynamics. a} 
    Comparing Coupler Ramsey decay of normlalized population difference (Eq.~\ref{eqn:norm_pop_diff}) with $g_{max} = 30 \text{MHz}$ to decay under $n=2$ and $n=4$ Coupler CPMG sequences. The x-axis is total evolution time, $t=n t_g$ for Ramsey and $t=2mnt_g$ for CPMG. The duration of a fixed $n$ CPMG sequence is modified by changing $m$. We see that the CMPG sequences effectively mitigate most of the decoherence, suggesting that most of the noise power is at low frequencies. The Gaussian shape of the Ramsey decay envelope is typical of 1/f-type noise (see Eq.~\ref{eqn:ramsey_decay_gauss}). When observed in detail, the CPMG decay envelopes display behavior not predicted by Gaussian theory. Increasing the number of CPMG pulses does not increase noise protection as predicted by Eq.~\ref{eqn:cpmg_decay_gauss}; the curves braid and have steps. All data points are the average of 10000 samples. \textbf{b} Ramsey decay rate $\Gamma_R$ vs $g_{\text{max}}$. We see that the decay rate is strongly dependent on $g_{\text{max}}$, crossing an order of magnitude in $30$ MHz. The $g_{\text{max}}$-dependence is well-predicted by Eq.~\ref{eqn:flux_sens_xi} given typical circuit parameters. }
    \label{fig:free_and_echoed_decay}
\end{figure*}

Coupler noise physically results from coupler frequency fluctuations. In our tunable coupler system depicted in Fig.~\ref{fig:circuit_pulse_diagrams}, the qubit frequencies $\omega_q$ and the coupler frequencies $\omega_c$ are controllable via the external fluxes, $\Phi_q$ and $\Phi_c$, respectively and the relation between $\omega$ and $\Phi$ is $\omega \approxeq \omega_{\text{max}}\sqrt{\left |\cos{\left( \frac{\pi \Phi}{\Phi_0}\right)}\right |}$,  where $\Phi_0$ is the flux quantum. The coupling $g$ developed between two qubits that are on resonance at $\omega_q$ is given by \cite{Yan2018}
\begin{equation}
\label{Eq:gmain}
    g \approxeq \left( k_d - k^2 \frac{\omega_q^2}{\omega_c^2 - \omega_q^2}\right)\frac{\omega_q}{2} \: ,
\end{equation}
where $k$ and $k_d$ are the indirect and direct coupling efficiencies that are functions of circuit parameters  (see supplementary material section \ref{sec:flux_noise}). The pseudo-qubit defined in Eq.~\ref{Eq:SpinHam} is therefore completely controllable via low-frequency manipulation of the qubit and coupler flux biases and no microwave control is necessary to implement dynamical decoupling of the entangled qubits. Another characteristic feature of our method is the ability to post-select experimental outcomes belonging only to the pseudo-qubit subspace and separate the processes of collective dephasing from those of energy relaxation. 

Fluctuations in $\Phi$ lead to fluctuations in frequency, i.e. to flux noise, which is ubiquitous in SQUIDs \cite{koch_1983}. During gates, the sensitivity of the coupler frequency to flux noise is substantially larger than that of the qubit, see Fig.~\ref{fig:circuit_pulse_diagrams} b. Noise in the coupler frequency leads to fluctuations in $g$. The fluctuation $\delta g(t)$ in the Hamiltonian \eqref{Eq:SpinHam} can be expressed through coupler flux fluctuations $\delta \Phi_c (t)$ as follows 
\begin{equation}\label{eqn:flux_sens_xi}
    \delta g(t)=2\pi \tilde\chi_\Phi(g) \delta\Phi_c(t)=\lambda (g) \xi(t),
\end{equation}
where the flux sensitivity of $g$ is defined as
\begin{equation}\label{eqn:flux_sens_chi}
   \tilde\chi_\Phi =\frac{1}{2\pi}\left|\frac{dg}{d\Phi_c}\right|\simeq
   \chi_\Phi^{(0)}+\chi_\Phi^{(1)}g+\chi_\Phi^{(2)}g^2 .
\end{equation}
Here $\xi(t)$ is a dimensionless classical random variable modeling flux fluctuations with characteristic amplitude $\delta\Phi_m$, and 
$\lambda(g)=2\pi\tilde\chi_\Phi(g) \delta\Phi_m$ is the amplitude of $g$-noise. It can be shown (see supplementary material section \ref{sec:flux_noise}) that in the studied parameter range the quadratic dependence of  $\tilde\chi_\Phi(g)$, displayed in Eq.~\ref{eqn:flux_sens_chi}, follows directly from Eq.~\ref{Eq:gmain}. 

The effect of $g$-noise on the pseudo-qubit defined in Eq.~\ref{Eq:SpinHam} may be characterized using what we call the Coupler CPMG pulse sequence. In this sequence, the pseudo-qubit is initialized in the state $\ket{01}$ via a microwave pulse. It is then exposed to $n$ repetitions of a spin echo-like pulse sequence \cite{Hahn1950}, each of which consists of a fast $\pi$-rotation around $z$-axis ($\sigma_z$ $\pi$-pulse) buffered before and after by exposure to $g$-noise for time $T_G$. The $\sigma_z$ pulse has the effect of refocusing the $\sigma_x$ $g$-noise. This exposure to $g$-noise is accomplished by $m$ repetitions of a Floquet-calibrated \cite{neill_2020} two-qubit gate with duration $t_g$ for which $\omega =0$ and $|g| > 0$, see Fig.~\ref{fig:circuit_pulse_diagrams} b. There are a total of $2m$ two-qubit gates between refocusing pulses; the total time between refocusing pulses is therefore $2m t_g$. After the $n$ echo sequences are completed, we can measure the pseudo-qubit observable $\langle \sigma_z \rangle$, which will decay due to $g$-noise. Studying the decay of this observable will reveal the character of the noise. The Coupler CPMG pulse sequence is shown in Fig.~\ref{fig:circuit_pulse_diagrams} d. This pulse sequence is analogous to standard, single qubit CPMG \cite{carr1954, gill1958}, with the main difference being that it takes place in the z-y plane of the Bloch sphere instead of the x-y plane, so the direction of refocusing pulses and measurements must be adjusted accordingly. It is also desirable to observe the $\sigma_z$ decay due to $g$-noise in the absence of the refocusing pulses. This may be done using the Coupler Ramsey pulse sequence, see Fig.~\ref{fig:circuit_pulse_diagrams} e.

The statistics of $\xi(t)$ dictate what type of decay we expect to see during these sequences. A common assumption is $\xi(t)$ is a Gaussian random process, which means that $\xi(t)$ has a jointly Gaussian distribution at all times. In the typical case where $\xi(t)$ is Gaussian  $1/f$ noise, for decay under the Coupler Ramsey sequence we would expect (up to logarithmic corrections, see supplementary material section \ref{sec:gaussian_decay}),
\begin{equation}\label{eqn:ramsey_decay_gauss}
    \langle \sigma_z(t) \rangle \approxeq e^{-(\Gamma_R t)^2} \cos{\left( G t\right)}, \: \Gamma_R \propto \lambda \: ,
\end{equation}
where $G$ is the coherent swap frequency. In the case of decay under an n-pulse CPMG sequence, 
\begin{equation}\label{eqn:cpmg_decay_gauss}
    \langle \sigma_z(t) \rangle \approxeq e^{-(\Gamma_C t)^2}, \: \Gamma_C \propto \frac{\lambda}{\sqrt{n}} \: .
\end{equation}

We experimentally characterize  $g$-noise on our superconducing qubit device \cite{Arute2019} by executing these sequences. We measure the observable
\begin{equation}\label{eqn:norm_pop_diff}
     \frac{\langle \sigma_z \rangle}{\langle I \rangle} = \frac{\bra{01} \rho(t) \ket{01} - \bra{10} \rho(t) \ket{10}}{\bra{01} \rho(t) \ket{01} + \bra{10} \rho(t) \ket{10}} \: ,
\end{equation}
as a function of time, number of CPMG cycles, and maximum coupling $g_{\text{max}}$. This normalization of $\langle \sigma_z \rangle$ eliminates the effect of $T_1$ noise in relevant cases, see supplementary material section \ref{sec:shaprio_loginov}.  We can compare the shapes of the measured decay envelopes with Eqs.~\ref{eqn:ramsey_decay_gauss} and \ref{eqn:cpmg_decay_gauss}, and the $g$-dependence of decay rates with Eq.~\ref{eqn:flux_sens_xi} to test the theory that our device is susceptible to Gaussian noise entering through the flux bias during two-qubit gates.

As can be seen in Fig.~\ref{fig:free_and_echoed_decay} a, the experimentally measured Ramsey decay envelopes are well predicted by Gaussian $1/f$ noise. Additionally, as shown in Fig.~\ref{fig:free_and_echoed_decay} b, the scaling of the Gaussian decay rate with $g_{\text{max}}$ agrees with the form of Eq.~\ref{eqn:flux_sens_xi}. Notably, the decay rate increases by an order of magnitude as $g_{\text{max}}$ is increased from 10 to 50 MHz, suggesting that this coupler noise heavily exceeds single qubit dephasing as an error mechanism during gates with large coupling, as predicted by Eq.~\ref{Eq:Larmor}. Additionally, the flux sensitivity function extracted matches the theory well. From the data we extract a value of $\chi_\Phi^{(2)}/\chi_\Phi^{(1)} \approxeq 0.078$ ns, while a purely theoretical calculation using typical circuit parameters yields $\chi_\Phi^{(2)}/\chi_\Phi^{(1)} \approxeq 0.08$ ns. This excellent agreement with theory strongly suggests that noise during two-qubit gates is dominated by flux noise in the coupler, as hypothesized.

The CPMG envelopes decay significantly slower than the Ramsey envelopes as would be predicted by Gaussian theory. However, as shown in Fig.~\ref{fig:free_and_echoed_decay} the details of these curves deviate significantly from what would be predicted by Gaussian $1/f$ noise. While Eq.~\ref{eqn:cpmg_decay_gauss} predicts smooth decay, we see very clear steps in the decay curves. Additionally, the model predicts that the decay rate $\Gamma_C$ should decrease proportionately to $\frac{1}{\sqrt{n}}$. This is not seen at all: the two curves "braid" and decay at the same rate.

\begin{figure}
    \centering
    \includegraphics[width=0.8\columnwidth]{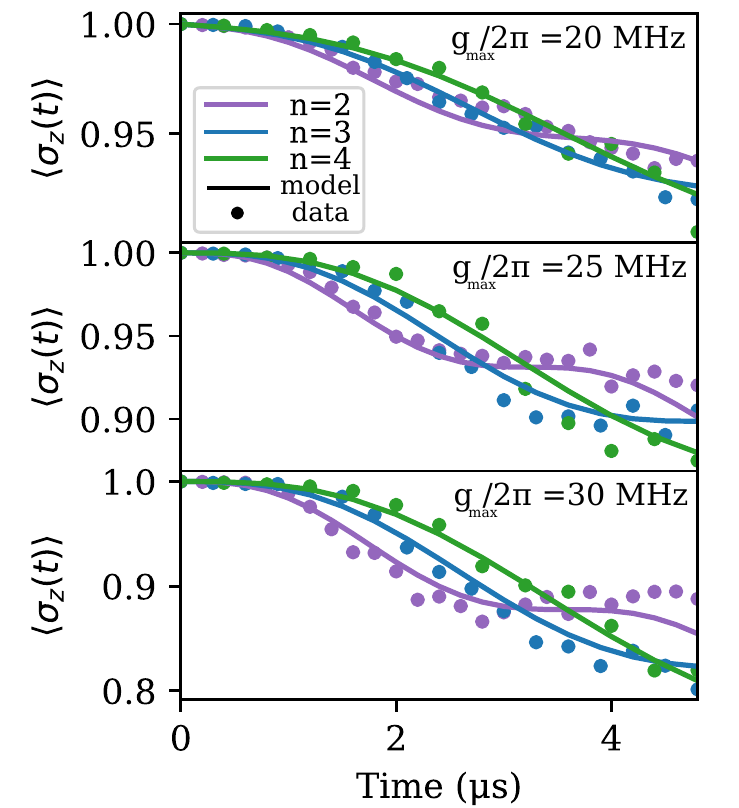}
    \caption{\textbf{Braiding in the CPMG decay envelopes} Fitting a single-fluctuator model to CPMG decay envelopes (Eq.~\ref{eqn:cpmg_driven_decay_env}) for different values of $n$ and $g_{max}$. Each set of 3 curves is fit using only 2 parameters, $\gamma$ and $\lambda$. Good agreement is found between theory and experiment. Fits for more values of $n$ can be found in supplementary material section \ref{sec:many_fits}. Typical values of $t_c = \frac{1}{\gamma} \approxeq 50 \mu\text{s}$, $\frac{\lambda}{2 \pi} \approxeq 0.1 - 1 \text{MHz}$ (value depends strongly on $g$, see Fig.~\ref{fig:cpmg_trnsfr}), and $\Gamma_{\phi}^{-1} \approxeq 100 \mu s$. All data points are the average of 10000 samples.}
    \label{fig:braiding_figure}
\end{figure}

The steps in the CPMG curves are difficult for any Gaussian noise model to produce (see supplementary material section \ref{sec:gaussian_smoothness} for further discussion on this). Therefore, these steps are signatures of non-Gaussian noise in our system.

It is reasonable to suggest that this non-Gaussian noise is the result of a small number of strongly coupled random telegraph noise (RTN) fluctuators, since Gaussian $1/f$ noise may be produced via a superposition of a large number of weakly coupled fluctuators \cite{Hooge1994} (see supplementary material section \ref{sec:1_f_from_tele}).  The CPMG decay curve associated with single RTN fluctuator with correlation time $t_c = \frac{1}{\gamma}$ is \cite{Ramon2015, Galperin2006, Faoro2004}, see supplementary material section \ref{sec:direct_averaging},
\footnotesize
\begin{equation}\label{eqn:cpmg_driven_decay_env}
    \langle \sigma_z(t=2 m n t_g) \rangle  = \left.
  \begin{cases}
    e^{-2 m n \gamma t_g} \left( q \frac{\cosh{\left( n \alpha\right)}}{\cosh{\left( \alpha \right)}} + \sinh{\left( n \alpha \right)}\right), & \text{n odd} \\
    e^{-2 m n \gamma t_g} \left( q \frac{\sinh{\left( n \alpha\right)}}{\cosh{\left( \alpha \right)}} + \cosh{\left( n \alpha \right)}\right), & \text{n even} \\
  \end{cases}
  \right . %\}
\end{equation}
\normalsize
where
\begin{align}
    \begin{split}
        &q = -\frac{4 \lambda^2}{\Omega^2} + \frac{\gamma^2 \cosh{\left(2 m \Omega t_g \right)}}{\Omega^2} \: , \\
        &\sinh{\left( \alpha\right)} = \frac{\gamma}{\Omega} \sinh{\left(2 m \Omega t_g \right)} \: ,\\
    \end{split}
\end{align}
and $\Omega =\sqrt{\gamma^2 - 4 \lambda^2}$ is the associated Rabi frequency. $\lambda$ is taken to be the average value of $\lambda\left( g(t)\right)$ over a gate, see supplementary material section \ref{sec:shaprio_loginov} for justification. It has also been assumed that $T_P << T_G$ (as in Fig.~\ref{fig:circuit_pulse_diagrams} d), such that the CPMG cycle time is $T_C = 2 m t_g$. The form of the solution depends heavily on $\Omega$. If $\Omega$ is real, the solution is over-damped, and decays smoothly. If $\Omega$ is imaginary, the solution is under-damped and has oscillatory components.

We can validate this model by repeating the previous CPMG measurements for more values of $n$ and attempting to fit the data simultaneously. The results of this are shown in Fig.~\ref{fig:braiding_figure}. The decay envelopes are excellently described by a single, under-damped RTN fluctuator alongside single qubit white noise dephasing, which adds a simple exponential prefactor $e^{-\frac{\Gamma_{\phi}}{4} t}$ to Eq.~\ref{eqn:cpmg_driven_decay_env}, see supplementary material section \ref{sec:shaprio_loginov}.

In each case, the fit fluctuator is strongly in the under-damped regime, $2 \lambda > \gamma$. In this regime, as shown in supplementary material section \ref{sec:echo_strongly_coupled}, Eq.~\ref{eqn:cpmg_driven_decay_env} is well approximated by,
\begin{equation}
    \langle \sigma_z(t=2 m n t_g) \rangle \approxeq e^{-2 m n t_g \gamma} e^{n \epsilon \sin{\left(2 m t_g \omega\right)}} \: ,
\end{equation}
where $\Omega = i \overline{\omega}$ and $\epsilon = \frac{\gamma}{\overline{\omega}}$. For the data shown in Fig.~\ref{fig:braiding_figure} a, $\epsilon \approxeq \frac{1}{100}$. If the number of CPMG cycles is also modest, $n \epsilon << 1$, and this further reduces to,

\begin{equation}\label{eqn:low_echo_decay_env}
    \langle \sigma_z(t=2 m n t_g) \rangle \approxeq e^{-2 m n t_g \gamma} \left( 1 + n \frac{\gamma}{\overline{\omega}} \sin{\left( 2 m t_g \overline{\omega}\right)} \right) \: .
\end{equation}

In this form, the dynamics are much more clear. The decay envelope will generally follow exponential decay and will produce steps with frequency $\frac{\overline{\omega}}{n}$. The implication of this is that it is difficult to dynamically suppress the decoherence caused by this kind of noise: more than $\frac{1}{\epsilon}$ echo pulses are required in time $t$ to cause the trajectory to deviate significantly from exponential decay. This is significantly different than what would be expected for Gaussian $1/f$ noise for example, for which protection increases monotonically and smoothly with $n$.

\begin{figure}
    \centering
    \includegraphics[width=0.9\columnwidth]{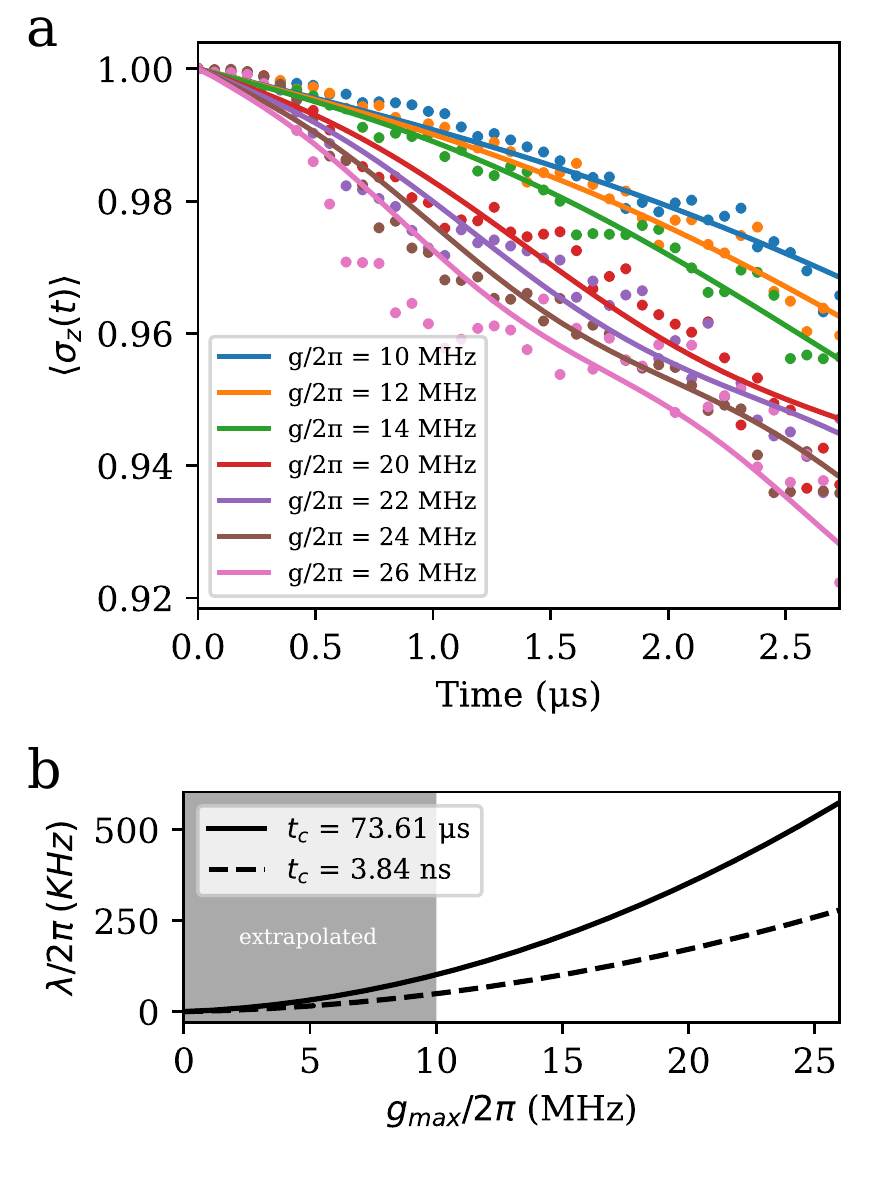}
    \caption{\textbf{Extracting the scaling of telegraph noise amplitude. a) } Experimental data (dots) vs fit model (lines) for $n=1$ CPMG sequences at various values of $g_{\text{max}}$. The fit value of $T_{\phi}$ is approximately  $90 \: \mu s$, which is a reasonable result for this device. \textbf{b )} The extracted noise amplitude $\lambda(g)$ for the two fluctuators. Note that the same function $\chi_{\Phi}$ was used for both fluctuators; the g-noise amplitudes were only allowed to differ by an overall scale.}
    \label{fig:cpmg_trnsfr}
\end{figure}

The scaling of the coupling strength of this single fluctuator with $g$ can be established by taking CPMG data on the same pair of qubits over a range of values of $g_{\text{max}}$. Fig.~\ref{fig:cpmg_trnsfr} shows the results of such an experiment. 

The data can be fit well with a model that includes one slow, strongly coupled fluctuator, white flux noise (emulated by a fast fluctuator), and single-qubit dephasing. The inclusion of white flux noise was critical to achieving a good fit, which is physically reasonable, as echo sequences do not suppress this kind of noise at all. The slow fluctuator has a correlation time of approximately $70 \mu s$. This is the strongly coupled, under-damped fluctuator that creates the steps seen in the data and the fit. The single-qubit dephasing rate represents white noise that does not scale with $g$, and the extracted value of $\frac{1}{\Gamma_{\phi}} \approxeq 90 \mu s$ is reasonable for this device. Ratio $\chi_\Phi^{(2)}/\chi_\Phi^{(1)} \approxeq 0.12$ ns for this data, which is also within expectation. 

It should be noted that this work alone is not enough to understand the physical origins of this non-Gaussian contribution to the noise. To do that, further study into the spatial dependence of the noise would have to be completed. Although this noise has been observed on several qubits in our system, this has not been studied systematically enough to determine if different qubits see fluctuators with similar parameters. Additionally, it would be impossible to tell if multiple pairs of qubits are seeing the same physical defect or just similar, independent defects with this kind of time-averaged, two-qubit measurement. These two situations may be discernible using time-averaged measurements taken after periodic pulse sequences on more than two qubits. This spatial dependence will be the subject of future study.

While the majority of this work was focused on the details of applying our technique to tunable-coupler Transmons, the basic methods transfer readily to other qubit architectures. As an example from trapped ion quantum computing, a similar technique could be used in the characterization of the effect of noise \cite{Hayes2012} on the coupling developed between ion electronic states during Mølmer-Sørensen gates \cite{Molmer1999}.

This work has elucidated the importance of studying noise via the physics of a specific device, especially as larger and larger quantum computers are built. Indeed, this approach is what allowed us to discover the dominant source of low-frequency noise that occurs during our two-qubit gates. Additionally, we have found very clear signatures of non-Gaussian, non $1/f$ noise in our solid-state device, which is quite atypical in the field. Further study of this kind of noise may reveal its physical origins, and yeild insight into better design, fabrication or control of quantum devices.

\begin{acknowledgments}
TM was supported in part by the U.S. Department of Energy, Office of Science, National Quantum Information Science Research Centers, Co-Design Center for Quantum Advantage under contract DE-SC0012704. TM would like to thank John Chiaverini and John Martyn for their comments on the manuscript.
\end{acknowledgments}

%TC:ignore
\bibliography{main}

\begin{thebibliography}{39}%
\makeatletter
\providecommand \@ifxundefined [1]{%
 \@ifx{#1\undefined}
}%
\providecommand \@ifnum [1]{%
 \ifnum #1\expandafter \@firstoftwo
 \else \expandafter \@secondoftwo
 \fi
}%
\providecommand \@ifx [1]{%
 \ifx #1\expandafter \@firstoftwo
 \else \expandafter \@secondoftwo
 \fi
}%
\providecommand \natexlab [1]{#1}%
\providecommand \enquote  [1]{``#1''}%
\providecommand \bibnamefont  [1]{#1}%
\providecommand \bibfnamefont [1]{#1}%
\providecommand \citenamefont [1]{#1}%
\providecommand \href@noop [0]{\@secondoftwo}%
\providecommand \href [0]{\begingroup \@sanitize@url \@href}%
\providecommand \@href[1]{\@@startlink{#1}\@@href}%
\providecommand \@@href[1]{\endgroup#1\@@endlink}%
\providecommand \@sanitize@url [0]{\catcode `\\12\catcode `\$12\catcode
  `\&12\catcode `\#12\catcode `\^12\catcode `\_12\catcode `\%12\relax}%
\providecommand \@@startlink[1]{}%
\providecommand \@@endlink[0]{}%
\providecommand \url  [0]{\begingroup\@sanitize@url \@url }%
\providecommand \@url [1]{\endgroup\@href {#1}{\urlprefix }}%
\providecommand \urlprefix  [0]{URL }%
\providecommand \Eprint [0]{\href }%
\providecommand \doibase [0]{http://dx.doi.org/}%
\providecommand \selectlanguage [0]{\@gobble}%
\providecommand \bibinfo  [0]{\@secondoftwo}%
\providecommand \bibfield  [0]{\@secondoftwo}%
\providecommand \translation [1]{[#1]}%
\providecommand \BibitemOpen [0]{}%
\providecommand \bibitemStop [0]{}%
\providecommand \bibitemNoStop [0]{.\EOS\space}%
\providecommand \EOS [0]{\spacefactor3000\relax}%
\providecommand \BibitemShut  [1]{\csname bibitem#1\endcsname}%
\let\auto@bib@innerbib\@empty
%</preamble>
\bibitem [{\citenamefont {Neill}\ \emph {et~al.}(2021)\citenamefont {Neill},
  \citenamefont {McCourt}, \citenamefont {Mi}, \citenamefont {Jiang},
  \citenamefont {Niu}, \citenamefont {Mruczkiewicz}, \citenamefont {Aleiner},
  \citenamefont {Arute}, \citenamefont {Arya}, \citenamefont {Atalaya},\ and\
  \citenamefont {et~al.}}]{neill_2020}%
  \BibitemOpen
  \bibfield  {author} {\bibinfo {author} {\bibfnamefont {C.}~\bibnamefont
  {Neill}}, \bibinfo {author} {\bibfnamefont {T.}~\bibnamefont {McCourt}},
  \bibinfo {author} {\bibfnamefont {X.}~\bibnamefont {Mi}}, \bibinfo {author}
  {\bibfnamefont {Z.}~\bibnamefont {Jiang}}, \bibinfo {author} {\bibfnamefont
  {M.~Y.}\ \bibnamefont {Niu}}, \bibinfo {author} {\bibfnamefont
  {W.}~\bibnamefont {Mruczkiewicz}}, \bibinfo {author} {\bibfnamefont
  {I.}~\bibnamefont {Aleiner}}, \bibinfo {author} {\bibfnamefont
  {F.}~\bibnamefont {Arute}}, \bibinfo {author} {\bibfnamefont
  {K.}~\bibnamefont {Arya}}, \bibinfo {author} {\bibfnamefont {J.}~\bibnamefont
  {Atalaya}}, \ and\ \bibinfo {author} {\bibnamefont {et~al.}},\ }\href
  {\doibase 10.1038/s41586-021-03576-2} {\bibfield  {journal} {\bibinfo
  {journal} {Nature}\ }\textbf {\bibinfo {volume} {594}},\ \bibinfo {pages}
  {508–512} (\bibinfo {year} {2021})}\BibitemShut {NoStop}%
\bibitem [{\citenamefont {Arute}\ \emph
  {et~al.}(2020{\natexlab{a}})\citenamefont {Arute}, \citenamefont {Arya},
  \citenamefont {Babbush}, \citenamefont {Bacon}, \citenamefont {Bardin},
  \citenamefont {Barends}, \citenamefont {Bengtsson}, \citenamefont {Boixo},
  \citenamefont {Broughton}, \citenamefont {Buckley}, \citenamefont {Buell},
  \citenamefont {Burkett}, \citenamefont {Bushnell}, \citenamefont {Chen},
  \citenamefont {Chen}, \citenamefont {Chen}, \citenamefont {Chiaro},
  \citenamefont {Collins}, \citenamefont {Cotton}, \citenamefont {Courtney},
  \citenamefont {Demura}, \citenamefont {Derk}, \citenamefont {Dunsworth},
  \citenamefont {Eppens}, \citenamefont {Eckl}, \citenamefont {Erickson},
  \citenamefont {Farhi}, \citenamefont {Fowler}, \citenamefont {Foxen},
  \citenamefont {Gidney}, \citenamefont {Giustina}, \citenamefont {Graff},
  \citenamefont {Gross}, \citenamefont {Habegger}, \citenamefont {Harrigan},
  \citenamefont {Ho}, \citenamefont {Hong}, \citenamefont {Huang},
  \citenamefont {Huggins}, \citenamefont {Ioffe}, \citenamefont {Isakov},
  \citenamefont {Jeffrey}, \citenamefont {Jiang}, \citenamefont {Jones},
  \citenamefont {Kafri}, \citenamefont {Kechedzhi}, \citenamefont {Kelly},
  \citenamefont {Kim}, \citenamefont {Klimov}, \citenamefont {Korotkov},
  \citenamefont {Kostritsa}, \citenamefont {Landhuis}, \citenamefont {Laptev},
  \citenamefont {Lindmark}, \citenamefont {Lucero}, \citenamefont {Marthaler},
  \citenamefont {Martin}, \citenamefont {Martinis}, \citenamefont {Marusczyk},
  \citenamefont {McArdle}, \citenamefont {McClean}, \citenamefont {McCourt},
  \citenamefont {McEwen}, \citenamefont {Megrant}, \citenamefont
  {Mejuto-Zaera}, \citenamefont {Mi}, \citenamefont {Mohseni}, \citenamefont
  {Mruczkiewicz}, \citenamefont {Mutus}, \citenamefont {Naaman}, \citenamefont
  {Neeley}, \citenamefont {Neill}, \citenamefont {Neven}, \citenamefont
  {Newman}, \citenamefont {Niu}, \citenamefont {O'Brien}, \citenamefont
  {Ostby}, \citenamefont {Pató}, \citenamefont {Petukhov}, \citenamefont
  {Putterman}, \citenamefont {Quintana}, \citenamefont {Reiner}, \citenamefont
  {Roushan}, \citenamefont {Rubin}, \citenamefont {Sank}, \citenamefont
  {Satzinger}, \citenamefont {Smelyanskiy}, \citenamefont {Strain},
  \citenamefont {Sung}, \citenamefont {Schmitteckert}, \citenamefont {Szalay},
  \citenamefont {Tubman}, \citenamefont {Vainsencher}, \citenamefont {White},
  \citenamefont {Vogt}, \citenamefont {Yao}, \citenamefont {Yeh}, \citenamefont
  {Zalcman},\ and\ \citenamefont {Zanker}}]{zhang_2020}%
  \BibitemOpen
  \bibfield  {author} {\bibinfo {author} {\bibfnamefont {F.}~\bibnamefont
  {Arute}}, \bibinfo {author} {\bibfnamefont {K.}~\bibnamefont {Arya}},
  \bibinfo {author} {\bibfnamefont {R.}~\bibnamefont {Babbush}}, \bibinfo
  {author} {\bibfnamefont {D.}~\bibnamefont {Bacon}}, \bibinfo {author}
  {\bibfnamefont {J.~C.}\ \bibnamefont {Bardin}}, \bibinfo {author}
  {\bibfnamefont {R.}~\bibnamefont {Barends}}, \bibinfo {author} {\bibfnamefont
  {A.}~\bibnamefont {Bengtsson}}, \bibinfo {author} {\bibfnamefont
  {S.}~\bibnamefont {Boixo}}, \bibinfo {author} {\bibfnamefont
  {M.}~\bibnamefont {Broughton}}, \bibinfo {author} {\bibfnamefont {B.~B.}\
  \bibnamefont {Buckley}}, \bibinfo {author} {\bibfnamefont {D.~A.}\
  \bibnamefont {Buell}}, \bibinfo {author} {\bibfnamefont {B.}~\bibnamefont
  {Burkett}}, \bibinfo {author} {\bibfnamefont {N.}~\bibnamefont {Bushnell}},
  \bibinfo {author} {\bibfnamefont {Y.}~\bibnamefont {Chen}}, \bibinfo {author}
  {\bibfnamefont {Z.}~\bibnamefont {Chen}}, \bibinfo {author} {\bibfnamefont
  {Y.-A.}\ \bibnamefont {Chen}}, \bibinfo {author} {\bibfnamefont
  {B.}~\bibnamefont {Chiaro}}, \bibinfo {author} {\bibfnamefont
  {R.}~\bibnamefont {Collins}}, \bibinfo {author} {\bibfnamefont {S.~J.}\
  \bibnamefont {Cotton}}, \bibinfo {author} {\bibfnamefont {W.}~\bibnamefont
  {Courtney}}, \bibinfo {author} {\bibfnamefont {S.}~\bibnamefont {Demura}},
  \bibinfo {author} {\bibfnamefont {A.}~\bibnamefont {Derk}}, \bibinfo {author}
  {\bibfnamefont {A.}~\bibnamefont {Dunsworth}}, \bibinfo {author}
  {\bibfnamefont {D.}~\bibnamefont {Eppens}}, \bibinfo {author} {\bibfnamefont
  {T.}~\bibnamefont {Eckl}}, \bibinfo {author} {\bibfnamefont {C.}~\bibnamefont
  {Erickson}}, \bibinfo {author} {\bibfnamefont {E.}~\bibnamefont {Farhi}},
  \bibinfo {author} {\bibfnamefont {A.}~\bibnamefont {Fowler}}, \bibinfo
  {author} {\bibfnamefont {B.}~\bibnamefont {Foxen}}, \bibinfo {author}
  {\bibfnamefont {C.}~\bibnamefont {Gidney}}, \bibinfo {author} {\bibfnamefont
  {M.}~\bibnamefont {Giustina}}, \bibinfo {author} {\bibfnamefont
  {R.}~\bibnamefont {Graff}}, \bibinfo {author} {\bibfnamefont {J.~A.}\
  \bibnamefont {Gross}}, \bibinfo {author} {\bibfnamefont {S.}~\bibnamefont
  {Habegger}}, \bibinfo {author} {\bibfnamefont {M.~P.}\ \bibnamefont
  {Harrigan}}, \bibinfo {author} {\bibfnamefont {A.}~\bibnamefont {Ho}},
  \bibinfo {author} {\bibfnamefont {S.}~\bibnamefont {Hong}}, \bibinfo {author}
  {\bibfnamefont {T.}~\bibnamefont {Huang}}, \bibinfo {author} {\bibfnamefont
  {W.}~\bibnamefont {Huggins}}, \bibinfo {author} {\bibfnamefont {L.~B.}\
  \bibnamefont {Ioffe}}, \bibinfo {author} {\bibfnamefont {S.~V.}\ \bibnamefont
  {Isakov}}, \bibinfo {author} {\bibfnamefont {E.}~\bibnamefont {Jeffrey}},
  \bibinfo {author} {\bibfnamefont {Z.}~\bibnamefont {Jiang}}, \bibinfo
  {author} {\bibfnamefont {C.}~\bibnamefont {Jones}}, \bibinfo {author}
  {\bibfnamefont {D.}~\bibnamefont {Kafri}}, \bibinfo {author} {\bibfnamefont
  {K.}~\bibnamefont {Kechedzhi}}, \bibinfo {author} {\bibfnamefont
  {J.}~\bibnamefont {Kelly}}, \bibinfo {author} {\bibfnamefont
  {S.}~\bibnamefont {Kim}}, \bibinfo {author} {\bibfnamefont {P.~V.}\
  \bibnamefont {Klimov}}, \bibinfo {author} {\bibfnamefont {A.~N.}\
  \bibnamefont {Korotkov}}, \bibinfo {author} {\bibfnamefont {F.}~\bibnamefont
  {Kostritsa}}, \bibinfo {author} {\bibfnamefont {D.}~\bibnamefont {Landhuis}},
  \bibinfo {author} {\bibfnamefont {P.}~\bibnamefont {Laptev}}, \bibinfo
  {author} {\bibfnamefont {M.}~\bibnamefont {Lindmark}}, \bibinfo {author}
  {\bibfnamefont {E.}~\bibnamefont {Lucero}}, \bibinfo {author} {\bibfnamefont
  {M.}~\bibnamefont {Marthaler}}, \bibinfo {author} {\bibfnamefont
  {O.}~\bibnamefont {Martin}}, \bibinfo {author} {\bibfnamefont {J.~M.}\
  \bibnamefont {Martinis}}, \bibinfo {author} {\bibfnamefont {A.}~\bibnamefont
  {Marusczyk}}, \bibinfo {author} {\bibfnamefont {S.}~\bibnamefont {McArdle}},
  \bibinfo {author} {\bibfnamefont {J.~R.}\ \bibnamefont {McClean}}, \bibinfo
  {author} {\bibfnamefont {T.}~\bibnamefont {McCourt}}, \bibinfo {author}
  {\bibfnamefont {M.}~\bibnamefont {McEwen}}, \bibinfo {author} {\bibfnamefont
  {A.}~\bibnamefont {Megrant}}, \bibinfo {author} {\bibfnamefont
  {C.}~\bibnamefont {Mejuto-Zaera}}, \bibinfo {author} {\bibfnamefont
  {X.}~\bibnamefont {Mi}}, \bibinfo {author} {\bibfnamefont {M.}~\bibnamefont
  {Mohseni}}, \bibinfo {author} {\bibfnamefont {W.}~\bibnamefont
  {Mruczkiewicz}}, \bibinfo {author} {\bibfnamefont {J.}~\bibnamefont {Mutus}},
  \bibinfo {author} {\bibfnamefont {O.}~\bibnamefont {Naaman}}, \bibinfo
  {author} {\bibfnamefont {M.}~\bibnamefont {Neeley}}, \bibinfo {author}
  {\bibfnamefont {C.}~\bibnamefont {Neill}}, \bibinfo {author} {\bibfnamefont
  {H.}~\bibnamefont {Neven}}, \bibinfo {author} {\bibfnamefont
  {M.}~\bibnamefont {Newman}}, \bibinfo {author} {\bibfnamefont {M.~Y.}\
  \bibnamefont {Niu}}, \bibinfo {author} {\bibfnamefont {T.~E.}\ \bibnamefont
  {O'Brien}}, \bibinfo {author} {\bibfnamefont {E.}~\bibnamefont {Ostby}},
  \bibinfo {author} {\bibfnamefont {B.}~\bibnamefont {Pató}}, \bibinfo
  {author} {\bibfnamefont {A.}~\bibnamefont {Petukhov}}, \bibinfo {author}
  {\bibfnamefont {H.}~\bibnamefont {Putterman}}, \bibinfo {author}
  {\bibfnamefont {C.}~\bibnamefont {Quintana}}, \bibinfo {author}
  {\bibfnamefont {J.-M.}\ \bibnamefont {Reiner}}, \bibinfo {author}
  {\bibfnamefont {P.}~\bibnamefont {Roushan}}, \bibinfo {author} {\bibfnamefont
  {N.~C.}\ \bibnamefont {Rubin}}, \bibinfo {author} {\bibfnamefont
  {D.}~\bibnamefont {Sank}}, \bibinfo {author} {\bibfnamefont {K.~J.}\
  \bibnamefont {Satzinger}}, \bibinfo {author} {\bibfnamefont {V.}~\bibnamefont
  {Smelyanskiy}}, \bibinfo {author} {\bibfnamefont {D.}~\bibnamefont {Strain}},
  \bibinfo {author} {\bibfnamefont {K.~J.}\ \bibnamefont {Sung}}, \bibinfo
  {author} {\bibfnamefont {P.}~\bibnamefont {Schmitteckert}}, \bibinfo {author}
  {\bibfnamefont {M.}~\bibnamefont {Szalay}}, \bibinfo {author} {\bibfnamefont
  {N.~M.}\ \bibnamefont {Tubman}}, \bibinfo {author} {\bibfnamefont
  {A.}~\bibnamefont {Vainsencher}}, \bibinfo {author} {\bibfnamefont
  {T.}~\bibnamefont {White}}, \bibinfo {author} {\bibfnamefont
  {N.}~\bibnamefont {Vogt}}, \bibinfo {author} {\bibfnamefont {Z.~J.}\
  \bibnamefont {Yao}}, \bibinfo {author} {\bibfnamefont {P.}~\bibnamefont
  {Yeh}}, \bibinfo {author} {\bibfnamefont {A.}~\bibnamefont {Zalcman}}, \ and\
  \bibinfo {author} {\bibfnamefont {S.}~\bibnamefont {Zanker}},\ }\href@noop {}
  {\enquote {\bibinfo {title} {Observation of separated dynamics of charge and
  spin in the fermi-hubbard model},}\ } (\bibinfo {year}
  {2020}{\natexlab{a}}),\ \Eprint {http://arxiv.org/abs/2010.07965}
  {arXiv:2010.07965 [quant-ph]} \BibitemShut {NoStop}%
\bibitem [{\citenamefont {Tan}\ \emph {et~al.}(2021)\citenamefont {Tan},
  \citenamefont {Sun}, \citenamefont {Tazhigulov}, \citenamefont {Chan},\ and\
  \citenamefont {Minnich}}]{tan2021}%
  \BibitemOpen
  \bibfield  {author} {\bibinfo {author} {\bibfnamefont {A.~T.~K.}\
  \bibnamefont {Tan}}, \bibinfo {author} {\bibfnamefont {S.-N.}\ \bibnamefont
  {Sun}}, \bibinfo {author} {\bibfnamefont {R.~N.}\ \bibnamefont {Tazhigulov}},
  \bibinfo {author} {\bibfnamefont {G.~K.-L.}\ \bibnamefont {Chan}}, \ and\
  \bibinfo {author} {\bibfnamefont {A.~J.}\ \bibnamefont {Minnich}},\
  }\href@noop {} {\enquote {\bibinfo {title} {Realizing symmetry-protected
  topological phases in a spin-1/2 chain with next-nearest neighbor hopping on
  superconducting qubits},}\ } (\bibinfo {year} {2021}),\ \Eprint
  {http://arxiv.org/abs/2112.10333} {arXiv:2112.10333 [quant-ph]} \BibitemShut
  {NoStop}%
\bibitem [{\citenamefont {Arute}\ \emph
  {et~al.}(2020{\natexlab{b}})\citenamefont {Arute}, \citenamefont {Arya},
  \citenamefont {Babbush}, \citenamefont {Bacon}, \citenamefont {Bardin},
  \citenamefont {Barends}, \citenamefont {Boixo}, \citenamefont {Broughton},
  \citenamefont {Buckley},\ and\ \citenamefont {et~al.}}]{rubin_2020}%
  \BibitemOpen
  \bibfield  {author} {\bibinfo {author} {\bibfnamefont {F.}~\bibnamefont
  {Arute}}, \bibinfo {author} {\bibfnamefont {K.}~\bibnamefont {Arya}},
  \bibinfo {author} {\bibfnamefont {R.}~\bibnamefont {Babbush}}, \bibinfo
  {author} {\bibfnamefont {D.}~\bibnamefont {Bacon}}, \bibinfo {author}
  {\bibfnamefont {J.~C.}\ \bibnamefont {Bardin}}, \bibinfo {author}
  {\bibfnamefont {R.}~\bibnamefont {Barends}}, \bibinfo {author} {\bibfnamefont
  {S.}~\bibnamefont {Boixo}}, \bibinfo {author} {\bibfnamefont
  {M.}~\bibnamefont {Broughton}}, \bibinfo {author} {\bibfnamefont {B.~B.}\
  \bibnamefont {Buckley}}, \ and\ \bibinfo {author} {\bibnamefont {et~al.}},\
  }\href {\doibase 10.1126/science.abb9811} {\bibfield  {journal} {\bibinfo
  {journal} {Science}\ }\textbf {\bibinfo {volume} {369}},\ \bibinfo {pages}
  {1084–1089} (\bibinfo {year} {2020}{\natexlab{b}})}\BibitemShut {NoStop}%
\bibitem [{\citenamefont {Kelly}\ \emph {et~al.}(2015)\citenamefont {Kelly},
  \citenamefont {Barends}, \citenamefont {Fowler}, \citenamefont {Megrant},
  \citenamefont {Jeffrey}, \citenamefont {White}, \citenamefont {Sank},
  \citenamefont {Mutus}, \citenamefont {Campbell}, \citenamefont {Chen},\ and\
  \citenamefont {et~al.}}]{kelly_2015}%
  \BibitemOpen
  \bibfield  {author} {\bibinfo {author} {\bibfnamefont {J.}~\bibnamefont
  {Kelly}}, \bibinfo {author} {\bibfnamefont {R.}~\bibnamefont {Barends}},
  \bibinfo {author} {\bibfnamefont {A.~G.}\ \bibnamefont {Fowler}}, \bibinfo
  {author} {\bibfnamefont {A.}~\bibnamefont {Megrant}}, \bibinfo {author}
  {\bibfnamefont {E.}~\bibnamefont {Jeffrey}}, \bibinfo {author} {\bibfnamefont
  {T.~C.}\ \bibnamefont {White}}, \bibinfo {author} {\bibfnamefont
  {D.}~\bibnamefont {Sank}}, \bibinfo {author} {\bibfnamefont {J.~Y.}\
  \bibnamefont {Mutus}}, \bibinfo {author} {\bibfnamefont {B.}~\bibnamefont
  {Campbell}}, \bibinfo {author} {\bibfnamefont {Y.}~\bibnamefont {Chen}}, \
  and\ \bibinfo {author} {\bibnamefont {et~al.}},\ }\href {\doibase
  10.1038/nature14270} {\bibfield  {journal} {\bibinfo  {journal} {Nature}\
  }\textbf {\bibinfo {volume} {519}},\ \bibinfo {pages} {66–69} (\bibinfo
  {year} {2015})}\BibitemShut {NoStop}%
\bibitem [{\citenamefont {Chen}\ \emph {et~al.}(2021)\citenamefont {Chen},
  \citenamefont {Satzinger}, \citenamefont {Atalaya}, \citenamefont {Korotkov},
  \citenamefont {Dunsworth}, \citenamefont {Sank}, \citenamefont {Quintana},
  \citenamefont {McEwen}, \citenamefont {Barends}, \citenamefont {Klimov},
  \citenamefont {Hong}, \citenamefont {Jones}, \citenamefont {Petukhov},
  \citenamefont {Kafri}, \citenamefont {Demura}, \citenamefont {Burkett},
  \citenamefont {Gidney}, \citenamefont {Fowler}, \citenamefont {Putterman},
  \citenamefont {Aleiner}, \citenamefont {Arute}, \citenamefont {Arya},
  \citenamefont {Babbush}, \citenamefont {Bardin}, \citenamefont {Bengtsson},
  \citenamefont {Bourassa}, \citenamefont {Broughton}, \citenamefont {Buckley},
  \citenamefont {Buell}, \citenamefont {Bushnell}, \citenamefont {Chiaro},
  \citenamefont {Collins}, \citenamefont {Courtney}, \citenamefont {Derk},
  \citenamefont {Eppens}, \citenamefont {Erickson}, \citenamefont {Farhi},
  \citenamefont {Foxen}, \citenamefont {Giustina}, \citenamefont {Gross},
  \citenamefont {Harrigan}, \citenamefont {Harrington}, \citenamefont {Hilton},
  \citenamefont {Ho}, \citenamefont {Huang}, \citenamefont {Huggins},
  \citenamefont {Ioffe}, \citenamefont {Isakov}, \citenamefont {Jeffrey},
  \citenamefont {Jiang}, \citenamefont {Kechedzhi}, \citenamefont {Kim},
  \citenamefont {Kostritsa}, \citenamefont {Landhuis}, \citenamefont {Laptev},
  \citenamefont {Lucero}, \citenamefont {Martin}, \citenamefont {McClean},
  \citenamefont {McCourt}, \citenamefont {Mi}, \citenamefont {Miao},
  \citenamefont {Mohseni}, \citenamefont {Mruczkiewicz}, \citenamefont {Mutus},
  \citenamefont {Naaman}, \citenamefont {Neeley}, \citenamefont {Neill},
  \citenamefont {Newman}, \citenamefont {Niu}, \citenamefont {O'Brien},
  \citenamefont {Opremcak}, \citenamefont {Ostby}, \citenamefont {Pató},
  \citenamefont {Redd}, \citenamefont {Roushan}, \citenamefont {Rubin},
  \citenamefont {Shvarts}, \citenamefont {Strain}, \citenamefont {Szalay},
  \citenamefont {Trevithick}, \citenamefont {Villalonga}, \citenamefont
  {White}, \citenamefont {Yao}, \citenamefont {Yeh}, \citenamefont {Zalcman},
  \citenamefont {Neven}, \citenamefont {Boixo}, \citenamefont {Smelyanskiy},
  \citenamefont {Chen}, \citenamefont {Megrant},\ and\ \citenamefont
  {Kelly}}]{chen_2021}%
  \BibitemOpen
  \bibfield  {author} {\bibinfo {author} {\bibfnamefont {Z.}~\bibnamefont
  {Chen}}, \bibinfo {author} {\bibfnamefont {K.~J.}\ \bibnamefont {Satzinger}},
  \bibinfo {author} {\bibfnamefont {J.}~\bibnamefont {Atalaya}}, \bibinfo
  {author} {\bibfnamefont {A.~N.}\ \bibnamefont {Korotkov}}, \bibinfo {author}
  {\bibfnamefont {A.}~\bibnamefont {Dunsworth}}, \bibinfo {author}
  {\bibfnamefont {D.}~\bibnamefont {Sank}}, \bibinfo {author} {\bibfnamefont
  {C.}~\bibnamefont {Quintana}}, \bibinfo {author} {\bibfnamefont
  {M.}~\bibnamefont {McEwen}}, \bibinfo {author} {\bibfnamefont
  {R.}~\bibnamefont {Barends}}, \bibinfo {author} {\bibfnamefont {P.~V.}\
  \bibnamefont {Klimov}}, \bibinfo {author} {\bibfnamefont {S.}~\bibnamefont
  {Hong}}, \bibinfo {author} {\bibfnamefont {C.}~\bibnamefont {Jones}},
  \bibinfo {author} {\bibfnamefont {A.}~\bibnamefont {Petukhov}}, \bibinfo
  {author} {\bibfnamefont {D.}~\bibnamefont {Kafri}}, \bibinfo {author}
  {\bibfnamefont {S.}~\bibnamefont {Demura}}, \bibinfo {author} {\bibfnamefont
  {B.}~\bibnamefont {Burkett}}, \bibinfo {author} {\bibfnamefont
  {C.}~\bibnamefont {Gidney}}, \bibinfo {author} {\bibfnamefont {A.~G.}\
  \bibnamefont {Fowler}}, \bibinfo {author} {\bibfnamefont {H.}~\bibnamefont
  {Putterman}}, \bibinfo {author} {\bibfnamefont {I.}~\bibnamefont {Aleiner}},
  \bibinfo {author} {\bibfnamefont {F.}~\bibnamefont {Arute}}, \bibinfo
  {author} {\bibfnamefont {K.}~\bibnamefont {Arya}}, \bibinfo {author}
  {\bibfnamefont {R.}~\bibnamefont {Babbush}}, \bibinfo {author} {\bibfnamefont
  {J.~C.}\ \bibnamefont {Bardin}}, \bibinfo {author} {\bibfnamefont
  {A.}~\bibnamefont {Bengtsson}}, \bibinfo {author} {\bibfnamefont
  {A.}~\bibnamefont {Bourassa}}, \bibinfo {author} {\bibfnamefont
  {M.}~\bibnamefont {Broughton}}, \bibinfo {author} {\bibfnamefont {B.~B.}\
  \bibnamefont {Buckley}}, \bibinfo {author} {\bibfnamefont {D.~A.}\
  \bibnamefont {Buell}}, \bibinfo {author} {\bibfnamefont {N.}~\bibnamefont
  {Bushnell}}, \bibinfo {author} {\bibfnamefont {B.}~\bibnamefont {Chiaro}},
  \bibinfo {author} {\bibfnamefont {R.}~\bibnamefont {Collins}}, \bibinfo
  {author} {\bibfnamefont {W.}~\bibnamefont {Courtney}}, \bibinfo {author}
  {\bibfnamefont {A.~R.}\ \bibnamefont {Derk}}, \bibinfo {author}
  {\bibfnamefont {D.}~\bibnamefont {Eppens}}, \bibinfo {author} {\bibfnamefont
  {C.}~\bibnamefont {Erickson}}, \bibinfo {author} {\bibfnamefont
  {E.}~\bibnamefont {Farhi}}, \bibinfo {author} {\bibfnamefont
  {B.}~\bibnamefont {Foxen}}, \bibinfo {author} {\bibfnamefont
  {M.}~\bibnamefont {Giustina}}, \bibinfo {author} {\bibfnamefont {J.~A.}\
  \bibnamefont {Gross}}, \bibinfo {author} {\bibfnamefont {M.~P.}\ \bibnamefont
  {Harrigan}}, \bibinfo {author} {\bibfnamefont {S.~D.}\ \bibnamefont
  {Harrington}}, \bibinfo {author} {\bibfnamefont {J.}~\bibnamefont {Hilton}},
  \bibinfo {author} {\bibfnamefont {A.}~\bibnamefont {Ho}}, \bibinfo {author}
  {\bibfnamefont {T.}~\bibnamefont {Huang}}, \bibinfo {author} {\bibfnamefont
  {W.~J.}\ \bibnamefont {Huggins}}, \bibinfo {author} {\bibfnamefont {L.~B.}\
  \bibnamefont {Ioffe}}, \bibinfo {author} {\bibfnamefont {S.~V.}\ \bibnamefont
  {Isakov}}, \bibinfo {author} {\bibfnamefont {E.}~\bibnamefont {Jeffrey}},
  \bibinfo {author} {\bibfnamefont {Z.}~\bibnamefont {Jiang}}, \bibinfo
  {author} {\bibfnamefont {K.}~\bibnamefont {Kechedzhi}}, \bibinfo {author}
  {\bibfnamefont {S.}~\bibnamefont {Kim}}, \bibinfo {author} {\bibfnamefont
  {F.}~\bibnamefont {Kostritsa}}, \bibinfo {author} {\bibfnamefont
  {D.}~\bibnamefont {Landhuis}}, \bibinfo {author} {\bibfnamefont
  {P.}~\bibnamefont {Laptev}}, \bibinfo {author} {\bibfnamefont
  {E.}~\bibnamefont {Lucero}}, \bibinfo {author} {\bibfnamefont
  {O.}~\bibnamefont {Martin}}, \bibinfo {author} {\bibfnamefont {J.~R.}\
  \bibnamefont {McClean}}, \bibinfo {author} {\bibfnamefont {T.}~\bibnamefont
  {McCourt}}, \bibinfo {author} {\bibfnamefont {X.}~\bibnamefont {Mi}},
  \bibinfo {author} {\bibfnamefont {K.~C.}\ \bibnamefont {Miao}}, \bibinfo
  {author} {\bibfnamefont {M.}~\bibnamefont {Mohseni}}, \bibinfo {author}
  {\bibfnamefont {W.}~\bibnamefont {Mruczkiewicz}}, \bibinfo {author}
  {\bibfnamefont {J.}~\bibnamefont {Mutus}}, \bibinfo {author} {\bibfnamefont
  {O.}~\bibnamefont {Naaman}}, \bibinfo {author} {\bibfnamefont
  {M.}~\bibnamefont {Neeley}}, \bibinfo {author} {\bibfnamefont
  {C.}~\bibnamefont {Neill}}, \bibinfo {author} {\bibfnamefont
  {M.}~\bibnamefont {Newman}}, \bibinfo {author} {\bibfnamefont {M.~Y.}\
  \bibnamefont {Niu}}, \bibinfo {author} {\bibfnamefont {T.~E.}\ \bibnamefont
  {O'Brien}}, \bibinfo {author} {\bibfnamefont {A.}~\bibnamefont {Opremcak}},
  \bibinfo {author} {\bibfnamefont {E.}~\bibnamefont {Ostby}}, \bibinfo
  {author} {\bibfnamefont {B.}~\bibnamefont {Pató}}, \bibinfo {author}
  {\bibfnamefont {N.}~\bibnamefont {Redd}}, \bibinfo {author} {\bibfnamefont
  {P.}~\bibnamefont {Roushan}}, \bibinfo {author} {\bibfnamefont {N.~C.}\
  \bibnamefont {Rubin}}, \bibinfo {author} {\bibfnamefont {V.}~\bibnamefont
  {Shvarts}}, \bibinfo {author} {\bibfnamefont {D.}~\bibnamefont {Strain}},
  \bibinfo {author} {\bibfnamefont {M.}~\bibnamefont {Szalay}}, \bibinfo
  {author} {\bibfnamefont {M.~D.}\ \bibnamefont {Trevithick}}, \bibinfo
  {author} {\bibfnamefont {B.}~\bibnamefont {Villalonga}}, \bibinfo {author}
  {\bibfnamefont {T.}~\bibnamefont {White}}, \bibinfo {author} {\bibfnamefont
  {Z.~J.}\ \bibnamefont {Yao}}, \bibinfo {author} {\bibfnamefont
  {P.}~\bibnamefont {Yeh}}, \bibinfo {author} {\bibfnamefont {A.}~\bibnamefont
  {Zalcman}}, \bibinfo {author} {\bibfnamefont {H.}~\bibnamefont {Neven}},
  \bibinfo {author} {\bibfnamefont {S.}~\bibnamefont {Boixo}}, \bibinfo
  {author} {\bibfnamefont {V.}~\bibnamefont {Smelyanskiy}}, \bibinfo {author}
  {\bibfnamefont {Y.}~\bibnamefont {Chen}}, \bibinfo {author} {\bibfnamefont
  {A.}~\bibnamefont {Megrant}}, \ and\ \bibinfo {author} {\bibfnamefont
  {J.}~\bibnamefont {Kelly}},\ }\href@noop {} {\enquote {\bibinfo {title}
  {Exponential suppression of bit or phase flip errors with repetitive error
  correction},}\ } (\bibinfo {year} {2021}),\ \Eprint
  {http://arxiv.org/abs/2102.06132} {arXiv:2102.06132 [quant-ph]} \BibitemShut
  {NoStop}%
\bibitem [{\citenamefont {Arute~\textit{et al.}}(2019)}]{Arute2019}%
  \BibitemOpen
  \bibfield  {author} {\bibinfo {author} {\bibfnamefont {F.}~\bibnamefont
  {Arute~\textit{et al.}}},\ }\href {\doibase 10.1038/s41586-019-1666-5}
  {\bibfield  {journal} {\bibinfo  {journal} {Nature}\ }\textbf {\bibinfo
  {volume} {574}},\ \bibinfo {pages} {505} (\bibinfo {year}
  {2019})}\BibitemShut {NoStop}%
\bibitem [{\citenamefont {Huang}\ \emph {et~al.}(2021)\citenamefont {Huang},
  \citenamefont {Broughton}, \citenamefont {Cotler}, \citenamefont {Chen},
  \citenamefont {Li}, \citenamefont {Mohseni}, \citenamefont {Neven},
  \citenamefont {Babbush}, \citenamefont {Kueng}, \citenamefont {Preskill},\
  and\ \citenamefont {McClean}}]{huang2021quantum}%
  \BibitemOpen
  \bibfield  {author} {\bibinfo {author} {\bibfnamefont {H.-Y.}\ \bibnamefont
  {Huang}}, \bibinfo {author} {\bibfnamefont {M.}~\bibnamefont {Broughton}},
  \bibinfo {author} {\bibfnamefont {J.}~\bibnamefont {Cotler}}, \bibinfo
  {author} {\bibfnamefont {S.}~\bibnamefont {Chen}}, \bibinfo {author}
  {\bibfnamefont {J.}~\bibnamefont {Li}}, \bibinfo {author} {\bibfnamefont
  {M.}~\bibnamefont {Mohseni}}, \bibinfo {author} {\bibfnamefont
  {H.}~\bibnamefont {Neven}}, \bibinfo {author} {\bibfnamefont
  {R.}~\bibnamefont {Babbush}}, \bibinfo {author} {\bibfnamefont
  {R.}~\bibnamefont {Kueng}}, \bibinfo {author} {\bibfnamefont
  {J.}~\bibnamefont {Preskill}}, \ and\ \bibinfo {author} {\bibfnamefont
  {J.~R.}\ \bibnamefont {McClean}},\ }\href@noop {} {\enquote {\bibinfo {title}
  {Quantum advantage in learning from experiments},}\ } (\bibinfo {year}
  {2021}),\ \Eprint {http://arxiv.org/abs/2112.00778} {arXiv:2112.00778
  [quant-ph]} \BibitemShut {NoStop}%
\bibitem [{\citenamefont {Sung}\ \emph {et~al.}(2021)\citenamefont {Sung},
  \citenamefont {Ding}, \citenamefont {Braum\"uller}, \citenamefont
  {Veps\"al\"ainen}, \citenamefont {Kannan}, \citenamefont {Kjaergaard},
  \citenamefont {Greene}, \citenamefont {Samach}, \citenamefont {McNally},
  \citenamefont {Kim}, \citenamefont {Melville}, \citenamefont {Niedzielski},
  \citenamefont {Schwartz}, \citenamefont {Yoder}, \citenamefont {Orlando},
  \citenamefont {Gustavsson},\ and\ \citenamefont {Oliver}}]{oliver_2021}%
  \BibitemOpen
  \bibfield  {author} {\bibinfo {author} {\bibfnamefont {Y.}~\bibnamefont
  {Sung}}, \bibinfo {author} {\bibfnamefont {L.}~\bibnamefont {Ding}}, \bibinfo
  {author} {\bibfnamefont {J.}~\bibnamefont {Braum\"uller}}, \bibinfo {author}
  {\bibfnamefont {A.}~\bibnamefont {Veps\"al\"ainen}}, \bibinfo {author}
  {\bibfnamefont {B.}~\bibnamefont {Kannan}}, \bibinfo {author} {\bibfnamefont
  {M.}~\bibnamefont {Kjaergaard}}, \bibinfo {author} {\bibfnamefont
  {A.}~\bibnamefont {Greene}}, \bibinfo {author} {\bibfnamefont {G.~O.}\
  \bibnamefont {Samach}}, \bibinfo {author} {\bibfnamefont {C.}~\bibnamefont
  {McNally}}, \bibinfo {author} {\bibfnamefont {D.}~\bibnamefont {Kim}},
  \bibinfo {author} {\bibfnamefont {A.}~\bibnamefont {Melville}}, \bibinfo
  {author} {\bibfnamefont {B.~M.}\ \bibnamefont {Niedzielski}}, \bibinfo
  {author} {\bibfnamefont {M.~E.}\ \bibnamefont {Schwartz}}, \bibinfo {author}
  {\bibfnamefont {J.~L.}\ \bibnamefont {Yoder}}, \bibinfo {author}
  {\bibfnamefont {T.~P.}\ \bibnamefont {Orlando}}, \bibinfo {author}
  {\bibfnamefont {S.}~\bibnamefont {Gustavsson}}, \ and\ \bibinfo {author}
  {\bibfnamefont {W.~D.}\ \bibnamefont {Oliver}},\ }\href {\doibase
  10.1103/PhysRevX.11.021058} {\bibfield  {journal} {\bibinfo  {journal} {Phys.
  Rev. X}\ }\textbf {\bibinfo {volume} {11}},\ \bibinfo {pages} {021058}
  (\bibinfo {year} {2021})}\BibitemShut {NoStop}%
\bibitem [{\citenamefont {Foxen}\ \emph {et~al.}(2020)\citenamefont {Foxen},
  \citenamefont {Neill}, \citenamefont {Dunsworth}, \citenamefont {Roushan},
  \citenamefont {Chiaro}, \citenamefont {Megrant}, \citenamefont {Kelly},
  \citenamefont {Chen}, \citenamefont {Satzinger}, \citenamefont {Barends},\
  and\ \citenamefont {et~al.}}]{foxen_2020}%
  \BibitemOpen
  \bibfield  {author} {\bibinfo {author} {\bibfnamefont {B.}~\bibnamefont
  {Foxen}}, \bibinfo {author} {\bibfnamefont {C.}~\bibnamefont {Neill}},
  \bibinfo {author} {\bibfnamefont {A.}~\bibnamefont {Dunsworth}}, \bibinfo
  {author} {\bibfnamefont {P.}~\bibnamefont {Roushan}}, \bibinfo {author}
  {\bibfnamefont {B.}~\bibnamefont {Chiaro}}, \bibinfo {author} {\bibfnamefont
  {A.}~\bibnamefont {Megrant}}, \bibinfo {author} {\bibfnamefont
  {J.}~\bibnamefont {Kelly}}, \bibinfo {author} {\bibfnamefont
  {Z.}~\bibnamefont {Chen}}, \bibinfo {author} {\bibfnamefont {K.}~\bibnamefont
  {Satzinger}}, \bibinfo {author} {\bibfnamefont {R.}~\bibnamefont {Barends}},
  \ and\ \bibinfo {author} {\bibnamefont {et~al.}},\ }\href {\doibase
  10.1103/physrevlett.125.120504} {\bibfield  {journal} {\bibinfo  {journal}
  {Physical Review Letters}\ }\textbf {\bibinfo {volume} {125}} (\bibinfo
  {year} {2020}),\ 10.1103/physrevlett.125.120504}\BibitemShut {NoStop}%
\bibitem [{\citenamefont {Uhrig}(2008)}]{Uhrig2008}%
  \BibitemOpen
  \bibfield  {author} {\bibinfo {author} {\bibfnamefont {G.~S.}\ \bibnamefont
  {Uhrig}},\ }\href {\doibase 10.1088/1367-2630/10/8/083024} {\bibfield
  {journal} {\bibinfo  {journal} {New Journal of Physics}\ }\textbf {\bibinfo
  {volume} {10}} (\bibinfo {year} {2008}),\ 10.1088/1367-2630/10/8/083024},\
  \Eprint {http://arxiv.org/abs/0803.1427} {arXiv:0803.1427} \BibitemShut
  {NoStop}%
\bibitem [{\citenamefont {Cywi{\'{n}}ski}\ \emph {et~al.}(2008)\citenamefont
  {Cywi{\'{n}}ski}, \citenamefont {Lutchyn}, \citenamefont {Nave},\ and\
  \citenamefont {{Das Sarma}}}]{Cywinski2008}%
  \BibitemOpen
  \bibfield  {author} {\bibinfo {author} {\bibfnamefont {{\L}.}~\bibnamefont
  {Cywi{\'{n}}ski}}, \bibinfo {author} {\bibfnamefont {R.~M.}\ \bibnamefont
  {Lutchyn}}, \bibinfo {author} {\bibfnamefont {C.~P.}\ \bibnamefont {Nave}}, \
  and\ \bibinfo {author} {\bibfnamefont {S.}~\bibnamefont {{Das Sarma}}},\
  }\href {\doibase 10.1103/PhysRevB.77.174509} {\bibfield  {journal} {\bibinfo
  {journal} {Physical Review B - Condensed Matter and Materials Physics}\
  }\textbf {\bibinfo {volume} {77}},\ \bibinfo {pages} {1} (\bibinfo {year}
  {2008})},\ \Eprint {http://arxiv.org/abs/0712.2225} {arXiv:0712.2225}
  \BibitemShut {NoStop}%
\bibitem [{\citenamefont {Bylander}\ \emph
  {et~al.}(2011{\natexlab{a}})\citenamefont {Bylander}, \citenamefont
  {Gustavsson}, \citenamefont {Yan}, \citenamefont {Yoshihara}, \citenamefont
  {Harrabi}, \citenamefont {Fitch}, \citenamefont {Cory}, \citenamefont
  {Nakamura}, \citenamefont {Tsai},\ and\ \citenamefont
  {Oliver}}]{Bylander2011}%
  \BibitemOpen
  \bibfield  {author} {\bibinfo {author} {\bibfnamefont {J.}~\bibnamefont
  {Bylander}}, \bibinfo {author} {\bibfnamefont {S.}~\bibnamefont
  {Gustavsson}}, \bibinfo {author} {\bibfnamefont {F.}~\bibnamefont {Yan}},
  \bibinfo {author} {\bibfnamefont {F.}~\bibnamefont {Yoshihara}}, \bibinfo
  {author} {\bibfnamefont {K.}~\bibnamefont {Harrabi}}, \bibinfo {author}
  {\bibfnamefont {G.}~\bibnamefont {Fitch}}, \bibinfo {author} {\bibfnamefont
  {D.~G.}\ \bibnamefont {Cory}}, \bibinfo {author} {\bibfnamefont
  {Y.}~\bibnamefont {Nakamura}}, \bibinfo {author} {\bibfnamefont {J.~S.}\
  \bibnamefont {Tsai}}, \ and\ \bibinfo {author} {\bibfnamefont {W.~D.}\
  \bibnamefont {Oliver}},\ }\href {\doibase 10.1038/nphys1994} {\bibfield
  {journal} {\bibinfo  {journal} {Nature Physics}\ }\textbf {\bibinfo {volume}
  {7}},\ \bibinfo {pages} {565} (\bibinfo {year}
  {2011}{\natexlab{a}})}\BibitemShut {NoStop}%
\bibitem [{\citenamefont {Biercuk}\ \emph {et~al.}(2011)\citenamefont
  {Biercuk}, \citenamefont {Doherty},\ and\ \citenamefont {Uys}}]{Biercuk2011}%
  \BibitemOpen
  \bibfield  {author} {\bibinfo {author} {\bibfnamefont {M.~J.}\ \bibnamefont
  {Biercuk}}, \bibinfo {author} {\bibfnamefont {A.~C.}\ \bibnamefont
  {Doherty}}, \ and\ \bibinfo {author} {\bibfnamefont {H.}~\bibnamefont
  {Uys}},\ }\href {\doibase 10.1088/0953-4075/44/15/154002} {\bibfield
  {journal} {\bibinfo  {journal} {Journal of Physics B: Atomic, Molecular and
  Optical Physics}\ }\textbf {\bibinfo {volume} {44}} (\bibinfo {year}
  {2011}),\ 10.1088/0953-4075/44/15/154002},\ \Eprint
  {http://arxiv.org/abs/1012.4262} {arXiv:1012.4262} \BibitemShut {NoStop}%
\bibitem [{\citenamefont {Norris}\ \emph {et~al.}(2016)\citenamefont {Norris},
  \citenamefont {Paz-Silva},\ and\ \citenamefont {Viola}}]{Norris2016}%
  \BibitemOpen
  \bibfield  {author} {\bibinfo {author} {\bibfnamefont {L.~M.}\ \bibnamefont
  {Norris}}, \bibinfo {author} {\bibfnamefont {G.~A.}\ \bibnamefont
  {Paz-Silva}}, \ and\ \bibinfo {author} {\bibfnamefont {L.}~\bibnamefont
  {Viola}},\ }\href {\doibase 10.1103/PhysRevLett.116.150503} {\bibfield
  {journal} {\bibinfo  {journal} {Physical Review Letters}\ }\textbf {\bibinfo
  {volume} {116}},\ \bibinfo {pages} {1} (\bibinfo {year} {2016})},\ \Eprint
  {http://arxiv.org/abs/1512.01575} {arXiv:1512.01575} \BibitemShut {NoStop}%
\bibitem [{\citenamefont {Sung}\ \emph {et~al.}(2019)\citenamefont {Sung},
  \citenamefont {Beaudoin}, \citenamefont {Norris}, \citenamefont {Yan},
  \citenamefont {Kim}, \citenamefont {Qiu}, \citenamefont {von L{\"{u}}pke},
  \citenamefont {Yoder}, \citenamefont {Orlando}, \citenamefont {Gustavsson},
  \citenamefont {Viola},\ and\ \citenamefont {Oliver}}]{Sung2019}%
  \BibitemOpen
  \bibfield  {author} {\bibinfo {author} {\bibfnamefont {Y.}~\bibnamefont
  {Sung}}, \bibinfo {author} {\bibfnamefont {F.}~\bibnamefont {Beaudoin}},
  \bibinfo {author} {\bibfnamefont {L.~M.}\ \bibnamefont {Norris}}, \bibinfo
  {author} {\bibfnamefont {F.}~\bibnamefont {Yan}}, \bibinfo {author}
  {\bibfnamefont {D.~K.}\ \bibnamefont {Kim}}, \bibinfo {author} {\bibfnamefont
  {J.~Y.}\ \bibnamefont {Qiu}}, \bibinfo {author} {\bibfnamefont
  {U.}~\bibnamefont {von L{\"{u}}pke}}, \bibinfo {author} {\bibfnamefont
  {J.~L.}\ \bibnamefont {Yoder}}, \bibinfo {author} {\bibfnamefont {T.~P.}\
  \bibnamefont {Orlando}}, \bibinfo {author} {\bibfnamefont {S.}~\bibnamefont
  {Gustavsson}}, \bibinfo {author} {\bibfnamefont {L.}~\bibnamefont {Viola}}, \
  and\ \bibinfo {author} {\bibfnamefont {W.~D.}\ \bibnamefont {Oliver}},\
  }\href {\doibase 10.1038/s41467-019-11699-4} {\bibfield  {journal} {\bibinfo
  {journal} {Nature Communications}\ }\textbf {\bibinfo {volume} {10}},\
  \bibinfo {pages} {1} (\bibinfo {year} {2019})},\ \Eprint
  {http://arxiv.org/abs/1903.01043} {arXiv:1903.01043} \BibitemShut {NoStop}%
\bibitem [{\citenamefont {Paz-Silva}\ \emph {et~al.}(2017)\citenamefont
  {Paz-Silva}, \citenamefont {Norris},\ and\ \citenamefont
  {Viola}}]{viola_2017}%
  \BibitemOpen
  \bibfield  {author} {\bibinfo {author} {\bibfnamefont {G.~A.}\ \bibnamefont
  {Paz-Silva}}, \bibinfo {author} {\bibfnamefont {L.~M.}\ \bibnamefont
  {Norris}}, \ and\ \bibinfo {author} {\bibfnamefont {L.}~\bibnamefont
  {Viola}},\ }\href {\doibase 10.1103/physreva.95.022121} {\bibfield  {journal}
  {\bibinfo  {journal} {Physical Review A}\ }\textbf {\bibinfo {volume} {95}}
  (\bibinfo {year} {2017}),\ 10.1103/physreva.95.022121}\BibitemShut {NoStop}%
\bibitem [{\citenamefont {Szankowski}\ \emph {et~al.}(2016)\citenamefont
  {Szankowski}, \citenamefont {Trippenbach},\ and\ \citenamefont
  {Cywinski}}]{euros_2016}%
  \BibitemOpen
  \bibfield  {author} {\bibinfo {author} {\bibfnamefont {P.}~\bibnamefont
  {Szankowski}}, \bibinfo {author} {\bibfnamefont {M.}~\bibnamefont
  {Trippenbach}}, \ and\ \bibinfo {author} {\bibfnamefont {L.}~\bibnamefont
  {Cywinski}},\ }\href {\doibase 10.1103/physreva.94.012109} {\bibfield
  {journal} {\bibinfo  {journal} {Physical Review A}\ }\textbf {\bibinfo
  {volume} {94}} (\bibinfo {year} {2016}),\
  10.1103/physreva.94.012109}\BibitemShut {NoStop}%
\bibitem [{\citenamefont {Krzywda}\ \emph {et~al.}(2019)\citenamefont
  {Krzywda}, \citenamefont {Szankowski},\ and\ \citenamefont
  {Cywinski}}]{euros_2019}%
  \BibitemOpen
  \bibfield  {author} {\bibinfo {author} {\bibfnamefont {J.}~\bibnamefont
  {Krzywda}}, \bibinfo {author} {\bibfnamefont {P.}~\bibnamefont {Szankowski}},
  \ and\ \bibinfo {author} {\bibfnamefont {L.}~\bibnamefont {Cywinski}},\
  }\href {\doibase 10.1088/1367-2630/ab0ce7} {\bibfield  {journal} {\bibinfo
  {journal} {New Journal of Physics}\ }\textbf {\bibinfo {volume} {21}}
  (\bibinfo {year} {2019}),\ 10.1088/1367-2630/ab0ce7}\BibitemShut {NoStop}%
\bibitem [{\citenamefont {Dutta}\ and\ \citenamefont
  {Horn}(1981)}]{dutta_1981}%
  \BibitemOpen
  \bibfield  {author} {\bibinfo {author} {\bibfnamefont {P.}~\bibnamefont
  {Dutta}}\ and\ \bibinfo {author} {\bibfnamefont {P.~M.}\ \bibnamefont
  {Horn}},\ }\href {\doibase 10.1103/RevModPhys.53.497} {\bibfield  {journal}
  {\bibinfo  {journal} {Rev. Mod. Phys.}\ }\textbf {\bibinfo {volume} {53}},\
  \bibinfo {pages} {497} (\bibinfo {year} {1981})}\BibitemShut {NoStop}%
\bibitem [{\citenamefont {Schoelkopf}\ \emph {et~al.}(1998)\citenamefont
  {Schoelkopf}, \citenamefont {Wahlgren}, \citenamefont {Kozhevnikov},
  \citenamefont {Delsing},\ and\ \citenamefont {Prober}}]{Schoelkopf1998}%
  \BibitemOpen
  \bibfield  {author} {\bibinfo {author} {\bibfnamefont {R.~J.}\ \bibnamefont
  {Schoelkopf}}, \bibinfo {author} {\bibfnamefont {P.}~\bibnamefont
  {Wahlgren}}, \bibinfo {author} {\bibfnamefont {A.~A.}\ \bibnamefont
  {Kozhevnikov}}, \bibinfo {author} {\bibfnamefont {P.}~\bibnamefont
  {Delsing}}, \ and\ \bibinfo {author} {\bibfnamefont {D.~E.}\ \bibnamefont
  {Prober}},\ }\href {\doibase 10.1126/science.280.5367.1238} {\bibfield
  {journal} {\bibinfo  {journal} {Science}\ }\textbf {\bibinfo {volume}
  {280}},\ \bibinfo {pages} {1238} (\bibinfo {year} {1998})}\BibitemShut
  {NoStop}%
\bibitem [{\citenamefont {An}\ \emph {et~al.}(2019)\citenamefont {An},
  \citenamefont {Matthiesen}, \citenamefont {Urban},\ and\ \citenamefont
  {H{\"{a}}ffner}}]{An2019}%
  \BibitemOpen
  \bibfield  {author} {\bibinfo {author} {\bibfnamefont {D.}~\bibnamefont
  {An}}, \bibinfo {author} {\bibfnamefont {C.}~\bibnamefont {Matthiesen}},
  \bibinfo {author} {\bibfnamefont {E.}~\bibnamefont {Urban}}, \ and\ \bibinfo
  {author} {\bibfnamefont {H.}~\bibnamefont {H{\"{a}}ffner}},\ }\href {\doibase
  10.1103/PhysRevA.100.063405} {\bibfield  {journal} {\bibinfo  {journal}
  {Physical Review A}\ }\textbf {\bibinfo {volume} {100}} (\bibinfo {year}
  {2019}),\ 10.1103/PhysRevA.100.063405},\ \Eprint
  {http://arxiv.org/abs/1906.06489} {arXiv:1906.06489} \BibitemShut {NoStop}%
\bibitem [{\citenamefont {Yoshihara}\ \emph
  {et~al.}(2014{\natexlab{a}})\citenamefont {Yoshihara}, \citenamefont
  {Nakamura}, \citenamefont {Yan}, \citenamefont {Gustavsson}, \citenamefont
  {Bylander}, \citenamefont {Oliver},\ and\ \citenamefont
  {Tsai}}]{Yoshihara2014}%
  \BibitemOpen
  \bibfield  {author} {\bibinfo {author} {\bibfnamefont {F.}~\bibnamefont
  {Yoshihara}}, \bibinfo {author} {\bibfnamefont {Y.}~\bibnamefont {Nakamura}},
  \bibinfo {author} {\bibfnamefont {F.}~\bibnamefont {Yan}}, \bibinfo {author}
  {\bibfnamefont {S.}~\bibnamefont {Gustavsson}}, \bibinfo {author}
  {\bibfnamefont {J.}~\bibnamefont {Bylander}}, \bibinfo {author}
  {\bibfnamefont {W.~D.}\ \bibnamefont {Oliver}}, \ and\ \bibinfo {author}
  {\bibfnamefont {J.-S.}\ \bibnamefont {Tsai}},\ }\href {\doibase
  10.1103/physrevb.89.020503} {\bibfield  {journal} {\bibinfo  {journal}
  {Physical Review B}\ }\textbf {\bibinfo {volume} {89}} (\bibinfo {year}
  {2014}{\natexlab{a}}),\ 10.1103/physrevb.89.020503}\BibitemShut {NoStop}%
\bibitem [{\citenamefont {Yan}\ \emph {et~al.}(2018)\citenamefont {Yan},
  \citenamefont {Krantz}, \citenamefont {Sung}, \citenamefont {Kjaergaard},
  \citenamefont {Campbell}, \citenamefont {Orlando}, \citenamefont
  {Gustavsson},\ and\ \citenamefont {Oliver}}]{Yan2018}%
  \BibitemOpen
  \bibfield  {author} {\bibinfo {author} {\bibfnamefont {F.}~\bibnamefont
  {Yan}}, \bibinfo {author} {\bibfnamefont {P.}~\bibnamefont {Krantz}},
  \bibinfo {author} {\bibfnamefont {Y.}~\bibnamefont {Sung}}, \bibinfo {author}
  {\bibfnamefont {M.}~\bibnamefont {Kjaergaard}}, \bibinfo {author}
  {\bibfnamefont {D.~L.}\ \bibnamefont {Campbell}}, \bibinfo {author}
  {\bibfnamefont {T.~P.}\ \bibnamefont {Orlando}}, \bibinfo {author}
  {\bibfnamefont {S.}~\bibnamefont {Gustavsson}}, \ and\ \bibinfo {author}
  {\bibfnamefont {W.~D.}\ \bibnamefont {Oliver}},\ }\href {\doibase
  10.1103/PhysRevApplied.10.054062} {\bibfield  {journal} {\bibinfo  {journal}
  {Physical Review Applied}\ }\textbf {\bibinfo {volume} {10}} (\bibinfo {year}
  {2018}),\ 10.1103/PhysRevApplied.10.054062},\ \Eprint
  {http://arxiv.org/abs/1803.09813} {arXiv:1803.09813} \BibitemShut {NoStop}%
\bibitem [{\citenamefont {Koch}\ \emph {et~al.}(1983)\citenamefont {Koch},
  \citenamefont {Clarke}, \citenamefont {Goubau}, \citenamefont {Martinis},
  \citenamefont {Pegrum},\ and\ \citenamefont {{van Harlingen}}}]{koch_1983}%
  \BibitemOpen
  \bibfield  {author} {\bibinfo {author} {\bibfnamefont {R.}~\bibnamefont
  {Koch}}, \bibinfo {author} {\bibfnamefont {J.}~\bibnamefont {Clarke}},
  \bibinfo {author} {\bibfnamefont {W.}~\bibnamefont {Goubau}}, \bibinfo
  {author} {\bibfnamefont {J.}~\bibnamefont {Martinis}}, \bibinfo {author}
  {\bibfnamefont {C.}~\bibnamefont {Pegrum}}, \ and\ \bibinfo {author}
  {\bibfnamefont {D.}~\bibnamefont {{van Harlingen}}},\ }\href {\doibase
  10.1007/BF00683423} {\bibfield  {journal} {\bibinfo  {journal} {Journal of
  Low Temperature Physics}\ }\textbf {\bibinfo {volume} {51}},\ \bibinfo
  {pages} {207} (\bibinfo {year} {1983})}\BibitemShut {NoStop}%
\bibitem [{\citenamefont {Hahn}(1950)}]{Hahn1950}%
  \BibitemOpen
  \bibfield  {author} {\bibinfo {author} {\bibfnamefont {E.~L.}\ \bibnamefont
  {Hahn}},\ }\href {\doibase 10.1103/PhysRev.80.580} {\bibfield  {journal}
  {\bibinfo  {journal} {Phys. Rev.}\ }\textbf {\bibinfo {volume} {80}},\
  \bibinfo {pages} {580} (\bibinfo {year} {1950})}\BibitemShut {NoStop}%
\bibitem [{\citenamefont {Carr}\ and\ \citenamefont
  {Purcell}(1954)}]{carr1954}%
  \BibitemOpen
  \bibfield  {author} {\bibinfo {author} {\bibfnamefont {H.~Y.}\ \bibnamefont
  {Carr}}\ and\ \bibinfo {author} {\bibfnamefont {E.~M.}\ \bibnamefont
  {Purcell}},\ }\href {\doibase 10.1103/PhysRev.94.630} {\bibfield  {journal}
  {\bibinfo  {journal} {Phys. Rev.}\ }\textbf {\bibinfo {volume} {94}},\
  \bibinfo {pages} {630} (\bibinfo {year} {1954})}\BibitemShut {NoStop}%
\bibitem [{\citenamefont {Meiboom}\ and\ \citenamefont
  {Gill}(1958)}]{gill1958}%
  \BibitemOpen
  \bibfield  {author} {\bibinfo {author} {\bibfnamefont {S.}~\bibnamefont
  {Meiboom}}\ and\ \bibinfo {author} {\bibfnamefont {D.}~\bibnamefont {Gill}},\
  }\href {\doibase 10.1063/1.1716296} {\bibfield  {journal} {\bibinfo
  {journal} {Review of Scientific Instruments}\ }\textbf {\bibinfo {volume}
  {29}},\ \bibinfo {pages} {688} (\bibinfo {year} {1958})},\ \Eprint
  {http://arxiv.org/abs/https://doi.org/10.1063/1.1716296}
  {https://doi.org/10.1063/1.1716296} \BibitemShut {NoStop}%
\bibitem [{\citenamefont {Hooge}(1994)}]{Hooge1994}%
  \BibitemOpen
  \bibfield  {author} {\bibinfo {author} {\bibfnamefont {F.}~\bibnamefont
  {Hooge}},\ }\href {\doibase 10.1109/16.333808} {\bibfield  {journal}
  {\bibinfo  {journal} {IEEE Transactions on Electron Devices}\ }\textbf
  {\bibinfo {volume} {41}},\ \bibinfo {pages} {1926} (\bibinfo {year}
  {1994})}\BibitemShut {NoStop}%
\bibitem [{\citenamefont {Ramon}(2015)}]{Ramon2015}%
  \BibitemOpen
  \bibfield  {author} {\bibinfo {author} {\bibfnamefont {G.}~\bibnamefont
  {Ramon}},\ }\href {\doibase 10.1103/PhysRevB.92.155422} {\bibfield  {journal}
  {\bibinfo  {journal} {Physical Review B - Condensed Matter and Materials
  Physics}\ }\textbf {\bibinfo {volume} {92}},\ \bibinfo {pages} {1} (\bibinfo
  {year} {2015})}\BibitemShut {NoStop}%
\bibitem [{\citenamefont {Galperin}\ \emph {et~al.}(2006)\citenamefont
  {Galperin}, \citenamefont {Altshuler}, \citenamefont {Bergli},\ and\
  \citenamefont {Shantsev}}]{Galperin2006}%
  \BibitemOpen
  \bibfield  {author} {\bibinfo {author} {\bibfnamefont {Y.~M.}\ \bibnamefont
  {Galperin}}, \bibinfo {author} {\bibfnamefont {B.~L.}\ \bibnamefont
  {Altshuler}}, \bibinfo {author} {\bibfnamefont {J.}~\bibnamefont {Bergli}}, \
  and\ \bibinfo {author} {\bibfnamefont {D.~V.}\ \bibnamefont {Shantsev}},\
  }\href {\doibase 10.1103/PhysRevLett.96.097009} {\bibfield  {journal}
  {\bibinfo  {journal} {Physical Review Letters}\ }\textbf {\bibinfo {volume}
  {96}},\ \bibinfo {pages} {1} (\bibinfo {year} {2006})},\ \Eprint
  {http://arxiv.org/abs/0511238} {arXiv:0511238 [cond-mat]} \BibitemShut
  {NoStop}%
\bibitem [{\citenamefont {Faoro}\ and\ \citenamefont
  {Viola}(2004)}]{Faoro2004}%
  \BibitemOpen
  \bibfield  {author} {\bibinfo {author} {\bibfnamefont {L.}~\bibnamefont
  {Faoro}}\ and\ \bibinfo {author} {\bibfnamefont {L.}~\bibnamefont {Viola}},\
  }\href {\doibase 10.1103/PhysRevLett.92.117905} {\bibfield  {journal}
  {\bibinfo  {journal} {Physical Review Letters}\ }\textbf {\bibinfo {volume}
  {92}},\ \bibinfo {pages} {1} (\bibinfo {year} {2004})},\ \Eprint
  {http://arxiv.org/abs/0312159} {arXiv:0312159 [quant-ph]} \BibitemShut
  {NoStop}%
\bibitem [{\citenamefont {Hayes}\ \emph {et~al.}(2012)\citenamefont {Hayes},
  \citenamefont {Clark}, \citenamefont {Debnath}, \citenamefont {Hucul},
  \citenamefont {Inlek}, \citenamefont {Lee}, \citenamefont {Quraishi},\ and\
  \citenamefont {Monroe}}]{Hayes2012}%
  \BibitemOpen
  \bibfield  {author} {\bibinfo {author} {\bibfnamefont {D.}~\bibnamefont
  {Hayes}}, \bibinfo {author} {\bibfnamefont {S.~M.}\ \bibnamefont {Clark}},
  \bibinfo {author} {\bibfnamefont {S.}~\bibnamefont {Debnath}}, \bibinfo
  {author} {\bibfnamefont {D.}~\bibnamefont {Hucul}}, \bibinfo {author}
  {\bibfnamefont {I.~V.}\ \bibnamefont {Inlek}}, \bibinfo {author}
  {\bibfnamefont {K.~W.}\ \bibnamefont {Lee}}, \bibinfo {author} {\bibfnamefont
  {Q.}~\bibnamefont {Quraishi}}, \ and\ \bibinfo {author} {\bibfnamefont
  {C.}~\bibnamefont {Monroe}},\ }\href {\doibase
  10.1103/PhysRevLett.109.020503} {\bibfield  {journal} {\bibinfo  {journal}
  {Phys. Rev. Lett.}\ }\textbf {\bibinfo {volume} {109}},\ \bibinfo {pages}
  {020503} (\bibinfo {year} {2012})}\BibitemShut {NoStop}%
\bibitem [{\citenamefont {Mølmer}\ and\ \citenamefont
  {Sørensen}(1999)}]{Molmer1999}%
  \BibitemOpen
  \bibfield  {author} {\bibinfo {author} {\bibfnamefont {K.}~\bibnamefont
  {Mølmer}}\ and\ \bibinfo {author} {\bibfnamefont {A.}~\bibnamefont
  {Sørensen}},\ }\href {\doibase 10.1103/physrevlett.82.1835} {\bibfield
  {journal} {\bibinfo  {journal} {Physical Review Letters}\ }\textbf {\bibinfo
  {volume} {82}},\ \bibinfo {pages} {1835–1838} (\bibinfo {year}
  {1999})}\BibitemShut {NoStop}%
\bibitem [{\citenamefont {Klyatskin}(2011)}]{Klyatskin2011}%
  \BibitemOpen
  \bibfield  {author} {\bibinfo {author} {\bibfnamefont {V.~I.}\ \bibnamefont
  {Klyatskin}},\ }\href@noop {} {\emph {\bibinfo {title} {Lectures on
  {{Dynamics}} of {{Stochastic Systems}}}}}\ (\bibinfo  {publisher} {{Elsevier,
  Amsterdam}},\ \bibinfo {year} {2011})\BibitemShut {NoStop}%
\bibitem [{\citenamefont {Bylander}\ \emph
  {et~al.}(2011{\natexlab{b}})\citenamefont {Bylander}, \citenamefont
  {Gustavsson}, \citenamefont {Yan}, \citenamefont {Yoshihara}, \citenamefont
  {Harrabi}, \citenamefont {Fitch}, \citenamefont {Cory}, \citenamefont
  {Nakamura}, \citenamefont {Tsai},\ and\ \citenamefont
  {Oliver}}]{bylander_11}%
  \BibitemOpen
  \bibfield  {author} {\bibinfo {author} {\bibfnamefont {J.}~\bibnamefont
  {Bylander}}, \bibinfo {author} {\bibfnamefont {S.}~\bibnamefont
  {Gustavsson}}, \bibinfo {author} {\bibfnamefont {F.}~\bibnamefont {Yan}},
  \bibinfo {author} {\bibfnamefont {F.}~\bibnamefont {Yoshihara}}, \bibinfo
  {author} {\bibfnamefont {K.}~\bibnamefont {Harrabi}}, \bibinfo {author}
  {\bibfnamefont {G.}~\bibnamefont {Fitch}}, \bibinfo {author} {\bibfnamefont
  {D.~G.}\ \bibnamefont {Cory}}, \bibinfo {author} {\bibfnamefont
  {Y.}~\bibnamefont {Nakamura}}, \bibinfo {author} {\bibfnamefont {J.-S.}\
  \bibnamefont {Tsai}}, \ and\ \bibinfo {author} {\bibfnamefont {W.~D.}\
  \bibnamefont {Oliver}},\ }\href {\doibase 10.1038/nphys1994} {\bibfield
  {journal} {\bibinfo  {journal} {Nature Physics}\ }\textbf {\bibinfo {volume}
  {7}},\ \bibinfo {pages} {565–570} (\bibinfo {year}
  {2011}{\natexlab{b}})}\BibitemShut {NoStop}%
\bibitem [{\citenamefont {Slichter}\ \emph {et~al.}(2012)\citenamefont
  {Slichter}, \citenamefont {Vijay}, \citenamefont {Weber}, \citenamefont
  {Boutin}, \citenamefont {Boissonneault}, \citenamefont {Gambetta},
  \citenamefont {Blais},\ and\ \citenamefont {Siddiqi}}]{siddiqi_12}%
  \BibitemOpen
  \bibfield  {author} {\bibinfo {author} {\bibfnamefont {D.~H.}\ \bibnamefont
  {Slichter}}, \bibinfo {author} {\bibfnamefont {R.}~\bibnamefont {Vijay}},
  \bibinfo {author} {\bibfnamefont {S.~J.}\ \bibnamefont {Weber}}, \bibinfo
  {author} {\bibfnamefont {S.}~\bibnamefont {Boutin}}, \bibinfo {author}
  {\bibfnamefont {M.}~\bibnamefont {Boissonneault}}, \bibinfo {author}
  {\bibfnamefont {J.~M.}\ \bibnamefont {Gambetta}}, \bibinfo {author}
  {\bibfnamefont {A.}~\bibnamefont {Blais}}, \ and\ \bibinfo {author}
  {\bibfnamefont {I.}~\bibnamefont {Siddiqi}},\ }\href {\doibase
  10.1103/physrevlett.109.153601} {\bibfield  {journal} {\bibinfo  {journal}
  {Physical Review Letters}\ }\textbf {\bibinfo {volume} {109}} (\bibinfo
  {year} {2012}),\ 10.1103/physrevlett.109.153601}\BibitemShut {NoStop}%
\bibitem [{\citenamefont {Yan}\ \emph {et~al.}(2013)\citenamefont {Yan},
  \citenamefont {Gustavsson}, \citenamefont {Bylander}, \citenamefont {Jin},
  \citenamefont {Yoshihara}, \citenamefont {Cory}, \citenamefont {Nakamura},
  \citenamefont {Orlando},\ and\ \citenamefont {Oliver}}]{yan_13}%
  \BibitemOpen
  \bibfield  {author} {\bibinfo {author} {\bibfnamefont {F.}~\bibnamefont
  {Yan}}, \bibinfo {author} {\bibfnamefont {S.}~\bibnamefont {Gustavsson}},
  \bibinfo {author} {\bibfnamefont {J.}~\bibnamefont {Bylander}}, \bibinfo
  {author} {\bibfnamefont {X.}~\bibnamefont {Jin}}, \bibinfo {author}
  {\bibfnamefont {F.}~\bibnamefont {Yoshihara}}, \bibinfo {author}
  {\bibfnamefont {D.~G.}\ \bibnamefont {Cory}}, \bibinfo {author}
  {\bibfnamefont {Y.}~\bibnamefont {Nakamura}}, \bibinfo {author}
  {\bibfnamefont {T.~P.}\ \bibnamefont {Orlando}}, \ and\ \bibinfo {author}
  {\bibfnamefont {W.~D.}\ \bibnamefont {Oliver}},\ }\href@noop {} {\bibfield
  {journal} {\bibinfo  {journal} {Nature communications}\ }\textbf {\bibinfo
  {volume} {4}},\ \bibinfo {pages} {2337} (\bibinfo {year} {2013})}\BibitemShut
  {NoStop}%
\bibitem [{\citenamefont {Yoshihara}\ \emph
  {et~al.}(2014{\natexlab{b}})\citenamefont {Yoshihara}, \citenamefont
  {Nakamura}, \citenamefont {Yan}, \citenamefont {Gustavsson}, \citenamefont
  {Bylander}, \citenamefont {Oliver},\ and\ \citenamefont {Tsai}}]{oliver_14}%
  \BibitemOpen
  \bibfield  {author} {\bibinfo {author} {\bibfnamefont {F.}~\bibnamefont
  {Yoshihara}}, \bibinfo {author} {\bibfnamefont {Y.}~\bibnamefont {Nakamura}},
  \bibinfo {author} {\bibfnamefont {F.}~\bibnamefont {Yan}}, \bibinfo {author}
  {\bibfnamefont {S.}~\bibnamefont {Gustavsson}}, \bibinfo {author}
  {\bibfnamefont {J.}~\bibnamefont {Bylander}}, \bibinfo {author}
  {\bibfnamefont {W.~D.}\ \bibnamefont {Oliver}}, \ and\ \bibinfo {author}
  {\bibfnamefont {J.-S.}\ \bibnamefont {Tsai}},\ }\href {\doibase
  10.1103/physrevb.89.020503} {\bibfield  {journal} {\bibinfo  {journal}
  {Physical Review B}\ }\textbf {\bibinfo {volume} {89}} (\bibinfo {year}
  {2014}{\natexlab{b}}),\ 10.1103/physrevb.89.020503}\BibitemShut {NoStop}%
\end{thebibliography}%
%\end{bibunit}

\onecolumngrid
\clearpage

\appendix

\setcounter{page}{1}
\setcounter{table}{0}
\setcounter{figure}{0}
\renewcommand{\thepage}{S\arabic{page}}  
\renewcommand{\thetable}{S\arabic{table}}  
\renewcommand{\thefigure}{S\arabic{figure}}
\appendixtitleoff
\renewcommand{\appendixtocname}{Supplementary material}
\begin{appendices}

\part{} % Start the appendix part
\parttoc
%\begin{bibunit}[ieeetr]
\section{Telegraph Noise}\label{sec:tele_noise}

Here the statistics of a single random telegraph noise (RTN) process are briefly reviewed. The RTN process is a Markovian process in which the variable $\xi(t)$ switches randomly between two values $\xi(t)=\pm 1$ with an average rate $\gamma$ \cite{Klyatskin2011}. We assume that the noise is symmetric, i.e. the probabilities of switching "up" and "down" are equal. The number of switches during time interval $(0,t)$ is described by a Poisson distribution. As  such, we can  obtain the differential equation for the probability distribution of $\xi (t)$:
\begin{equation}
\label{Eq:PdfEquation}
\frac{d}{dt}P^\xi_{\sigma_0,\sigma}(t)=-\gamma(P^\xi_{\sigma_0,\sigma}(t)-P^\xi_{\sigma_0,-\sigma})=-\gamma(2P^\xi_{\sigma_0,\sigma}(t)-1)
\end{equation}
Here $P_{\sigma_0,\sigma}^\xi(t)$ is the probability that the discrete variable $\xi(t)$ assumes the value $\sigma$ conditioned on initial value $\xi(0)
=\sigma_0$ and we assumed that $P_{\sigma_0,+}(t)+P_{\sigma_0,-}(t)=1$. For example, $P_{+-}^\xi (t)$ is the probability that the variable $\xi(t)=-1$  conditioned upon  $\xi(0)
=+1$.
Eq.~\eqref{Eq:PdfEquation} can be readily solved, and we obtain:
\begin{equation}
P^\xi_{\sigma_0,\sigma}(t)=\left(\delta_{\sigma_0,\sigma}-\frac{1}{2}\right)e^{-2\gamma t}+\frac{1}{2}
\end{equation}
The two-point time-domain correlator of RTN and its Fourier transform can be obtained as well:
\begin{equation}
\label{Eq:correlator}
\langle \xi(t)\xi(0)\rangle=e^{-2\gamma|t|}
\end{equation} 
and
\begin{equation}
\label{Eq:Sf}
S(f)=\int_{-\infty}^{\infty}dt e^{i2\pi f t}\langle \xi(t)\xi(0)\rangle=
%\frac{4\gamma}{\omega^2+4\gamma^2}=
\frac{\gamma}{\pi^2f^2 +\gamma^2}
\end{equation} 

There also exists a useful recurrence relationship between higher order correlators and the second order correlator \cite{Klyatskin2011},

\begin{equation}\label{eq:tele_recurr}
    \langle \xi(t_1) \dots \xi(t_n) \rangle = \langle \xi(t_1) \xi(t_2) \rangle \langle \xi(t_3) \dots \xi(t_n) \rangle 
\end{equation}
or $t_1\geq t_2> t_3,...,t_n$.

\section{$1/f$ Noise From Telegraph Fluctuators}\label{sec:1_f_from_tele}

Consider the case of the noise that drives the qubit coming from the coupling to many independent random telegraph fluctuators. If the probability densities of the coupling strength $\lambda$ and of the inverse correlation times of the fluctuators $\gamma$ are $g_{\lambda}(\lambda)$ and $g_{\gamma}(\gamma)$, respectively, the power spectrum of the noise is,

\begin{equation}
    S(f) = \int_{-\infty}^{\infty} d\lambda \: g_{\lambda}(\lambda) \int_{-\infty}^{\infty } d\gamma  \:  g_{\gamma}(\gamma) \frac{\lambda^2 \gamma}{\pi^2 f^2 + \gamma^2}
\end{equation}

In the particular case where $\lambda$ is the same for all fluctuators, $g_{\lambda_0}(\lambda_0) = \delta(\lambda_0-\lambda)$, and $\gamma$ has a log-uniform density of states,

\begin{equation}
    g_{\gamma}(\gamma) = \begin{cases}
    \frac{1}{\left(\ln{(\gamma_{\text{max}})} - \ln{(\gamma_{\text{min}})} \right)\gamma}, & \gamma_{\text{min}} < \gamma < \gamma_{\text{max}}\\
    0, & \text{otherwise} \\
  \end{cases}
\end{equation}
 
$S(f)$ takes the form,

\begin{equation}
    S(f) = \lambda^2 \frac{\cot^{-1}{\left(\frac{f \pi}{\gamma_{\text{max}}}\right)} - \cot^{-1}{\left(\frac{f \pi}{\gamma_{\text{min}}}\right)}}{f \pi \left(\ln{(\gamma_{\text{max}})} - \ln{(\gamma_{\text{min}})} \right)}
\end{equation}

When $\gamma_{\text{min}} << f << \gamma_{\text{max}}$, the numerator is well approximated by $\frac{\pi}{2}$, and the noise spectrum is approximately $S(f) = A/f$. It should be noted that there are many possible other choices of $g_{\lambda}(\lambda)$ and $g_{\gamma}(\gamma)$ that yield a similar result.

\section{Flux Sensitivity in  Tunable Coupler Systems}\label{sec:flux_noise}

We are interested in understanding the dependance of the flux sensitivity $\tilde\chi_\Phi$ on the coupling $g$. To do this we start with the formula for the total coupling $g$ between the qubits:
\begin{equation}
\label{Eq:g}
g=\left(k_d-k^2\frac{\omega_q^2}{\omega_c^2-\omega_q^2}\right)\frac{\omega_q}{2},
\end{equation}
where $\omega_q$ is the qubit frequency (we assume that both qubits are on resonance), $k_d$ and $k^2$ are dimensionless parameters that we will call coupling efficiencies.
These parameters can be expressed through effective capacitances of the qubit-coupler system \cite{Yan2018}. $\omega_c$ is the coupler frequency:
\begin{equation}
\label{Eq:omega_c}
\omega_c\simeq\frac{1}{\hbar}\left(\sqrt{8E_JE_c\left |\cos(\pi\Phi/\Phi_0)\right |}-E_C\right)
\simeq \omega_{\max}\sqrt{\left |\cos(\pi\Phi/\Phi_0)\right |},
\end{equation}
where $E_J$ and $E_C$ are the Josephson and charging energies respectively and $\omega_{\max}$ is the frequency of a
tunable transmon at the flux insensitive point $\Phi=0$. Expression~\eqref{Eq:omega_c} is valid if the strong inequality $E_J/E_C \gg 1$
is satisfied.
Using Eqs. \eqref{Eq:g} and \eqref{Eq:omega_c} we can express the flux sensitivity as
\begin{equation}
\label{Eq:chi_Phi}
\tilde\chi_\Phi=\left|\frac{\partial g}{\partial \Phi}\right|= \frac{ k^2 \omega_q^3   \omega_{\max }^2 
   \sin (\pi 
    \Phi/\Phi_0)}{4 \Phi_0 \left(\omega_{\max }^2
   \cos (\pi  \Phi/\Phi_0)-\omega_q^2\right)^2}
\end{equation}

Solving the system of equations~\eqref{Eq:g} and \eqref{Eq:omega_c} for $\Phi$ and $\omega_c$ and substituting the result into Eq.~\eqref{Eq:chi_Phi} we obtain the 
desired relation between the flux sensitivity and $g$: 
\begin{equation}
\label{Eq:chi(g)}   
\tilde\chi _{\Phi}(g)=\frac{\left(k_d
   \omega _q-2
   g\right)\sqrt{\omega_{\max}^4\left(k_d
   \omega _q-2
   g\right)^2-\omega _q^4
   \left(k_{qq}
   \omega _q-2
   g\right)^2}}{4 \Phi_0 k^2 \omega _q^3},
\end{equation} 
where $k_{qq}=k_d+k^2$ is the total coupling efficiency, which includes both direct and indirect interactions between the qubits. The flux sensitivity in \eqref{Eq:chi(g)} 
is positive since we are considering only those $\omega_c$ for which the indirect coupling prevails, i.e.  according to  Eq.~\eqref{Eq:g} the value of $g$ is negative; we also assume $\omega_c>\omega_q$.  
In the parameter range of  interest for the experiment, and in particular with the account taken of the smallness of the dimensionless parameter $k$, $\tilde\chi_\Phi(g)$ is extremely well approximated by a quadratic polynomial in $g$ as $\tilde\chi_\Phi(g)=\chi_\Phi^{(0)}+\chi_\Phi^{(1)}g+\chi_\Phi^{(2)}g^2$.  We note that in the vicinity of $g=0$ the $g$-noise is negligible and the total qubit dephasing time $T_\varphi$ is dominated by other noise sources. As such, we will treat the first term $\chi_\Phi^{(0)}$ of
the polynomial as a free parameter.

\section{Direct averaging over Noise Trajectories}\label{sec:direct_averaging}

Here the average of the stochastic dynamics of the qubit will be found by first analytically solving for the dynamics for a single trajectory of the RTN and then directly averaging over all possible trajectories.

First, we will consider the case of Coupler Ramsey decay under noise described by a single RTN process entering through the flux bias of the coupler. If the frequencies of the both qubits coincide, the complete Hamiltonian is,

\begin{equation}\label{eq:free_ham}
    H(t) = (g(t) + \lambda(t) \xi(t))\sigma_x
\end{equation}
Here, $g(t)$ is a periodic sequence of pulses with period $T_{\text{gate}}$. From the analysis completed in section \ref{sec:flux_noise}, this also implies that $\lambda(t)$ is periodic with the same period.  The associated propagator is,

\begin{equation}\label{eq:coupler_pulse}
    U(t) = e^{-i \sigma_x G(t, 0)}
\end{equation}

\begin{equation}
    G(t_2, t_1) = \int_{t_1}^{t_2} [g(\tau) + \lambda(\tau) \xi(\tau)] d \tau
\end{equation}

If the system is initialized in the state $\rho(0) = \ket{01}\bra{01}$, the z-component of the one-excitation subspace Bloch vector $N(t)\equiv Tr[\sigma_z\rho(t)]$ is given by the expression

\begin{align}\label{eq:unitary_Bloch}
\begin{split}
    N(t) &= \bra{01} U(t) \rho(0)U^{\dagger}(t)\ket{01} -
    \bra{10} U(t) \rho(0)U^{\dagger}(t)
    \ket{10} \\
    &= \text{Re}(e^{2 i \int_{0}^t \lambda(\tau) \xi(\tau) d \tau} e^{2 i \int_{0}^t g(\tau) d \tau} )
    \end{split}
\end{align}

The average Bloch vector component can be found by averaging Eq.~\ref{eq:unitary_Bloch} over noise trajectories. Taking into account that he measurement is done after an even  number of $g$-gates, $\int_0^t g(t')dt'  k\pi$, we have 
\begin{equation}
    \langle N(t) \rangle = \text{Re}\left( \chi(t) e^{2 i \int_{0}^t g(\tau) d \tau}\right)
\end{equation}

\begin{equation}
    \chi(t) = \langle e^{2 i \int_{0}^t \lambda(\tau) \xi(\tau) d \tau} \rangle
\end{equation}
The averaging can be  accomplished by expanding the functional $\chi(t)$ in a time-ordered Taylor series

\begin{equation}\label{eq:direct_av_time_ordered}
    \chi(t)  = \sum_{k=0}^{\infty} (2i)^k \int_{0}^{t} d t_1 \int_{0}^{t_1} d t_2 \dots \int_{0}^{t_{k-1}} d t_k \langle \xi(t_1) \dots \xi(t_k) \rangle \lambda(t_1) \dots \lambda(t_k) 
\end{equation}

In the considered problem $\lambda(t)$ is a periodic function of time. The typical period is the periodicity of the gate (moreover, $\lambda(t)$ is very nonsinusoidal, for much of the gate duration it is constant). The period of $\lambda(t)$ is much shorter than the typical time on which $\chi(t)$ varies. Therefore the major contribution to $\chi(t)$ comes from the term in $\lambda(t)$ that is independent of time. A justification of approximating $\lambda(t)$ by a constant can be done using the master equation formulation in Sec. V. One can see there that the fast-oscillating terms in $\lambda(t)$ lead to fast oscillating terms in the density matrix, which are small. 

For $\lambda(t)\equiv \lambda = \frac{1}{t_g}\int_0^{t_g} \lambda(\tau) d\tau$ we can use the recurrence relation for the moments given in Eq.~\ref{eq:tele_recurr} to find a second order linear differential equation for $ \chi(t)  $,

\begin{equation}\label{eq:direct_ramsey}
    \frac{d^2 \chi(t)  }{d t^2} + 2 \gamma \frac{d \chi(t)  }{d t} + 4 \lambda^2 \chi(t)  = 0
\end{equation}
We can infer the initial conditions $\chi(0) = 1$ and $\chi'(0)  = 0$ from Eq.~\ref{eq:direct_av_time_ordered}. Then

\begin{equation}\label{eq:free_z}
    \chi(t)  = e^{- \gamma t} \left ( \cosh{\left(  t \Omega \right)} + \frac{\gamma \sinh{\left(  t \Omega\right)}}{\Omega} \right)
\end{equation}

\begin{equation}\label{eq:omega}
    \Omega = \sqrt{\gamma^2 - 4 \lambda^2}
\end{equation}

We can now consider the case of Coupler-CPMG decay. Here we apply a periodic sequence of $\sigma_z$ gates to suppress the coupler noise. In the limit that the $\sigma_z$ gates are very short, the Hamiltonian that describes the Coupler-CPMG reads:
\begin{align}
    \label{eq:H_CPMG}
H_\mathrm{CPMG} = H(t) - \frac{\pi}{2}\sigma_z\sum_k\delta\left[t-(k+1/2)T_C\right]
\end{align}
Here the period $T_C$ is the duration of a sequence of $2m$ two-qubit gates. The pulses $\propto \sigma_z$ are the pulses of the difference of the qubit frequencies, and during these pulses in the experiment $g(t)=0$, so that the coupled noise is not accumulated. As in the conventional CPMG, the first refocusing pulse is applied at $T_C/2$, and the measurement is at $nT_C$, that is, the time interval between the last refocusing pulse and the measurement is $T_C/2$.

The time evolution operator of the system is
\begin{align}
    \label{eq:CPMG_propagator_general}
U_\mathrm{CPMG}(nT_C,0) = \mathcal{T}\exp\left[-i\int_0^{nT_C}dt H_\mathrm{CPMG}(t)\right] 
\end{align}
($\mathcal{T}$ is the time ordering operator). The operator $U_\mathrm{CPMG}$ can be simplified if one takes into account that $\sigma_z\sigma_x=-\sigma_x\sigma_z$, and therefore $\sigma_z\exp[-i\int H(t) dt]= \exp[i\int H(t) dt] \sigma_z$. One can then use in Eq.~(\ref{eq:CPMG_propagator_general}) that $\exp(-i\pi\sigma_z/2) = -i\sigma_z$ and move in the time-ordered operator $U_\mathrm{CPMG}$ all $\sigma_z$ at times $(2k+1)T_C$ to $(2k+2)T_C$ ($k=0,...,\lfloor(n-2)/2\rfloor$; for odd $n-1$ the last $\sigma_z$ is moved to $nT_C$). This gives
\begin{align}
    \label{eq:simplified_U_CPMG}
U_\mathrm{CPMG}(nT_C) =(-i)^{n-1}\exp\left[-i\int_{0}^{nT_C}h(t)H(t)\right],
\end{align}
where $h(t)$ is a filter function. It changes sign depending on $t$ being in the interval preceded by an even or odd number of refocusing pulses,
\begin{align}\label{eq:filter_func}
  h(t) = 1+2\sum_{m=1}^n(-1)^m \Theta[t-(m-1/2)T_C]
  \end{align}

We now consider the expectation value of the z-component of the Bloch vector $N_\mathrm{CPMG}(nT_C)= \langle Tr[\sigma_z U_\mathrm{CPMG}(nT_C)\rho(0)U_\mathrm{CPMG}^\dagger(nT_C)]\rangle$. Taking into account that $[H(t),H(t')]=0$, we can write the general expression for the observable in the form similar to that in the absence of the CPMG pulses   

\begin{equation}
    \label{eq:N_CPMG_general}
    N_\mathrm{CPMG}(n T_C) = \text{Re} \left(  \chi_\mathrm{CPMG}(n T_C)  \right)
\end{equation}
where
\begin{equation}\label{eq:cpmg_decay}
    \chi_\mathrm{CPMG}(t) = \langle e^{2 i \int_{0}^t h(\tau) \lambda(\tau) \xi(\tau) d \tau} \rangle
\end{equation}

There are several ways to calculate the average (\ref{eq:cpmg_decay}).They take advantage of  $\xi(t)$ being a Markov random process and of the property (\ref{eq:tele_recurr}). Here we start by  employing time ordered expansion of the telegraph noise to find an integro-differential equation for  $\chi_\mathrm{CPMG}(t)$,  

\begin{equation}
    \frac{d \chi_\mathrm{CPMG}(t)}{ d t} = -4 \lambda(t) h(t) \int_{0}^{t} e^{- 2 \gamma (t - t_1)} \lambda(t_1) h(t_1) \chi_\mathrm{CPMG}(t_1) d t_1 
\end{equation}
We define $\Lambda(t)$,
\begin{equation}
    \Lambda(t) = 2 \int_{0}^{t} e^{- 2 \gamma (t - t_1)} \lambda(t_1) h(t_1) \chi_\mathrm{CPMG}(t_1) d t_1 
\end{equation}
and arrive at the system of equations,
\begin{align}
\label{eq:matrix_differential}
    \begin{split}
    \frac{d}{dt} \begin{pmatrix} \chi_\mathrm{CPMG}(t)  \\\Lambda(t)  \end{pmatrix} =& A(t) \begin{pmatrix} \chi_\mathrm{CPMG}(t) \\\Lambda(t)  \end{pmatrix} \\
    A(t) =& \begin{pmatrix} 0 &  - 2 \lambda(t) h(t) \\ 2 \lambda(t) h(t) & -2 \gamma  \end{pmatrix}
\end{split}
\end{align}
Given that we consider the case  $\lambda(t)$=const and $h(t)=\pm 1$, the solution of Eq.~(\ref{eq:matrix_differential}) can be obtained using the transfer matrix approach based on the piece-wise solution within an interval where $h(t)=$const.

For $h=1$ we have
\begin{align}
    \label{eq:transfer_h_1}
    &\left(\begin{array}{c}
    \chi_\mathrm{CPMG}(t)\\
    \Lambda(t)
    \end{array}\right) = \hat X_{h=1}(t-t_1)
    \left(\begin{array}{c}
    \chi_\mathrm{CPMG}(t_1)\\
    \Lambda(t_1)
    \end{array}\right),\nonumber\\
    &\hat X_{h=1}(t) = e^{- \gamma t} \begin{pmatrix} \cosh{\left( \Omega t\right)} + \frac{\gamma}{\Omega} \sinh{\left( \Omega t\right)} & -\frac{2 \lambda \sinh{\left( \Omega t\right)}}{\Omega} \\ \frac{2 \lambda \sinh{\left( \Omega t\right)}}{\Omega} & \cosh{\left( \Omega t\right)} - \frac{\gamma}{\Omega} \sinh{\left( \Omega t\right)}\end{pmatrix},
\end{align}
where $\Omega$ is given by Eq.~\ref{eq:omega}.

The solution of Eq.~(\ref{eq:matrix_differential}) for $h=-1$ has the same form, except that $X_{h=1}$ has to be replaced with $\hat X_{h=-1}$. The expression for $\hat X_{h=-1}$ can be obtained from Eq.~(\ref{eq:transfer_h_1}) by replacing $\lambda \to -\lambda$. Alternatively, it can be written as
\begin{align}
    \label{eq:transfer_h_-1}
    \hat X_{h=-1}(t) =\hat Z \hat X_{h=1}(t)\hat Z, \quad 
  \hat Z = \begin{pmatrix}1 & 0 \\ 0 & -1\end{pmatrix}
\end{align}

The introduction of the matrix $\hat Z$ and the form of the solution for $h=\pm 1$ allows us to write the expression for the function $\chi_\mathrm{CPMG}(nT_C)$ in the form

\begin{equation}
\label{eq:direct_driven_env_evo}
    \chi_\mathrm{CPMG}(n T_C)  = \begin{pmatrix} 1 & 0\end{pmatrix}\hat Z{}^{\,(n\,\mathrm{mod}\, 2)} \left( \hat X_{h=1}(T_C/2) \hat Z \hat X_{h=1}(T_C/2)\right)^n \begin{pmatrix} 1 \\ 0\end{pmatrix} 
\end{equation}
We have
%\
\begin{align}
    \label{eq:XZX_matrix}
\hat X_{h=1}(T_C/2) \hat Z \hat X_{h=1}(T_C/2)=
e^{-\gamma T_C}\begin{pmatrix}
(\gamma/\Omega)\sinh \Omega T_C + q & -(4\lambda\gamma/\Omega^2)\sinh^2(\Omega T_C/2)\\
(4\lambda\gamma/\Omega^2)\sinh^2(\Omega T_C/2)&(\gamma/\Omega)\sinh \Omega T_C - q
\end{pmatrix},
\end{align}
where 
\begin{equation}
    q = \frac{-4 \lambda^2}{\Omega^2} + \frac{\gamma^2 \cosh(\Omega T_C)}{\Omega^2}
\end{equation}
The matrix (\ref{eq:XZX_matrix}) is not skew-Hermitian, although the off-diagonal matrix elements have opposite signs, since $\Omega$ is either real or imaginary. Therefore its eigenvectors are not orthogonal.

The eigenvalues of the matrix $X Z X$ are $- e^{-\alpha}$ and $e^{\alpha}$ where $\alpha$ is a solution of the equation 
\begin{equation}
    \sinh\alpha = \frac{\gamma}{\Omega}\sinh \Omega T_C
\end{equation}
The expression  for $ \chi_\mathrm{CPMG} $ is then easily found,

\begin{equation}\label{eq:driven_decay_env}
    \chi_\mathrm{CPMG}(n T_C)  = \left.
  \begin{cases}
    e^{-n \gamma T_C} \left( q \frac{\cosh{\left( n \alpha\right)}}{\cosh{\left( \alpha \right)}} + \sinh{\left( n \alpha \right)}\right), & \text{n odd} \\
    e^{-n \gamma T_C} \left( q \frac{\sinh{\left( n \alpha\right)}}{\cosh{\left( \alpha \right)}} + \cosh{\left( n \alpha \right)}\right), & \text{n even} \\
  \end{cases}
  \right\}
\end{equation}

It is clear how to extend the results to a superposition of independent RTN processes. For example, in the case of CPMG evolution,

\begin{align}
\label{eq:multi_TLS_trivia}
    \begin{split}
    \chi_\mathrm{CPMG}(t) =& \langle e^{2 i \int_{0}^t h(\tau) \sum_k \lambda_k(\tau) \xi_k(\tau) d \tau} \rangle \\
    =& \prod_k \langle e^{2 i \int_{0}^t h(\tau)  \lambda_k(\tau)\xi_k(\tau) d \tau} \rangle\\
    =& \prod_k\chi_\mathrm{CPMG}^{(k)}(t)
\end{split}
\end{align}
with $\chi_\mathrm{CPMG}^{(k)}(nT_C)$ given by Eq.~(\ref{eq:driven_decay_env}) for the parameters $\gamma,\lambda$ referred to the $k$th fluctuator.

If we set in Eq.~(\ref{eq:multi_TLS_trivia}) $h(t)=1$, it describes the result in the absence of refocusing pulses.

\section{The Shapiro-Loginov formula}\label{sec:shaprio_loginov}

The direct averaging approach used in section \ref{sec:direct_averaging} is useful when a simple, closed form solution for single trajectories of the noise is available. When this is not possible, other techniques must be used. One example of such a technique is the Shapiro-Loginov formula,  which is valid for  random processes with exponential correlators: 

\begin{equation}
\label{Eq:Shapiro}
\langle\xi(t)\frac{d}{dt}R[\xi(t),t]\rangle=\frac{d}{dt}\langle\xi(t)R[\xi(t),t]\rangle+2\gamma\langle\xi(t)R[\xi(t),t]\rangle 
\end{equation}
Here $R[\xi(t),t]$ is any functional of all histories $\{\xi(t'), t' \le t\}$ that lead to value $\xi(t')=\xi(t)$ at $t'=t$ and the angular brackets mean averaging over all possible noise instances.

First we consider the case of Coupler Ramsey decay. Let us write the Liouville - Von Neumann equation for the evolution of a density matrix of a qubit under a single source of the telegraph noise, and average over the noise instances using Shapiro-Loginov formula. We write this equation in the vectorized form, which for a two-level system is a well known Bloch equation 
for three components of the Bloch vector. For a two-qubit system in a single excitation subspace the x,y and z components are respectively: $P(t)=\bra{01}\rho\ket{10}+\bra{10}\rho\ket{01}$,  
$Q(t)=i (\bra{01}\rho\ket{10}-\bra{10}\rho\ket{01})$ and  $N(t)=\bra{01}\rho\ket{01}-\bra{10}\rho\ket{10}$. The Bloch equation reads:
\begin{equation}
\label{Eq:Bloch}
\dot\rho(t)=(L_q(t)+\lambda(t) \xi(t) L_x)\rho(t)
\end{equation}
where
\begin{equation}
L_q(t)=\left(\begin{array}{ccc}
0&\omega(t)&0\\
-\omega(t)&0&g(t)\\
0&-g(t)&0
\end{array}\right)
\end{equation}
is the regular part of the Liouvillian  with $\omega(t)$ and $g(t)$ being the detuning and coupling between the qubits, respectively. The second term represents a stochastic part of the Liouvillian and describes coupling of the qubit  with the noise of amplitude $\lambda(t)$. The operator
 \begin{equation}
 L_x=\left(\begin{array}{ccc}
0&0&0\\
0&0&1\\
0&-1&0
\end{array}\right)
\end{equation}
represents the noise coupling. The important property
of the telegraph noise is that $\xi^2(t)=1$, which substantially simplifies calculations. The second simplification comes from the Shapiro-Loginov formula.
As a first step, we average the Liouville equation and obtain:
\begin{equation}\label{eq:Bloch_ship_log}  
\frac{d}{dt}\langle\rho(t)\rangle=L_q(t)\langle\rho(t)\rangle+\lambda(t) L_x \langle \xi(t) \rho(t)\rangle
\end{equation}
Now we need to come up with an equation for  $\mu(t)=\langle \xi(t) \rho(t)\rangle$. To obtain this equation we multiply the Bloch equation
\eqref{Eq:Bloch} by $\xi(t)$ and replace $\langle\xi(t) \dot \rho(t)\rangle$ using Shapiro-Loginov formula \eqref{Eq:Shapiro}.  
After multiplication by $\xi(t)$ the second term in Eq.~\eqref{Eq:Bloch} will be proportional to $\langle\rho(t)\rangle\times \text{const}$  because  $\xi^2(t)=1$. This yields
\begin{equation}
\frac{d}{dt}\mu(t)+2\gamma\mu(t)=L_q(t)\mu(t)+\lambda(t) L_x \langle\rho(t)\rangle
\end{equation}
Therefore we obtained a closed system of  linear differential equations which can be expressed compactly in a matrix form
as
\begin{equation}
\label{Eq:Shapiro2q}
\left(\begin{array}{c}
\dot\rho(t)\\
\dot\mu(t)
\end{array}
\right)
={\cal L}(t)
\left(\begin{array}{c}
\rho(t)\\
\mu(t)
\end{array}
\right)
\end{equation}
where
\begin{equation}
{\cal L}(t)=\left(\begin{array}{cc}
L_q(t)&\lambda(t) L_x \\
 \lambda(t) L_x &L_q(t)-2\gamma I_q
 \end{array}
\right)
\end{equation}
is $6\times 6$ matrix comprised of four $3\times 3$ blocks, $I_q$ is a $3\times 3$ unit matrix. The system of linear differential equations \eqref{Eq:Shapiro2q} 
must be solved with the initial conditions $\mu(0)=\langle\xi(0)\rho(0)\rangle=\langle\xi(0)\rangle\langle\rho(0)\rangle=0$. From now on we will omit angular brackets for averaged quantities and assume that $\rho(t)$ is the averaged density matrix (like in Eq.~\eqref{Eq:Shapiro2q}) unless specified otherwise. 

It is seen from this equation that, as mentioned earlier, if $\lambda(t)$ has an oscillating component with typical period  $T_\lambda \ll \gamma^{-1},|\lambda|^{-1}$, this component leads to the terms in $\rho(t),\mu(t)$ oscillating with the same period. The amplitude of these terms is $\propto (\gamma T)^{-1}, |\lambda T|^{-1} \ll 1$. This justifies keeping only the time-independent term in $\lambda$.

The Liouvillian ${\cal L}(t)$ can be expressed in the operator form 
using Pauli matrices:
\begin{equation}
\label{Eq:CentralSpin2q}
{\cal L}(t)=I_{2} \otimes L_q(t)+\lambda\sigma_x\otimes L_x +\gamma\left(\sigma_z-I_2\right)\otimes I_q
\end{equation}
This form will be important for generalization to a  multi-fluctuator case. Let us split Liouvillian~\eqref{Eq:CentralSpin2q} into two parts such that ${\cal L}={\cal L}_q+{\cal L}_{noise}$,
where ${\cal L}_q$ is the first term in Eq.~\eqref{Eq:CentralSpin2q} and ${\cal L}_{noise}$ is the sum of the second and third terms.
If $\omega(t)=0$ the matrices ${\cal L}_q$ and ${\cal L}_{noise}$ commute and can be diagonalized separately.
\begin{equation} 
\varrho(t)=e^{{\cal L}_q G(t)}e^{{\cal L}_{noise} t}\varrho(0)
\end{equation}

\begin{equation}
    G(t) = \int_0^t g(\tau) d \tau
\end{equation}
Here $\varrho(t)$ is the generalized density matrix such that $\varrho^T(t)=(\rho(t),\mu(t))$. The Liouvillian 
${\cal L}_q$ describes uniform rotation of the Bloch vector in $yz$ plane while ${\cal L}_{noise}$ describes its dynamics in the corresponding rotating frame.
The explicit form of the matrix matrix ${\cal L}_{noise}$ is:
\begin{equation}
{\cal L}_{noise}=2 \left(
\begin{array}{cccccc}
 -\gamma  & 0 & 0 & 0 & 0 & 0 \\
 0 & -\gamma  & 0 & 0 & 0 & -\lambda  \\
 0 & 0 & -\gamma  & 0 & \lambda  & 0 \\
 0 & 0 & 0 & 0 & 0 & 0 \\
 0 & 0 & -\lambda  & 0 & 0 & 0\\
 0 & \lambda  & 0 & 0 & 0 & 0 \\
\end{array}
\right)
\end{equation}
The eigenvalues and eigenvectors of ${\cal L}_{noise}$ can be obtained by solving two quadratic equations. As a result, for  the initial condition $N(0)=1$ we obtain:
\begin{align}
N(t)&=\cos(2 G(t))\chi(t)\\
Q(t)&=\sin(2 G(t))\chi(t)\\
P(t)&=0
\end{align}
where
\begin{equation}
\chi(t)=e^{-\gamma t}\left(\cosh\left(\Omega t\right)+\frac{\gamma}{\Omega }\sinh\left(\Omega t\right)\right)
\end{equation}

Where $\Omega$ is as given in Eq.~\ref{eq:omega}. As required, this solution is identical to that found using the direct averaging approach.

The solution in the case of CPMG decay is found nearly identically to that in the case of Coupler Ramsey decay. It is most conveneint to break the evolution into stages. Identically to the previous section, we can use the Shapiro-Loginov equation to find the evolution during $g$ pulses for the generalized density matrix $\varrho$,

\begin{equation}
    \frac{d \varrho}{d t} = (i H_G(t) + L_G) \varrho
\end{equation}

Where $H_G(t)$ represents the driving from the coupler and $L_G$ the coupler noise. These matricies commute, so the solution is given in terms of two commuting propagators,

\begin{equation}
    \varrho(t + T_G) = X_G U_G \varrho(t)
\end{equation}

As above, the eigenvectors of $H_G(t)$ are time independant, so the propegator is found easily. Similarly, the evolution during the frequency pulses is given by,

\begin{equation}
    \varrho(t + T_P) = X_P U_P \varrho(t)
\end{equation}

We know that the coupler noise is approximately zero when the coupler is off, so the the operator $U_P$ represents the frequency pulse itself and $X_P$ represents the evolution of the uncoupled fluctuator during the pulse. Therefore, we can write the solution after $n$ repetitions of the drive pulse via,

\begin{equation}
    \varrho(n T_C) = (X_G U_G X_P U_P X_G U_G)^n \varrho(0)
\end{equation}

In the case where $U_P$ is a $\pi$ pulse and $T_P \to 0$, the above matrix reduces to a block diagonal form where at most 2 elements are coupled to each other. The evolution of the z component of the Bloch vector is given by,

\begin{equation}
    \frac{d }{d t} \begin{pmatrix} \langle N(t) \rangle \\ \langle \xi(t) P(t) \rangle \end{pmatrix} = (XZX)^n \begin{pmatrix} \langle N(t) \rangle \\ \langle \xi(t) P(t) \rangle \end{pmatrix}
\end{equation}

Where $P(t)$ is the x-component of the Bloch vector, and the matrix $XZX$ identical to the one found in Eq.~\ref{eq:direct_driven_env_evo}. The solution is then also given by Eq.~\ref{eq:driven_decay_env}, as required.

The problem of many fluctuators coupling to a single qubit must also be considered. Before we proceed with this we need to investigate  statistical properties of
of a product of many RTN variables $\xi_1(t)\xi_2(t)\dots \xi_N(t)$.
Let us consider a product of two independent telegraph variables $\xi(t)=\xi_1(t)\xi_2(t)$ and find its distribution based on the distributions of 
$\xi_1(t)$ and $\xi_2(t)$. It is sufficient to concentrate on only one initial condition, e.g. $\xi(0)=+1$. Based on the probability calculus for discrete and independent random variables we can express the probabilities $P_{++}^\xi(t)$ and $P_{+-}^\xi(t)$ as follows:
\begin{align}
P_{++}^\xi(t)&=\frac{1}{2}\left(P_{++}^{\xi_1}(t)P_{++}^{\xi_2}(t)+P_{--}^{\xi_1}(t)P_{--}^{\xi_2}(t)+P_{+-}^{\xi_1}(t)P_{+-}^{\xi_2}(t)+P_{-+}^{\xi_1}(t)P_{-+}^{\xi_2}(t)\right)
\label{Eq:P++}\\
P_{+-}^\xi(t)&=\frac{1}{2}\left(P_{++}^{\xi_1}(t)P_{+-}^{\xi_2}(t)+P_{--}^{\xi_1}(t)P_{-+}^{\xi_2}(t)+P_{+-}^{\xi_1}(t)P_{++}^{\xi_2}(t)+P_{-+}^{\xi_1}(t)P_{--}^{\xi_2}(t)\right),
\label{Eq:P+-}
\end{align}
where the factor $1/2$ is due to the fact that there are two equally probable and indistinguishable cases ($\xi_1(0),\xi_2(0)=+1,+1$ and $\xi_1(0),\xi_2(0)=-1,-1$ )
satisfying initial condition $\xi(0)=+1$, and the distributions functions for individual fluctuators are:
\begin{equation}
\label{Eq:Pi}
P_{\sigma_0,\sigma}^{\xi_i}(t)=\frac{1}{2}+\left(\delta_{\sigma_0,\sigma}-\frac{1}{2}\right)e^{-2\gamma_i t}
\end{equation}
Substitution of Eq.~\eqref{Eq:Pi} into Eqs.~\eqref{Eq:P++} and ~\eqref{Eq:P+-} yields:
\begin{align}
P_{++}^\xi(t)&=\frac{1}{2}\left(1+e^{-2(\gamma_1+\gamma_2)t}\right)
\label{P++2}\\
P_{+-}^\xi(t)&=\frac{1}{2}\left(1-e^{-2(\gamma_1+\gamma_2)t}\right)
\label{P+-2}
\end{align}
Eqs.~\eqref{P++2} and ~\eqref{P+-2}  ensure that any product of independent telegraph variables is also a telegraph variable with the switching rate $\gamma=\sum_i\gamma_i$. As such, we can apply Shapiro-Loginov formula to a product of any number of telegraph variables and repeat the procedure described in the previous section.

Now we are ready to describe a set of $M$ fluctuators coupled to our two-qubit  system. To understand the structure of the master equations let us imagine that $L_q=0$ and both  $L_x$ and $L_q$ are scalars. We make these assumptions only for instructive purposes because they are nonsensical in the context of Bloch equations. For the sake of simplicity we assume that $M=2$. Repeating the single fluctuator procedure described above, i.e. multiplying "master equations" sequentially by $\xi_1(t)$, $\xi_2(t)$,
and $\xi_1(t)\xi_2(t)$, using $\xi_i^2(t)=1$ and applying Shapiro-Loginov formula we obtain:
\begin{equation}
\label{Eq:Instructive}
\frac{d}{dt}\left(\begin{array}{c}\rho \\ \mu _1\\  \mu _2\\ \mu _{12}\end{array}\right)=
\left(
\begin{array}{cccc}
 0 & \lambda _2 & \lambda _1 & 0 \\
 \lambda _2 & -2 \gamma _2 & 0 & \lambda _1 \\
 \lambda _1 & 0 & -2 \gamma _1 & \lambda _2 \\
 0 & \lambda _1 & \lambda _2 & -2 \gamma _1-2 \gamma _2 \\
\end{array}
\right)
\left(\begin{array}{c}\rho \\ \mu _1\\  \mu _2\\ \mu _{12}\end{array}\right)
\end{equation}
where $\mu_1(t)=\langle \xi_1(t)\rho(t) \rangle$, $\mu_2(t)=\langle \xi_2(t)\rho(t) \rangle$ and
$\mu_{12}(t)=\langle \xi_1(t) \xi_2(t)\rho(t) \rangle$. It is straightforward now to rewrite the matrix in Eq.~\eqref{Eq:Instructive} 
in the operator form:
\begin{equation}
{\cal L}=\lambda_1\sigma_x\otimes I_2+\lambda_2 I_2 \otimes\sigma_x+2\gamma_1\sigma_z\otimes I_2+2\gamma_2 I_2 \otimes\sigma_z -2(\gamma_1+\gamma_2)I_4
\end{equation}
 As such we can associate the matrices  $\sigma_\alpha\otimes I_2$ and $I_2\otimes \sigma_\alpha$ with fluctuators 1 and 2 
 respectively. Generalization to any number of fluctuators and any qubit Liouvillian is  straightforward and we arrive at the 
 following Liouvillian of the system of $M$ fluctuators coupled to a two-qubit system:
\begin{equation}
\label{Eq:CentralSpin}
{\cal L}=I_{2^M} \otimes L_q(t)+ \sum_{i=1}^M\lambda_i\sigma_x(i)\otimes L_x + \sum_{i=1}^M\gamma_i\left(\sigma_z(i)-I_{2^M}\right)\otimes I_d, 
\end{equation}
where
\begin{equation} 
\sigma_\alpha(i)=I_2\otimes I_2\otimes  \dots \otimes \sigma_\alpha\otimes  \dots \otimes I_2
\end{equation}
and the Pauli matrix $\sigma_\alpha$ is exactly at $i_{th}$ position in this product. Eq.~\eqref{Eq:CentralSpin} can be interpreted as a central spin problem describing Ising-type interaction of the individual fluctuators (peripheral spins) with the two-qubit system
(central spin). In the case $\Delta(t)=0$ when the qubit Liouvillian commutes with the noise Hamiltonian this problem can be solved exactly. 

At first, one need to transform  equations of motion to a rotating frame associated with $g$. Then in the absence of the the z-component of the magnetic field ($\Delta(t)=0$) the fluctuators do not interact with each  other and eigenvalues of the operator $L_x$  are good quantum numbers. As such, each peripheral spin senses only the local ``magnetic field'' with $x$-component 
0, or $\pm 4\pi \lambda_i$,  imaginary $z$-component $2 i\gamma_i$. Therefore the Liouvillian can be diagonalized by rotating quantization axis of each spin to a local frame defined by this magnetic field. Since one of the filed components is imaginary this transformation is described by a hyperbolic rotation for each qubit:
\begin{equation}
z _i=\left(
\begin{array}{cc}
 \cosh \left(\frac{\beta _i}{2}\right) & i \sinh
   \left(\frac{\beta _i}{2}\right) \\
 -i \sinh \left(\frac{\beta _i}{2}\right) & \cosh
   \left(\frac{\beta _i}{2}\right) \\
\end{array}
\right),
\end{equation}
where 
\begin{equation}
\cosh(\beta_i)=\frac{\gamma_i}{\sqrt{\gamma_i^2-16\pi^2\lambda_i^2}}
\end{equation}
To diagonalize the Liouvillian \eqref{Eq:CentralSpin} we construct $2^M\times 2^M$ rotation matrices
\begin{equation}
\zeta(i)=I_2\otimes I_2\otimes  \dots \otimes z_i\otimes  \dots \otimes I_2
\end{equation}
where $z_i$ is at $i_{th}$ position in this product.
It is convenient to rewrite Eq.~\eqref{Eq:CentralSpin} as
\begin{equation}
\label{Eq:CentralSpin2}
{\cal L}_r(t)= L_q(t) \otimes I_{2^M} +L_x  \otimes \sum_{i=1}^M\lambda_i\sigma_x(i)  +I_d \otimes \sum_{i=1}^M\gamma_i\left(\sigma_z(i)-I_{2^M}\right) , 
\end{equation}
which corresponds to the master equation
\begin{equation}
\dot\varrho(t)=\varrho(t){\cal L}_r(t)
\end{equation}
Then we diagonalize $L_x$ to separate the blocks corresponding to its eigenvalues $\pm i$ and 0. 
The transformation, which completely diagonalizes ${\cal L}_r(t)$ in the rotating frame takes the form:
\begin{equation}
{\cal L}_{diag}=Z^\dagger{\cal U}_x^\dagger{\cal L}_r{\cal U}_xZ,
\end{equation}
where 
\begin{equation}
Z=(P_++P_-)\otimes\prod_{i=1}^M\zeta(i)+P_0\otimes I_{2^M},
\end{equation}
$P_\pm$  and $P_0$ are the projectors onto the eigenstates of $L_x$ with eigenvalues $\pm i$ and $0$, respectively,
${\cal U}_x=u_x\otimes I_{2^M}$ and 
\begin{equation}
u_x=\left(
\begin{array}{ccc}
 0 & 0 & 1 \\
 -\frac{i}{\sqrt{2}} & \frac{i}{\sqrt{2}} & 0 \\
 \frac{1}{\sqrt{2}} & \frac{1}{\sqrt{2}} & 0 \\
\end{array}
\right)
\end{equation}
is the matrix diagonalizing $L_x$. After this transformation the problem is reduced to solving $3\times 2^M$ linear differential equations and the final result corresponding to the initial condition $N(0)=1$ reads:
\begin{align}
N(t)&=\cos(2 G(t))\chi(t)\\
Q(t)&=\sin(2 G(t))\chi(t)\\
P(t)&=0
\end{align}
where
\begin{equation}
\chi(t)=\prod_k e^{-\gamma_k t}\left(\cosh\left(\Omega_k t\right)+\frac{\gamma_k}{\Omega_k}\sinh\left(\Omega_k t\right)\right)
\end{equation}

In real experiments, there is noise present other than the coupler noise we are focused on. For example, independent single qubit decay and dephasing may contribute significantly to what is seen in experimental measurements. Single qubit decay and dephasing can be modeled alongside classical coupler nosie via the Lindblad equation,

\begin{align}\label{eq:full_lindblad}
\begin{split}
    \frac{d \rho}{d t} =& -i [H(t), \rho] + L_t[\rho] +  L_l[\rho]\\
    L_t[\rho] =& \sum_{m=1}^2  \Gamma_m^{\phi}  \left( n_m n_m \rho(t) - \frac{1}{2} \{n_m n_m,  \rho(t)\}\right) \\
    L_l[\rho] =& \sum_{m=1}^2 \Gamma_m^{1}  \left( \sigma_m^- \rho(t) \sigma_m^+ - \frac{1}{2} \{ \rho(t), \sigma_m^+ \sigma_m^- \}\right)
\end{split}
\end{align}

Where $n_m$ is the number operator for qubit m and $\sigma_m^-$ and $\sigma_m^+$ are the annihilation and creation operators for qubit m, respectively. First we will focus on the free evolution problem. It is most convenient to divide the total Hamiltonian (\ref{eq:free_ham}) into two parts,

\begin{align}
\begin{split}
    H(t) =& H_C(t) + H_N(t) \\
    H_C(t)=& g(t) \sigma_x \\
    H_N(t) =& \lambda(t) \xi(t) \sigma_x
\end{split}
\end{align}

Let $U_C(t)$ be the unitary generated by $H_C$.

\begin{align}
    \begin{split}
        U_C(t) =& e^{-i \sigma_x G(t)} \\
        G(t) =& \int_0^t g(\tau) d \tau
    \end{split}
\end{align}

Eq.~\ref{eq:full_lindblad} can then be moved into the interaction picture of $U_C(t)$. Defining $\Tilde{\rho} = U^{\dagger}_C \rho U_C$ and noting that $[U_C, H_N] = 0$,

\begin{align}
\begin{split}
    \frac{d \Tilde{\rho}}{d t} =& -i [H_N(t), \Tilde{\rho}] + \Tilde{L}_t[\Tilde{\rho}] + \Tilde{L}_l[\Tilde{\rho}] \\
   \Tilde{L}_t[\Tilde{\rho}] =& \sum_{m=1}^2  \Gamma_m^{\phi}  \left( \Tilde{n}_m \Tilde{n}_m \Tilde{\rho} - \frac{1}{2} \{\Tilde{n}_m \Tilde{n}_m,  \Tilde{\rho}\}\right) \\
    \Tilde{L}_l[\Tilde{\rho}] =& \sum_{m=1}^2 \Gamma_m^{1}  \left( \Tilde{\sigma}_m^- \Tilde{\rho} \Tilde{\sigma}_m^+ - \frac{1}{2} \{ \Tilde{\rho}, \Tilde{\sigma}_m^+ \Tilde{\sigma}_m^- \}\right)
\end{split}
\end{align}

Where the notation $\Tilde{a} = U_C^{\dagger} a U_C$ is used for the jump operators. If this equation is expanded in the vectorized $\sigma_x$ basis , $\Tilde{\rho} = \begin{pmatrix}
\bra{+}\Tilde{\rho} \ket{+}, \bra{+}\Tilde{\rho} \ket{-}, \bra{-}\Tilde{\rho} \ket{+}, \bra{-}\Tilde{\rho} \ket{-}
\end{pmatrix}^T$,

\begin{equation}\label{eq:lindblad_vec}
    \frac{d \Tilde{\rho}}{d t} = \mathcal{L_{\text{lind}}}(t) \Tilde{\rho}
\end{equation}

\small
\begin{equation}
\mathcal{L_{\text{lind}}}(t)  = \frac{1}{4} \left(
\begin{array}{cccc}
  \left(-\Gamma _{\phi }-2 \Gamma _1\right) &  \Delta \Gamma _1 e^{-2 i G(t)} &  \Delta \Gamma _1 e^{2 i G(t)} & \Gamma _{\phi } \\
  \Delta \Gamma _1 e^{2 i G(t)} &  \left(-8 i \xi(t)  \lambda(t) -\Gamma _{\phi }-2 \Gamma _1\right) & e^{4 i G(t)} \Gamma _{\phi } &  \Delta \Gamma _1 e^{2 i G(t)} \\
 \Delta \Gamma _1 e^{-2 i G(t)} &  e^{-4 i G(t)} \Gamma _{\phi } &  \left(8 i \xi(t)  \lambda(t) -\Gamma _{\phi }-2 \Gamma _1\right) &  \Delta \Gamma _1 e^{-2 i G(t)} \\
 \Gamma _{\phi } &  \Delta \Gamma _1 e^{-2 i G(t)} &  \Delta \Gamma _1 e^{2 i G(t)} &  \left(-\Gamma _{\phi }-2 \Gamma _1\right) \\
\end{array}
\right)
\end{equation}
\normalsize

Where $\Gamma_1 = \Gamma_1^{1} + \Gamma_2^{1}$, $\Delta\Gamma_1 = \Gamma_2^{1} - \Gamma_1^{1}$ , and $\Gamma_{\phi} = \Gamma_1^{\phi} + \Gamma_2^{\phi}$. Note that this equation will yield the solution for a single trajectory of $\xi(t)$, since $\xi(t)$ is a random process. 

Several simplifications can be made to this equation. First of all, if the driving is periodic, than $\lambda(t)$ may be replaced with it's average value, as justified previously. The time dependance of $G(t)$ can be handled in several ways. Depending on the applied control, it may be justified to assume $g(t)$ is a constant $g$, in which case $G(t) \to g t$. This is useful because it would allow for analytical solution of the averaged equations. Here, we have a different physical limit that allows even further simplification. In our experiments, $g(t)$ is generally on the order of $10 \text{MHz}$, while $\lambda$, $\gamma$ and $\Gamma$ are generally less than 1 MHz. Therefore, these quickly rotating terms can be dropped which yields a set of ODEs for $\Tilde{\rho}$ in which the only time dependence in the coefficients comes from $\xi(t)$.

Next $\Tilde{\rho}$ must be averaged over $\xi(t)$. This can be done using the Shapiro-Loginov equation (as in section \ref{sec:shaprio_loginov}). We can arrive at a set of constant coefficient linear ODEs for the averaged $ \Tilde{\rho} $ and $\mu$ (as defined near Eq.~\ref{eq:Bloch_ship_log}),

\begin{equation}
\label{Eq:shap_white}
\left(\begin{array}{c}
\Dot{\Tilde{\rho}}\\
\dot\mu
\end{array}
\right)
={\mathcal{L}_{SL}}
\left(\begin{array}{c}
\Tilde{\rho}\\
\mu
\end{array}
\right)
\end{equation}

If $\Tilde{\rho}$ is given in the vectorized energy basis, $\Tilde{\rho} = \begin{pmatrix}
\bra{01}\Tilde{\rho} \ket{01}, \bra{01}\Tilde{\rho} \ket{10}, \bra{10}\Tilde{\rho} \ket{01}, \bra{10}\Tilde{\rho} \ket{10}
\end{pmatrix}^T$, $\cal L$ is given by,

\begin{equation}
        \mathcal{L}_{SL} = \begin{pmatrix}
    A & B \\ B & A - \gamma I 
    
    \end{pmatrix}
\end{equation}

\begin{equation}
    A =\frac{1}{8} \begin{pmatrix}
    - 4 \Gamma_1 - \Gamma_{\phi} & 0 & 0 & \Gamma_{\phi} \\ 
    0 & -4 \Gamma_1 - 3 \Gamma_{\phi} & -\Gamma_{\phi} & 0 \\
    0 & -\Gamma_{\phi} &  -4 \Gamma_1 -3 \Gamma_{\phi} & 0 \\
    \Gamma_{\phi} & 0 & 0 & -4 \Gamma_1 - \Gamma_{\phi} \\
    \end{pmatrix}
\end{equation}

\begin{equation}
    B = i \lambda \begin{pmatrix}
    0 & 1 & -1 & 0 \\ 1 & 0 & 0 & -1 \\ -1 & 0 & 0 & 1 \\ 0 & -1 & 1 & 0
    \end{pmatrix}
\end{equation}

$\mathcal{ L}_{SL}$ is analytically diagonalizable, so the dynamics can easily be computed in closed form. Particularly relevant to experiments is the normalized $\langle \sigma_z \rangle$ observable,

\begin{align}\label{eq:free_z_lind}
\begin{split}
    \frac{\bra{01}\Tilde{\rho}\ket{01} - \bra{10}\Tilde{\rho}\ket{10}}{\bra{01}\Tilde{\rho}\ket{01} + \bra{10}\Tilde{\rho}\ket{10}} =& e^{-\frac{1}{4} t (4 \gamma + \Gamma_{\phi})} \left( \cosh{\left(  t \Omega \right)} + \frac{\gamma \sinh{\left(  t \Omega\right)}}{\Omega}\right) \\
    \Omega =& \sqrt{\gamma^2 - 4 \lambda^2}
\end{split}
\end{align}

Note that this is simply the solution without including white noise (Eq.~\ref{eq:direct_ramsey}) with an additional exponentially decaying prefactor.

\begin{figure}[H]
    \centering
    \includegraphics[width=0.7\textwidth]{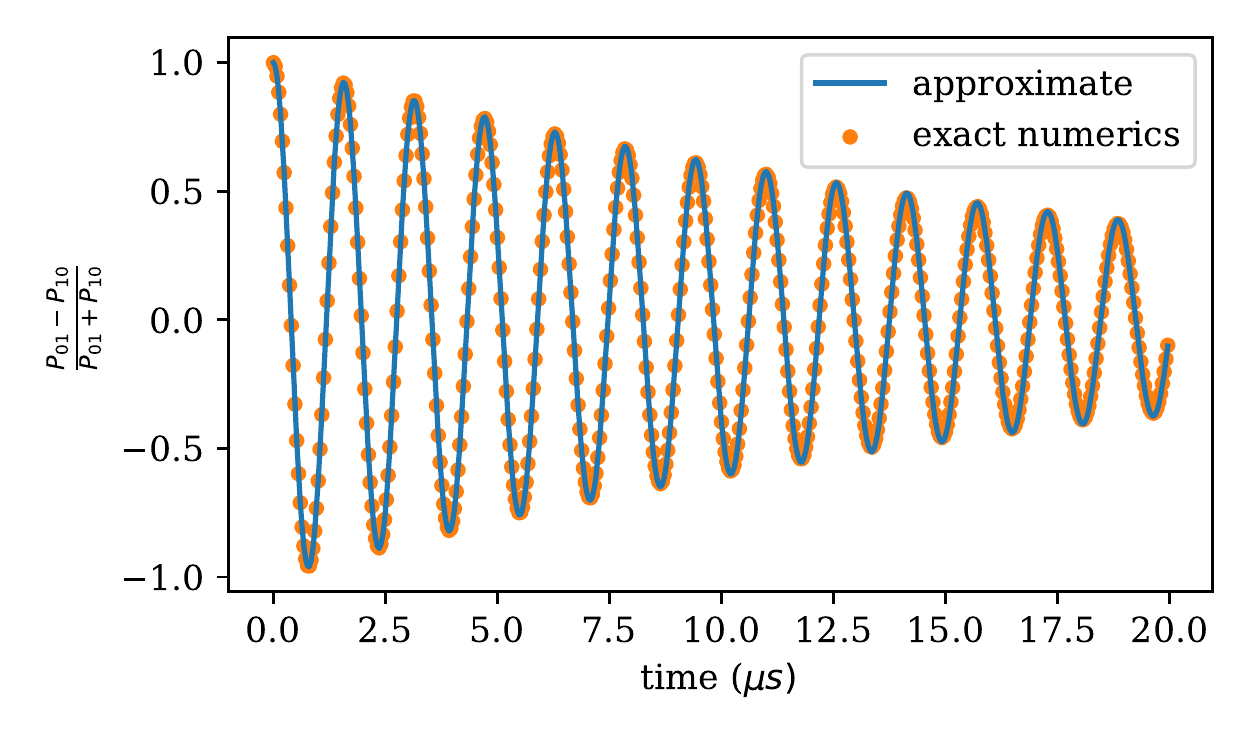}
    \caption{Comparison of the approximate solution for the decay envelope of the normalized $\langle \sigma_z \rangle$ in the case of free decay (Eq.~\ref{eq:free_z_lind}) to exact numerical simulation of the time dependant master equation constructed using eq. \ref{eq:lindblad_vec}. $g(t)$ was taken to be a $40 \text{ns}$ smoothed trapezoidal pulse with a maximum of $g_{max}$.  Parameter values used here are $\gamma = 0.05 \text{MHz}$, $\lambda = 0.3 \text{MHz}$, $g_{max} = 10 \text{MHz}$, $\Gamma_{\phi} = 0.1 \text{MHz}$, $\Gamma_{1} = 0.15 \text{MHz}$, and $\Delta\Gamma_{1} = 0.05 \text{MHz}$. This represents the strongest noise we ever see experimentally at an artifically lowered $g$ value, which should be a good stress test the approximation. The approximation seems to work well over a wide range of parameter values. The envelope function oscillates because these parameter values lead to an under-damped solution.}
    \label{fig:index_plot}
\end{figure}

The solution for CPMG pulse sequences can be found using the free decay solution. In line with section \ref{sec:direct_averaging}, if the qubit frequency pulses are instantaneous the effective Hamiltonian could be written $H(t) = h(t) (H_C(t) + H_N(t))$, where $h(t)$ is the filter function defined in Eq.~\ref{eq:filter_func}. In this case, the dynamics for a CPMG sequence of a given length would be given by

\begin{equation}
\label{eq:Shapiro2q}
\left(\begin{array}{c}
\Dot{\Tilde{\rho}}\\
\dot\mu
\end{array}
\right)
=\cal \dots L_{\text{even}} L_{\text{odd}} L_{\text{odd}} L_{\text{even}}
\left(\begin{array}{c}
\Tilde{\rho}\\
\mu
\end{array}
\right)
\end{equation}

Where $\cal L_{\text{even}} = \cal L(\lambda = \lambda)$ and $\cal L_{\text{odd}} = \cal L(\lambda = -\lambda)$

If we compute the normalized $\langle \sigma_z \rangle$ observable (as in Eq.~\ref{eq:free_z}) using this scheme, we find that it is equivalent to Eq.~\ref{eq:driven_decay_env} with a prefactor of $e^{-\frac{\Gamma_{\phi} t}{4}}$.

\section{Decay Under Gaussian Noise}\label{sec:gaussian_decay}

A bounded integral of  a Gaussian random process $x(t)$ is a Gaussian random variable $X(t)$,

\begin{equation}
    X(t) = \int_{0}^{t} g(\tau) x(\tau) d\tau
\end{equation}

\begin{equation}
    \int_0^t |g(\tau)| d\tau < \infty
\end{equation}
Therefore, if $\xi(t)$ is Gaussian noise, decay functions $\chi(t)$ or $\chi_\mathrm{CPMG}(t)$, as in Eq.~(\ref{eq:cpmg_decay}), are given by

\begin{equation}
    \chi(t) = \langle e^{2 i \int_{0}^t h(\tau) \xi(\tau) d \tau} \rangle =
    %e^{-\frac{\sigma^2}{2} } = }
    e^{-\Gamma(t)}
\end{equation}

\begin{equation}
    %\sigma^2}
    \Gamma(t) = 2 \int_{0}^t d \tau_1  \int_{0}^t d \tau_2 h(\tau_1) h(\tau_2) \langle \xi(\tau_1) \xi(\tau_2) \rangle
\end{equation}

For stationary $\xi(t)$ the correlator depends only on the time difference, $\langle \xi(\tau_1) \xi(\tau_2) \rangle =c(\tau_1 - \tau_2)$, and then $\Gamma(t)$ can be expressed in terms of the noise power spectrum $S(\omega)=S(-\omega)$,

\begin{equation}\label{eqn:gauss_decay}
    \Gamma(t) =  2 \int_{0}^{\infty} \frac{S(\omega)}{\pi} F(\omega, t) d \omega
\end{equation}

\begin{equation}
    c(\tau_1 - \tau_2) = \int_{-\infty}^{\infty} \frac{S(\omega)}{2 \pi} e^{i \omega (\tau_1 - \tau_2)} d \omega
\end{equation}

\begin{equation}
    F(\omega, t) = \int_0^t h(\tau_1) e^{i \omega \tau_1} d \tau_1 \int_0^t h(\tau_2) e^{-i \omega \tau_2} d \tau_2 
\end{equation}

In the case of Coupler Ramsey measurement $h(t) = 1$, and $F(\omega, t)$ is

\begin{equation}
    F(\omega, t) = 4 \frac{\sin^2{\left( \frac{\omega t}{2}\right)}}{\omega^2}
\end{equation}

In the case of 1/f noise $S(\omega) = \frac{\lambda^2}{\omega}$, and $\Gamma(t)$ is

\begin{equation}
    \Gamma(t) = - \frac{2 \lambda^2 \left( -1 + \cos{\left( \omega_m t\right) + \omega_m t (\omega_m t \: \text{Ci}(\omega_m t) - \sin{\left( \omega_m t\right)})}\right)}{\pi \omega_m^2}
\end{equation}
\begin{equation}
    \text{Ci}(z) = - \int_z^{\infty} \frac{\cos{(t)}}{t} d t 
\end{equation}
where  $\omega_m$ is the low-frequency cutoff of $S(\omega)$. 
In the experimentally relevant limit where $\omega_m t$ is small,

\begin{equation}\label{eq:gauss_free_decay}
    \Gamma(t) \approxeq \frac{\lambda^2 t^2}{\pi} (3 - 2 \gamma_{\text{Euler}} + 2 \ln{\left(\frac{1}{\omega_m t} \right)})
\end{equation}
Since $\ln{\left(\frac{1}{\omega_m t} \right)}$ is a slowly changing function, Eq.~\ref{eq:gauss_free_decay} produces approximately Gaussian decay,

\begin{equation}
    \chi(t) \approxeq e^{- \Gamma_G t^2}
\end{equation}

\begin{equation}
    \Gamma_g \propto \lambda^2
\end{equation}

We can now discuss the case of  decay in the presence of CPMG filtering. Let us consider a Gaussian noise with the power spectral density equal to that of a single RTN fluctuator:
 \begin{equation}
 \label{Eq:S(omega)}
 S(\omega)=\frac{4\lambda^2\gamma}{\omega^2+4\gamma^2}
 \end{equation}
 For an arbitrary CPMG sequence with $n$ echo pulses the decay envelope function is given by the standard expression:
 \begin{equation}
 \chi(n, T_C)=e^{-\Gamma(n,T_C)},
 \end{equation}

\begin{figure}[htbp]
\begin{center}
\includegraphics[width=0.5\linewidth]{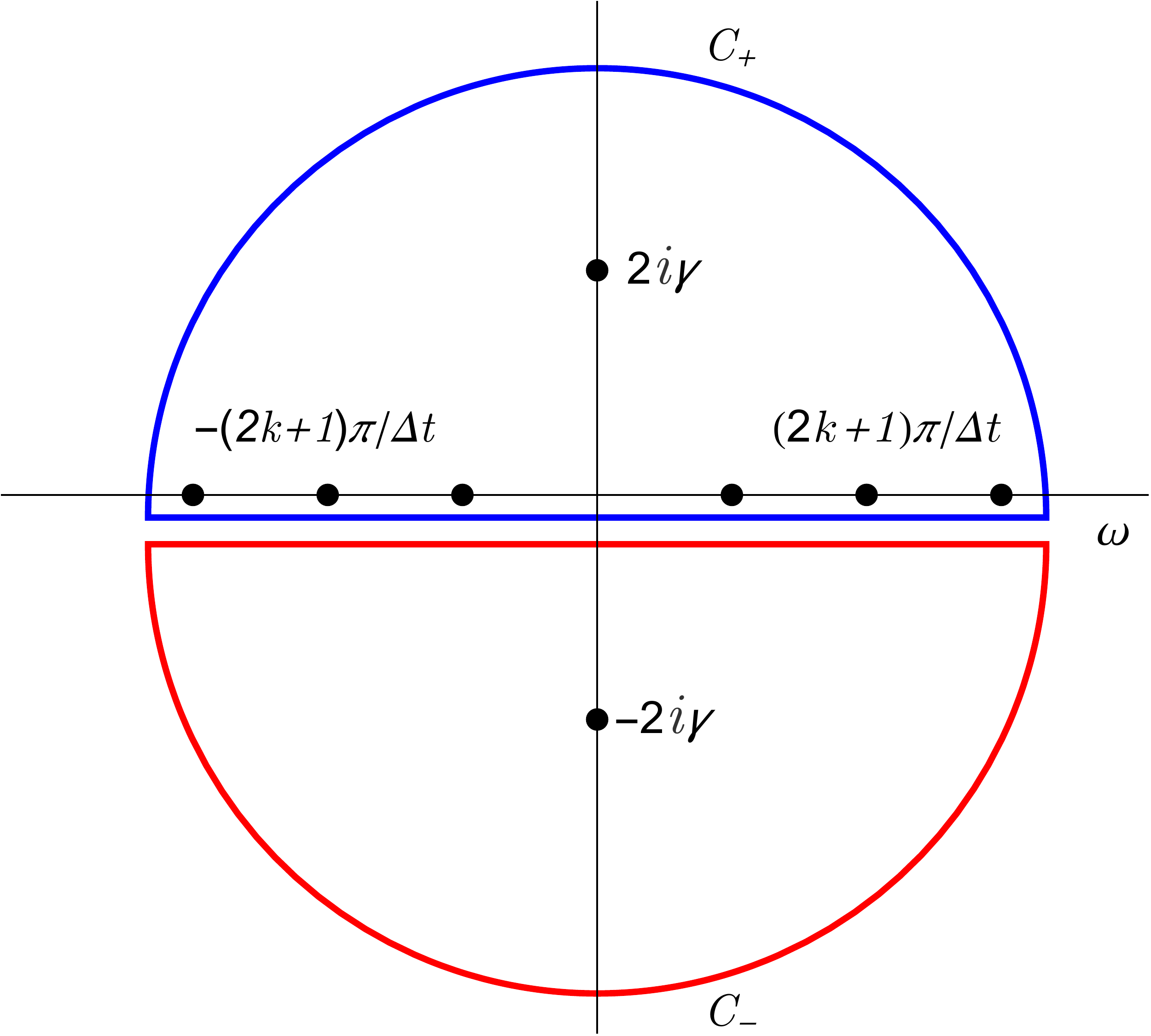} 
\end{center}
\caption{Integration contours for computing $\Gamma(n,T_C =\Delta t)$}
\label{Fig:contours}
\end{figure}  
The filter function in this case is:
\begin{equation}
\label{Eq:CPMG-Filter}
F(n,\omega T_C)=\frac{1}{\omega^2}\left(1-\frac{1}{\cos(\omega T_C/2 )}\right)^2\times\left\{\begin{array}{c} 
\sin^2(n \omega  T_C/2) \hskip 0.5 cm \text{if } n \text{ even}\\
\cos^2(n \omega  T_C/2)  \hskip 0.5 cm \text{if } n \text{ odd}
\end{array}
\right .
\end{equation}
Here $ T_C$ is time interval between two consecutive $\pi$-pulses of the $n$-pulse sequence such that the total sequence time $t=n T_C$.  Using Eqs.~\eqref{Eq:S(omega)} and~\eqref{Eq:CPMG-Filter} we can split the integral in Eq.~\eqref{eqn:gauss_decay} as follows: 
\begin{equation}
\label{Eq:Gamma-split}
\Gamma(n, T_C)=I_0( T_C)-\frac{1}{2}(-1)^n\left(I_+(n, T_C)+I_-(n, T_C)\right),
\end{equation}
where
\begin{align}
\label{Eq:I0}
I_0( T_C)&=\frac{1}{2\pi}\int_{-\infty}^\infty d\omega\left(\frac{16 \lambda ^2 \gamma 
}{\omega ^2 \left(4
   \gamma ^2+\omega ^2\right)}+ \frac{16 \lambda ^2 \gamma\left(
  \left(-1+\sec \left({\omega   T_C}/{2}\right)\right)^2-1\right)}{\omega ^2 \left(4
   \gamma ^2+\omega ^2\right)}\right)\\
\label{Eq:I+}   
I_+(n, T_C)&=\frac{1}{2\pi}\int_{-\infty}^\infty d\omega\frac{16 \lambda ^2 \gamma  e^{i
  n \omega T_C} \left(-1+\sec \left({\omega   T_C}/{2}\right)\right)^2}{\omega ^2 \left(4 \gamma
   ^2+\omega ^2\right)}\\
\label{Eq:I-}
I_-(n, T_C)&=\frac{1}{2\pi}\int_{-\infty}^\infty d\omega\frac{16 \lambda ^2 \gamma  e^{-i
  n \omega T_C} \left(-1+\sec \left({\omega   T_C}/{2}\right)\right)^2}{\omega ^2 \left(4 \gamma
   ^2+\omega ^2\right)}
\end{align}
The integrands in Eqs.~\eqref{Eq:I0}-\eqref{Eq:I-} have infinite series of poles at frequencies $\omega_k=\pm(2k+1)\pi/ T_C$ and additional two poles at $\pm 2i\gamma$ (see Fig~\ref{Fig:contours}). To evaluate the integrals $I_+$ and $I_-$ we use the contours
$C_+$ and $C_-$ respectively, as shown in Fig.~\ref{Fig:contours}, and obtain:
\begin{equation}
\label{Eq:I+final}
I_+(n, T_C) =\sum _{k=0}^{\infty } \frac{(-1)^{n+1}32 n \lambda^2  T_C}{\pi^2\gamma(2 k +1)^2 
   \left(1+{(2 k +1)^2 \pi ^2}/{(2 \gamma \Delta
   t})^2\right)}-\frac{4 \lambda ^2 e^{-2 n
   \gamma   T_C} \sinh ^4\left({\gamma 
    T_C}/{2}\right)}{\gamma ^2 \cosh ^2(\gamma 
    T_C)}
\end{equation}  
and
\begin{equation}
\label{Eq:I-final}
I_-(n, T_C)= -\frac{4 \lambda ^2 e^{-2 n
   \gamma   T_C} \sinh ^4\left({\gamma 
    T_C}/{2}\right)}{\gamma ^2 \cosh ^2(\gamma 
    T_C)}
\end{equation}
The integral $I_0$ can be computed by integrating the first term in Eq.~\eqref{Eq:I0} along the horizontal line proximate to the real axis and lying in the lower half-plane. Then we can close the contour ($C_-$) in the lower half-plane to evaluate the remaining integral
along this contour. This 
yields:
\begin{equation}
\label{Eq:I0final}
I_0(n, T_C)=-\frac{4 \lambda ^2  \sinh ^4\left({\gamma 
    T_C}/{2}\right)}{\gamma ^2 \cosh ^2(\gamma 
    T_C)}
\end{equation}
After evaluating the sum in Eq.~\eqref{Eq:I+final} and substituting Eqs.~\eqref{Eq:I+final}-~\eqref{Eq:I0final} into Eq.~\eqref{Eq:Gamma-split} we finally obtain:
\begin{equation}
\label{Eq:Gamma-final}
\Gamma (n,  T_C)=\frac{2 \lambda ^2 n}{\gamma ^2}
  \left (\gamma   T_C-\tanh (\gamma 
   T_C)\right)-\delta \Gamma
   (n, T_C),
\end{equation}
where
\begin{equation}
\label{Eq:deltaGamma}
\delta \Gamma 
   (n, T_C)=\frac{8 \lambda ^2 e^{-n\gamma T_C} \sinh ^4\left({\gamma 
    T_C}/{2}\right)}{\gamma ^2 \cosh ^2(\gamma 
    T_C)}\times\left\{\begin{array}{c} 
\sinh(n \gamma T_C) \hskip 0.5 cm \text{if } n \text{ even}\\
\cosh(n \gamma  T_C)  \hskip 0.5 cm \text{if } n \text{ odd}
\end{array}
\right .
\end{equation}
Eqs.~\eqref{Eq:Gamma-final} and~\eqref{Eq:deltaGamma} are exact for any Gaussian noise with a Lorentzian power spectrum and for any $n$. They also match  the weak-coupling limit of the CPMG formula for a non-Gaussian noise induced by a single random telegraph noise source. For large $n$ only the first term, proportional to $n$ is important. This term describes the well-known for the CPMG exponential decay for large $n$, with $\Gamma(n, T_C)\propto n$.

\begin{figure}[H]
    \centering
    \includegraphics[width=0.5\textwidth]{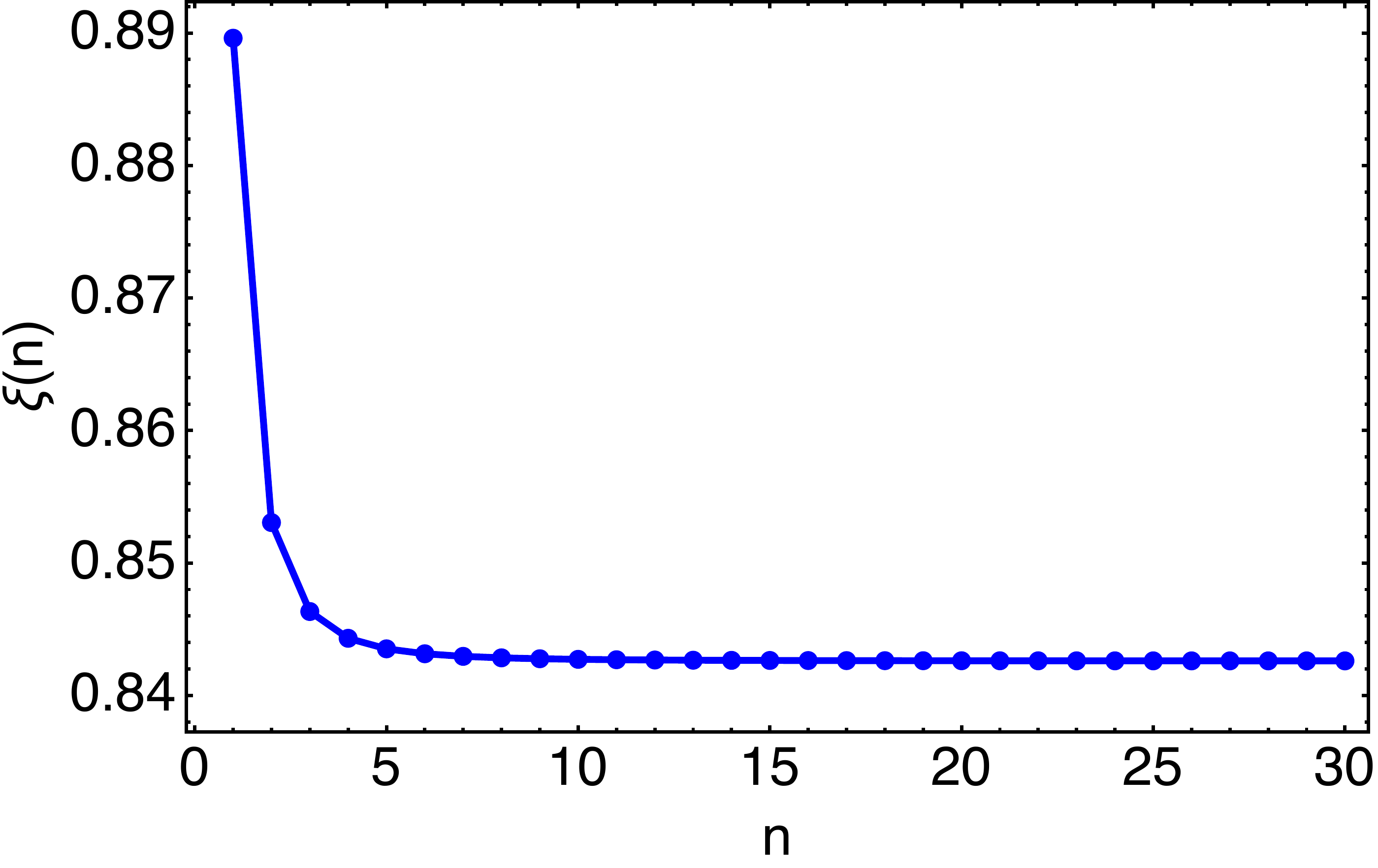}
   \caption{Dependence of $\Gamma'(n, T_C)$  on $n$ for $1/f$ noise}
   \label{Fig:xi_n}
\end{figure} 

For $1/f$-noise with $S(\omega) = \frac{\lambda^2}{\omega}$ we find
\begin{equation}
\frac{\partial\Gamma(n, T_C)}{\partial n}=2  T_C^2 \lambda ^2 \xi(n)\propto\omega_c^{-2},
\end{equation}
where
\begin{equation}
\xi(n)=
\int_0^{\infty }
   \left(\frac{x-\tanh (x)}{x^3}+\frac{8 e^{-2 n x}
   \text{sech}^2(x) \sinh
   ^4\left(\frac{x}{2}\right)}{x^2}\right) \, dx
\end{equation}
only weakly depends on $n$  (see Fig.~\ref{Fig:xi_n}) and quickly saturates at $\xi(\infty)=\int_0^\infty  \left(x-\tanh (x)\right)/{x^3}=0.8525$. Therefore, letting $T_C = \frac{t}{n}$

\begin{equation}
    \Gamma(t) \approxeq \frac{2 t^2 \lambda^2 \xi(n)}{n}
\end{equation}

This describes Gaussian decay with a rate inversely proportional to $n$. Therefore, under a Gaussian noise model we would expect that increasing the number of echo pulses that occur in time $t$ should always increase the amount of noise protection.

\section{The Smoothness of Decay Under Gaussian Noise}\label{sec:gaussian_smoothness}

The time derivative of the decay envelope under Gaussian noise is given by,

\begin{equation}
    \frac{d \chi(t)}{d t} = -e^{-\Gamma(t)} \frac{d \Gamma(t)}{d t}
\end{equation}
The first two terms in this expression are always positive for non-zero $S(\omega)$, as $F(\omega, t)$ is non-negative. Therefore, if zeros are to be present in the derivative of $\chi(t)$, the following condition must be met,

\begin{equation}
    0 = \frac{d \Gamma(t)}{d t}  = \int_{0}^{\infty} S(\omega) \frac{d F(\omega, t)}{d t} d \omega
\end{equation}

For an even CPMG sequence, $\frac{d F(\omega, t)}{d t}$ is given by,

\begin{equation}
    \frac{d F(\omega, t)}{d t} = \frac{\sin \left(\frac{t \omega }{2}\right) \left(\sec \left(\frac{t \omega }{2 n}\right)-1\right) \left(n \cos \left(\frac{t \omega }{2}\right) \left(\sec \left(\frac{t \omega }{2 n}\right)-1\right)+\sin \left(\frac{t \omega }{2}\right) \tan \left(\frac{t \omega }{2 n}\right) \sec \left(\frac{t \omega }{2 n}\right)\right)}{n \omega }
\end{equation}

\begin{figure}
    \centering
    \includegraphics[width=0.5\textwidth]{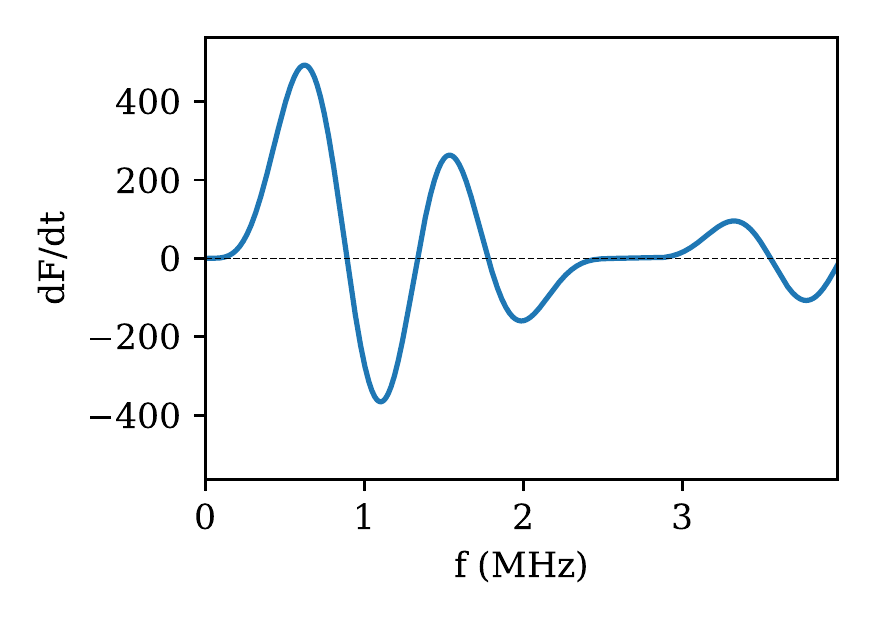}
    \caption{Plot of $ \frac{d F(\omega, t)}{d t}$ for n=2 and $t_z=1500 \text{ns}$, $f = \frac{\omega}{2 \pi}$.}
    \label{fig:deriv_ff}
\end{figure}

Where $F(\omega, t)$ was obtained by substituting $T_C = \frac{t}{n}$ into equation \ref{Eq:CPMG-Filter}. Since $F(\omega,t)$ has the form of a product of $\omega^{-2}$ and a function of $\omega t/n$, the derivative $dF/dt \propto (\omega n)^{-1}$. For $n=2$, the function $dF/dt$ has zeros at,

\begin{equation}
    \omega = \frac{8 \pi m}{t}, \: \frac{4 \left( \frac{-2 \pi}{3} + 2 \pi m\right)}{t}, \: \frac{4 \left( \frac{2 \pi}{3} + 2 \pi m\right)}{t}, \: m \in \mathbb{Z}
\end{equation}
The first positive zero is at $\omega_0 = \frac{8 \pi}{3 t_z}$, and $F(\omega, t_Z) > 0$ for $\omega < \omega_0$, as shown in figure \ref{fig:deriv_ff}.
It is clear from figure \ref{fig:deriv_ff} that in order for there to be a single zero in $ \frac{d F(\omega, t)}{d t}$ at $t = t_z \approxeq 1500 \text{ns}$, it is necessary that the noise must have much more power at frequencies larger than $\omega_0 \approxeq 1 \text{MHz}$ than at low frequencies. This requirement directly contradicts modern experimental observations of flux noise in SQUID-based devices \cite{bylander_11, siddiqi_12, yan_13, oliver_14} which find $1/f$ noise over a range extending well past $1 \text{MHz}$. Additionally, this high frequency power must be concentrated in the regions of frequency where $ \frac{d F(\omega, t)}{d t}$ is positive, which would mean $S(\omega)$ could not be smooth on the MHz scale. The additional requirements of periodic zeros and non-positivity of the derivative (as seen in the data and predicted by our best non-Gaussian model) make it harder still for a Gaussian model to explain this behavior.

\section{Fitting to Many CPMG Curves Simultaniously}\label{sec:many_fits}
 
\begin{figure}
    \centering
    \includegraphics[width=0.7\textwidth]{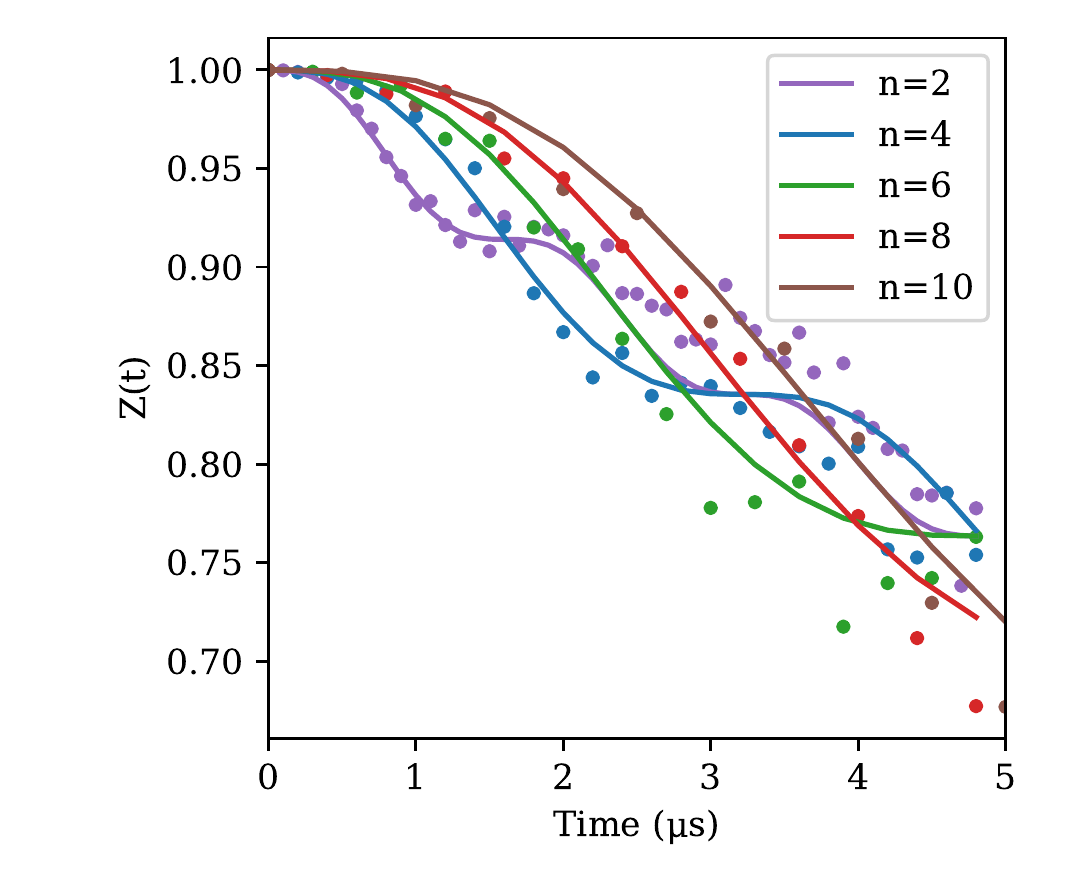}
    \caption{Fit of the 2 parameter single fluctuator model to 5 CPMG curves simultaneously. As in the main text, the x-axis is real time, $t=2mnt_g$. $m$ is varied to change the duration of a constant $n$ CPMG sequence.}
    \label{fig:5_fit}
\end{figure}

Figure \ref{fig:5_fit} shows a 2 parameter fit to CPMG curves for 5 different values of $n$.

\section{Echoing Strongly Coupled Fluctuators}\label{sec:echo_strongly_coupled}

Consider a single, far underdamped fluctuator, $\lambda >> \gamma$. Then,

\begin{equation}
    \Omega = i \overline{\omega}
\end{equation}

Where $\overline{\omega} = \sqrt{4 \lambda^2 - \gamma^2}\approx 2\lambda$ is a real number. If we then make the substitution $\frac{\gamma}{\overline{\omega}} = \epsilon$, equation \ref{eq:driven_decay_env} reduces to,
\begin{equation}
     \chi(n T_C)  \approxeq e^{-n T_C \gamma} e^{n \epsilon \sin{\left(T_C \overline{\omega}\right)}} 
\end{equation}

\begin{figure}
\begin{center}
\includegraphics[width=0.4\linewidth]{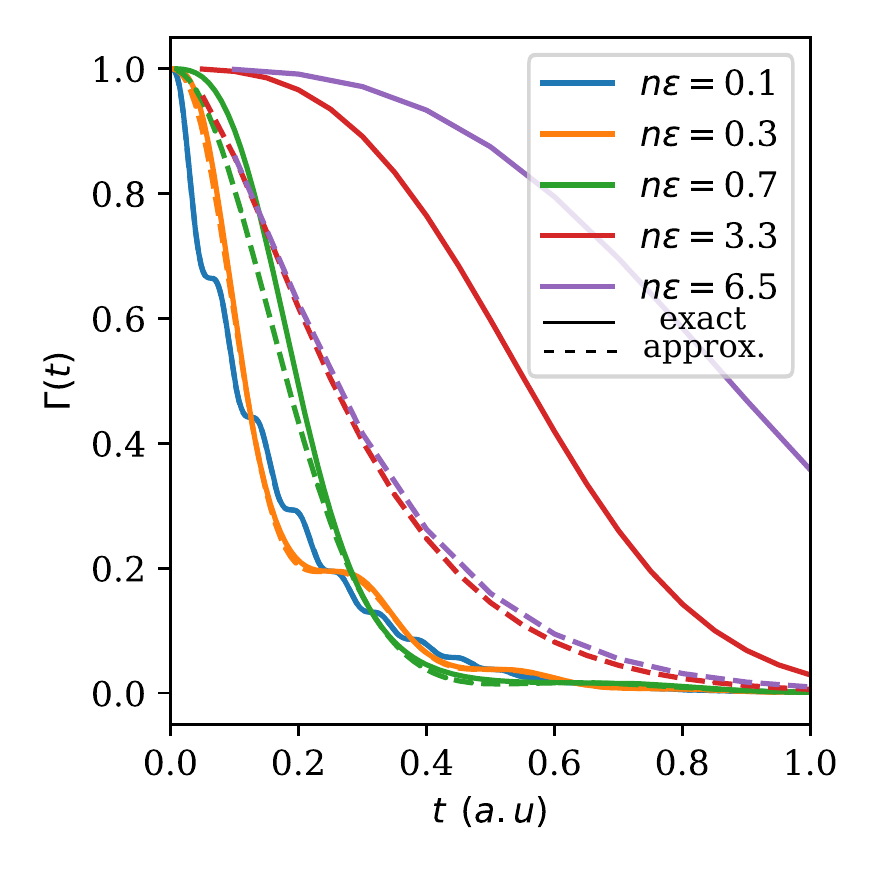}
\end{center}
\caption{Comparing exact and analytical solutions for the echoed decay envelope. Exact solutions are calculated using Eq.~\ref{eq:driven_decay_env} and are shown in solid lines, while the approximate solutions are calculated using Eq.~\ref{eq:approx_cpmg} and are shown by dashed lines. The approximation is almost exact for the two lowest values of $n \epsilon$, but has signifigant error for $n \epsilon > 1$.}
\label{fig:approx_vs_exact_cpmg} 
\end{figure}

If $n \epsilon << 1$, we can expand to leading order in $n \epsilon \sin{\left(\frac{t \omega}{n}\right)}$,
\begin{equation}\label{eq:approx_cpmg}
      \chi(n T_C) \approxeq e^{-n T_C \gamma} \left( 1 + n \epsilon \sin{\left( T_C \overline{\omega}\right)} \right)
\end{equation}
The implication of this equation is that for modest n, increasing the number of echoing pulses will not protect qubits from the dephasing effects of the strongly coupled fluctuator given that we keep $t=nT_C$ constant. Only when $n \epsilon > 1$ do we start to see significant protection from the noise. This is shown in figure \ref{fig:approx_vs_exact_cpmg}. This behaviour is qualitatively quite different from what happens with gaussian $1/f$-type noise, where increasing the number of echo pulses that occur in time $t$ will move the center of the filter function to higher frequencies and monotonically reduce dephasing.

\section{Alternative Explanation of Coupler CPMG}\label{sec:add_cpmg}

\begin{figure}
\begin{center}
\includegraphics[width=\linewidth]{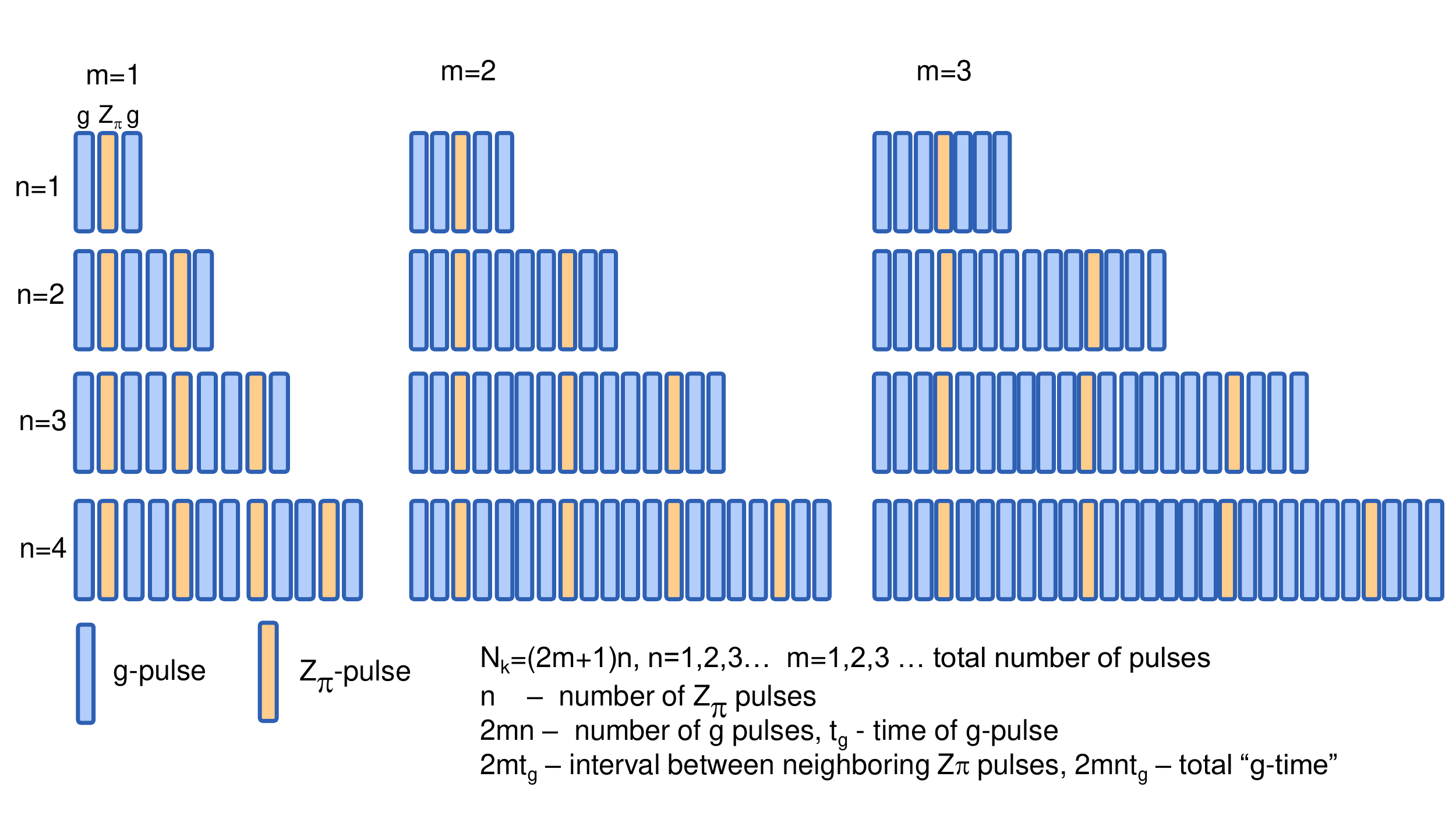}
\end{center}
\caption{Examples of the coupler CPMG sequence for different values of $n$ and $m$.}
\label{fig:cpmg_examples} 
\end{figure}
 %( link to the Appendix %)
%\putbib[sup]
%\end{bibunit}

\end{appendices}
%TC:endignore
\end{document}